\documentclass[12pt]{article}
\usepackage[a4paper, bindingoffset=0cm, hmargin={2.5cm, 2cm},vmargin={2.5cm, 2.5cm}]{geometry}
\usepackage[utf8]{inputenc}
\usepackage{bm}
\usepackage{comment}
\usepackage{amsmath}
\usepackage{amssymb}
\usepackage{amsthm}
\usepackage{amsfonts}
\usepackage{mathtools}
\usepackage{listings}
\usepackage{float}
\usepackage{stackengine}
\usepackage{tikz}
\usetikzlibrary{patterns}
\usetikzlibrary{decorations.markings}
\usepackage{graphicx}
\usepackage[sc,small]{caption}
\usepackage{subcaption}
\usepackage[retainorgcmds]{IEEEtrantools}
\usepackage{dsfont}

\usepackage{tikz-feynman}

\usepackage{url}

\usepackage{lineno,hyperref}

\usepackage{stackengine}
\def\subrangle#1{\stackengine{5pt}{}{$\!\scriptstyle #1$}{U}{l}{F}{F}{L}}

\renewcommand{\theequation}{\thesection.\arabic{equation}}

\newcommand{\req}[1]{(\ref{#1})}

\def\fc#1#2{\frac{#1}{#2}}

\newcommand{\nwc}{\newcommand}
\nwc{\ba}  {\begin{array}}
\nwc{\ea}  {\end{array}}
\nwc{\bdm} {\begin{displaymath}}
\nwc{\edm} {\end{displaymath}}

\nwc{\bea} {\begin{equation}\ba{lcl}}
\nwc{\eea} {\ea\end{equation}}

\nwc{\be} {\begin{equation}}
\nwc{\ee} {\end{equation}}

\nwc{\bda} {\bdm\ba{lcl}}
\nwc{\eda} {\ea\edm}

\nwc{\bc}  {\begin{center}}
\nwc{\ec}  {\end{center}}

\nwc{\ds}  {\displaystyle}

\nwc{\nn} {\nonumber}
\nwc{\nnn} {\nonumber \vspace{.2cm} \\ }
\nwc{\ra}{\rightarrow}
\nwc{\lra}{\longrightarrow}
\def\lf{\left}\def\ri{\right}

\nwc{\p} {\partial}

\def\ap{\alpha'}

\def\Ic{{\cal I}}

\def\z{\zeta}

\def\Oc{{\cal O}}

\def\si{\sigma}

\numberwithin{equation}{section}

\makeatletter
\let\save@rangle\rangle
\def\rangle{\save@rangle\@ifnextchar_{\expandafter\subrangle\@gobble}{}}
\makeatother

\allowdisplaybreaks

\let\originalleft\left
\let\originalright\right
\renewcommand{\left}{\mathopen{}\mathclose\bgroup\originalleft}
\renewcommand{\right}{\aftergroup\egroup\originalright}

\DeclareMathOperator{\Tr}{Tr}
\DeclareMathOperator{\sign}{sign}
\DeclareMathOperator{\KN}{KN}

\tikzset{->-/.style={decoration={
			markings,
			mark=at position .5 with {\arrow{>}}},postaction={decorate}}}

\tikzset{-<-/.style={decoration={
			markings,
			mark=at position .5 with {\arrow{<}}},postaction={decorate}}}
\newcommand{\tikzmark}[2]{\tikz[remember picture,baseline,inner sep=0pt,outer sep=0pt,anchor=base] \node (#1) {\ensuremath{#2}};}

\usetikzlibrary{decorations.pathreplacing}

\begin{document}

	\begin{center}
		{\bf\LARGE
			Scattering three closed strings  \\[3mm]
			off a D$\bm{p}$-brane in pure spinor formalism}
		
		\vspace{1cm}
		
		{\large\sc
			{ Andreas Bischof$^1$, Michael Haack$^2$ and Stephan Stieberger$^1$}
			
			\vspace{1cm}

			{\it\small
				$^1$Max--Planck--Institut f\"ur Physik,
Werner--Heisenberg--Institut, \\
				80805 M\"unchen, Germany\\\vspace{0.5cm}
				$^2$Arnold Sommerfeld Center for Theoretical Physics,\\ 
				 Ludwig--Maximilians--Universit\"at München, 80333 M\"unchen, Germany
			}
		}
		
	\end{center}
	\vspace{1.5cm}
	
	\begin{center}
		{\bf Abstract}\\
	\end{center}
	We compute the disk amplitude of three closed strings in the pure spinor formalism.
	Among others, this amplitude probes tree--level gravitational interactions in the presence of  D$p$-branes.
	After disentangling holomorphic and anti--holomorphic closed string coordinates on the disk by means of introducing monodromy phases we find a compact expression for the disk amplitude of three closed strings in terms of open superstring six--point amplitudes.
	Furthermore, we provide the low--energy expansion (in the inverse string tension) of our amplitude and discuss some relevant D$p$-brane couplings associated to it. 
	Finally, we write down 
	an expression for the general structure of the disk amplitude of any number $n_c$ of closed strings in terms of pure  open string  amplitudes involving $2n_c$ open strings.

	\vspace{1cm}
	\begin{flushright}
{\small  MPP--2023--79}\\
\small LMU-ASC 28/23
\end{flushright}
	
	\clearpage
	\tableofcontents
	\break
	\section{Introduction}
Disk amplitudes are especially interesting, because they give the first quantum corrections to the effective action in string theory. One example are higher derivative gravitational corrections to the Dirac-Born-Infeld action from disk amplitudes \cite{Bachas:1999um}. Hence, there is already a considerable body of literature on disk amplitudes. Given that our main interest in this paper is in closed string disk amplitudes, let us just recall some relevant computations involving closed strings. 
For instance, for the bosonic string the dilaton one-point function was calculated in \cite{Douglas:1986eu,Liu:1987nz} and a generalization to the superstring in the RNS and the pure spinor formalisms was given in \cite{Ohta:1987nq} and \cite{2pt}, respectively. Furthermore, the scattering process of two closed strings on the disk was already performed in
\cite{Klebanov:1995ni,Gubser:1996wt,Garousi:1996ad,Hashimoto:1996bf} in the RNS formalism -- see also \cite{Lust:2004cx,Becker:2011bw,Garousi:2017fbe,Aldi:2020dvw} for a detailed review -- and in \cite{2pt} in the pure spinor formalism. Disk amplitudes of three closed strings were considered for instance in \cite{Becker:2011ar,Becker:2016bzb,Mousavi:2018evq,ovsc}. However, most of these works (except \cite{ovsc}) consider the special case of scattering one RR field and two NSNS fields and all of them are formulated within the RNS framework. Finally, there are several computations of disk amplitudes involving both closed and open strings, see for example \cite{Garousi:1998fg,Garousi:2000ea,ovsc,Stieberger:2015vya} or \cite{Alencar:2008fy,Alencar:2011tk} in the RNS and pure spinor formalism, respectively (cf.\ also \cite{Stieberger:2014cea,Stieberger:2015qja,Stieberger:2016lng,Srisangyingcharoen:2020lhx}). 
\par

In this work we generalize the existing calculations in several ways. First, we use the pure spinor formalism for our closed string three-point disk amplitude. Thus, a large part of our results holds for NSNS, RR, RNS and NSR states (using the language of the RNS formalism), even though at some point we will specify our findings to the scattering of only NSNS states. Apart from the broad validity of its results, the pure spinor formalism offers further advantages. As in \cite{ovsc,Stieberger:2015vya} we make use of the relation between closed string $n$-point functions on the disk and open string $2n$-point functions on the disk, obtained via contour deformations of the corresponding world-sheet integrals (generalizing the sphere calculation of \cite{KLT} to the disk). The pure spinor formalism has proven very powerful when considering open string disk amplitudes, as its BRST cohomology structure allows to obtain very compact expressions \cite{6pt,npt_1,npt_2}. Hence, there is an extended literature on open string disk amplitudes in the pure spinor formalism that we can build on. 
\par

A second important generalization of our present analysis is that we work on the upper half plane while in \cite{ovsc} the calculations are set up on its double cover, i.e.\ the sphere. In \cite{Stieberger:2015vya} computations are established on the upper half plane but  only the case of one closed string coupling to an arbitrary number of open strings is considered explicitely. Working on the upper half plane has two consequences. On the one hand, working on the double cover implies that certain poles are missing in the final amplitude. In contrast, our result on the upper half plane displays all the expected poles. Moreover, we find a formula expressing the closed string three-point function in terms of only two independent partial open string amplitudes instead of six as originally anticipated in \cite{ovsc}, cf.\ \eqref{eq::amp_mon_4a} below.
\par

One motivation for our work, apart from the more formal aspects mentioned above, was phenomenological. We were wondering whether it might be possible to infer disk level corrections to the four-dimensional Einstein-Hilbert term from the three-point function of gravitons with only external polarizations. These could arise from the compactification of a disk level $e^{-\Phi} \epsilon_{10} \epsilon_{10} R^4$-term on a Calabi-Yau manifold with non-vanishing Euler number. Such a term is expected in the world-volume theory of D9-branes \cite{Green:2016tfs}. However, our results do not show any hints of such a disk level correction to the Einstein-Hilbert term. We have a few more comments on this point at the end of section \ref{InterpretLowEXP}.

Our work is organized as follows: In section \ref{sec::psf} we give a short introduction into the pure spinor formalism and review the relevant aspects for the computation of the closed string disk three-point function. In particular, we review the worldsheet degrees of freedom of the pure spinor formalism, define the massless vertex operators and give a scattering amplitude prescription for three closed strings on the disk. In section \ref{sec::integration} we use the monodromy relations of the worldsheet, analytical continuation and $PSL(2, \mathbb{R})$ transformations to express the closed string amplitude in terms of open string amplitudes. We then continue in section \ref{LowEXP} with an interpretation of our results. We perform an $\alpha'$-expansion, compare some of the leading terms to the DBI action as a consistency check and comment on the absence of disk-corrections to the Einstein-Hilbert action. We end with some concluding remarks and an outlook on how our three-point results generalize to higher points. Finally, five appendices contain some technical details. In appendix \ref{sec::block} we present building blocks to efficiently organize the kinematic content of the CFT correlation function using the cohomology of the BRST operator of the pure spinor formalism, following a similar analysis in \cite{6pt,npt_1} for the open string. In appendix \ref{sec::correlator} we explicitly perform the contractions in the three-point function and express the result in the building blocks discussed in appendix \ref{sec::block}. After performing an appropriate $PSL(2,\mathbb{R})$ transformation we reproduce the result of section \ref{sec::integration} in the main text. Hence, the calculation of this appendix serves as a consistency check of our findings. In appendix \ref{sec::phase} we give some details of the contour deformations. Appendix \ref{sec::psl2r} discusses the invariance of a disk correlator under $PSL(2, \mathbb{R})$ transformations and, finally, in appendix \ref{sec:symmetrization} we give arguments why certain poles are missing in the amplitude on the double cover. 

	\section{The pure spinor formalism}
\label{sec::psf}

Let us begin with a short introduction to some aspects of the pure spinor formalism. In this section, we present the worldsheet degrees of freedom and moreover outline the calculation of closed string tree level amplitudes on the disk $D_2$ in the pure spinor approach to superstring theory. We follow closely the introduction into the pure spinor formalism of \cite{2pt}.


\subsection{Matter and ghost CFT of the pure spinor formalism}\label{sec::pdf_matter}

In the pure spinor formalism the type IIB action of the worldsheet degrees of freedom is given by
\begin{IEEEeqnarray}{rCl}
	S=\frac{1}{2\pi}\int\mathrm d^2z\,\left(\frac12\partial X^m\overline\partial X_m+p_\alpha\overline\partial\theta^\alpha+\overline p_\alpha\partial\overline\theta^\alpha - w_\alpha\overline\partial\lambda^\alpha-\overline w_\alpha\partial\overline\lambda^\alpha\right)\ ,\label{eq::psfaction2}
\end{IEEEeqnarray}
where $X^m(z,\overline z),\theta^\alpha(z),p_\alpha(z);\overline\theta^\alpha(\overline z), \overline p_\alpha(\overline z)$ are the matter variables \cite{GS_cov,Siegel:1985xj} and the pure spinor ghosts $\lambda^\alpha(z), w_\alpha(z);\overline\lambda^\alpha(\overline z), \overline w_\alpha(\overline z)$ are the ghosts introduced by Berkovits \cite{Berkovits}. In order to ensure that the theory has vanishing central charges in $D=10$, the ghost field $\lambda^\alpha$, which is a bosonic spinor, satisfies the pure spinor constraint
\begin{IEEEeqnarray}l
	\lambda^\alpha\gamma_{\alpha\beta}^m\lambda^\beta=0\ ,\qquad m=0,\ldots,9\ ,\quad \alpha,\beta=1,\ldots,16\ ,\label{eq::ps_const}
\end{IEEEeqnarray}
where $\gamma^m_{\alpha\beta}$ are the symmetric $16\times16$ Pauli matrices in $D=10$. The right-moving spinors for type IIA have opposite chirality, which in this notation would be expressed by lowering or raising their indices.\par
It is convenient to introduce the composite fields
\begin{IEEEeqnarray}{rCl}
	\Pi^m&=&\partial X^m+\frac 1 2 (\theta\gamma^m\partial\theta)\ , \label{Pi_def} \\
	d_\alpha&=&p_\alpha-\frac 1 2 \left(\partial X^m+\frac 1 4 (\theta\gamma^m\partial\theta)\right)(\gamma_m\theta)_\alpha \ , \label{dalpha_def}
\end{IEEEeqnarray}
which are the supersymmetric momentum and the Green-Schwarz constraint. These conformal primaries of weight $h=1$ appear in the vertex operators of massless fields and, thus, play an important role in the calculation of scattering amplitudes in the pure spinor formalism, as we will review below. Because $\lambda^{\alpha}$ is a commuting $SO(1,9)$ Weyl spinor, there is a further $h=1$ field $N^{mn}(z)=\frac12 (\lambda\gamma^{mn}w)$, which is the ghost contribution to the Lorentz current.\par
Furthermore, the holomorphic energy momentum tensor $T$ is given by
\begin{IEEEeqnarray}{rCl}
	T(z)=-\frac12\partial X^m\partial X_m-p_\alpha\partial\theta^\alpha+w_\alpha\partial\lambda^\alpha \ ,
	\label{eq::psfemt}
\end{IEEEeqnarray}
with a similar expression for the anti-holomorphic energy momentum tensor.\footnote{Of course, all the formulas in the rest of this subsection have an obvious antiholomorphic counterpart.}

For the computation of any scattering amplitude we will need the following OPEs \cite{Berkovits, Berkovits:2002zk}:
\begin{IEEEeqnarray}{C}
	\begin{IEEEeqnarraybox}[][c]{l}	
	\IEEEstrut
	\begin{IEEEeqnarraybox}[][c]{rClCrCl}
		\IEEEstrut
		X^m(z,\overline z)X^n(w,\overline w)&=&-\eta^{mn}\ln{|z-w|^2}\ , & \qquad & p_\alpha(z)\theta^\beta(w) & = & \frac{\delta_\alpha^{\hphantom\alpha\beta}}{z-w}\ , \\
		\Pi^m(z)\Pi^n(w) & = & \frac{-\eta^{mn}}{(z-w)^2}\ , & \quad & d_\alpha(z) d_\beta(w) & = & -\frac{\gamma^m_{\alpha\beta}\Pi_m(w)}{z-w}\ , \\
		d_\alpha(z)\Pi^m(w) & = & \frac{(\gamma^m\partial\theta)_{\alpha}(w)}{z-w}\ , & \quad & d_\alpha(z)\theta^\beta(w) & = & \frac{\delta^{\hphantom\alpha\beta}_\alpha}{z-w}\ ,
		\IEEEstrut
	\end{IEEEeqnarraybox}\\
	\IEEEstrut
	N^{mn}(z)N^{pq}(w)= 2 \frac{\eta^{p[n}N^{m]q}(w)-\eta^{q[n}N^{m]p}(w)}{z-w}-6\frac{\eta^{m[q}\eta^{p]n}}{(z-w)^2}\ .
	\end{IEEEeqnarraybox}
\label{psfOPE}
\end{IEEEeqnarray}
Furthermore, we can obtain the physical spectrum of the pure spinor superstring from the cohomology of the BRST operator, which takes the rather simple form \cite{Berkovits}
\begin{IEEEeqnarray}{C}
	Q=\oint\frac{\mathrm{d}z}{2\pi i}\,\lambda^\alpha(z) d_\alpha(z)\ .
	\label{BRST}
\end{IEEEeqnarray}
The BRST charge is nilpotent $Q^2=0$ as can be verified by using the OPE of $d_\alpha$ with $d_\beta$ in \eqref{psfOPE} and the fact that the ghost field $\lambda$ satisfies the pure spinor constraint \eqref{eq::ps_const}.\par
We can write the action of the conformal weight one fields $\Pi^m$ and $d_\alpha$ on a generic superfield $\mathcal V$ as
\begin{IEEEeqnarray}{rCl}
	d_\alpha(z) \mathcal V (X(w),\theta(w))&=&\frac{D_\alpha \mathcal V(X(w),\theta(w))}{z-w}\ ,\\
	\Pi^m(z)\mathcal V(X(w),\theta(w))&=&-\frac{ik^m\mathcal V(X(w),\theta(w))}{z-w}\ ,
\end{IEEEeqnarray}
where $D_\alpha=\partial_\alpha+\frac12(\gamma^m\theta)_\alpha\partial_m$. Therefore, the BRST charge acts on a superfield $\mathcal V(X,\theta)$ as  $Q\mathcal V=\lambda^\alpha D_\alpha\mathcal V.$


\subsection{Massless vertex operators in the pure spinor formalism}
\label{sec_Vops}
In string theory a scattering process can be described by a punctured Riemann surface, where each puncture represents a vertex operator position that corresponds to the creation or annihilation of a string state. Here, we are interested in the scattering amplitudes of closed strings on the disk $D_2$, which can be mapped to the upper half plane $\mathbb H_+$ by a conformal transformation. So we actually always mean the upper half plane $\mathbb H_+$ when talking about the disk. \par
Moreover, we focus on massless states of closed strings, which are subject to 
\begin{IEEEeqnarray}l
	k^2=0\ ,
\end{IEEEeqnarray}
where $k$ is the momentum of the string state.
The vertex operator for such a closed string state at the position $z$ splits into a direct product of left- and right-moving open string vertex operators
\begin{IEEEeqnarray}{C}
	V^{(a,b)}(z, \overline z) =  V^{(a)}\left(z\right) \overline V \, \! ^{(b)}\left(\overline z\right)\ , \qquad a,b \in \{ 0,1 \}\ ,
	\label{closed_vops}
\end{IEEEeqnarray}
where \cite{Berkovits}
\begin{IEEEeqnarray}{rCl}
	V^{(0)}(z)\equiv V(z)&=&\left[\lambda^\alpha A_\alpha(X,\theta)\right](z)\ , \label{V0} \\
	V^{(1)}(z)\equiv U(z)&=&\biggl[\partial\theta^\alpha A_\alpha(X,\theta)+\Pi^m A_m(X,\theta)+d_\alpha W^\alpha(X,\theta)+\frac12N^{mn}\mathcal F_{mn}(X,\theta)\biggl](z)\nonumber\\\label{V1}
\end{IEEEeqnarray}
are related to the massless open string vertex operators. The vertex operator $V^{(0)}$ is BRST closed, which means that
\begin{IEEEeqnarray}{rCl}
	Q V^{(0)} = 0\ .
	\label{QV0}
\end{IEEEeqnarray}
This is equivalent to putting the superfield $A_\alpha$ on-shell. Moreover, the vertex operator $V^{(1)}$ is BRST exact and therefore fulfils 
\begin{IEEEeqnarray}{rCl}
	QV^{(1)}=\partial V^{(0)}\ .
	\label{QV1}
\end{IEEEeqnarray}
Hence, it is in the BRST cohomology once we integrate it over the world-sheet. Thus, we call $V^{(0)}$ and $V^{(1)}$ the unintegrated and integrated vertex operator, respectively.\footnote{In the literature the unintegrated and integrated vertex operator are often denoted by $V$ and $U$, respectively, but in \cite{2pt} a different notation was introduced. There the open string vertex operators are denoted by $V^{(0)}$ and $V^{(1)}$ and the closed string vertex operator $V^{(a,b)}$ can simultaneously contain an integrated and unintegrated factor after gauge fixing. \label{VUV0V1}} Analogous statements hold for the right-moving part of \eqref{closed_vops}.\par
In \eqref{V0} and \eqref{V1}, the massless modes of the vertex operator are described by the spacetime superfields  $A_\alpha, A_m, W^\alpha$ and $\mathcal F_{mn}$ (the superfields of super-Maxwell theory). 
The super Yang-Mills (SYM) fields $A_m, W^\alpha$ and $\mathcal F_{mn}$ in \eqref{V1} are not independent. Rather, they are the field strengths given by \cite{Berkovits:2002zk}
\begin{IEEEeqnarray}l
	\begin{IEEEeqnarraybox}[][c]{rCl}
	\IEEEstrut
	A_m & = & \frac18\gamma^{\alpha\beta}_mD_\alpha A_\beta \ ,\\
	W^\alpha & = & \frac1{10}\gamma_m^{\alpha\beta}(D_\beta A^m-\partial^mA_\beta)\ ,\\
	\mathcal F_{mn} & = & \partial_mA_n-\partial_nA_m\ .
	\IEEEstrut
	\end{IEEEeqnarraybox}
\end{IEEEeqnarray}
The equations of motion of the superfields are given by the following set of equations \cite{Berkovits:2002zk,Witten:1985nt}
\begin{IEEEeqnarray}{l}
	\begin{IEEEeqnarraybox}[][c]{rClCrCl}
		\IEEEstrut
		2 D_{(\alpha} A_{\beta)} & = & \gamma^m_{\alpha \beta} A_m\ , & \qquad & D_\alpha A_m & = & (\gamma_m W)_\alpha + \partial_m A_\alpha\ , \\
		D_\alpha {\cal F}_{mn} & = & 2 \partial_{[m} ( \gamma_{n]} W)_\alpha\ , & \quad & D_\alpha W^\beta & = & \frac14 (\gamma^{mn})_\alpha \, \! ^\beta  {\cal F}_{mn} \ .
		\IEEEstrut
	\end{IEEEeqnarraybox}
	\label{eq::eoms_superfields}
\end{IEEEeqnarray}
Moreover, each of the superfields is a functional of $X$ and $\theta$. We can organize the $X^m$-dependence of the superfields into plane waves with momentum $k^m$ and we can perform an expansion of the superfields in $\theta$. 
Using the gauge choice $\theta^\alpha A_\alpha(X,\theta)=0$, the expansions can be found for instance in \cite{Schlotterer:2011psa}. For example, for the bosonic spacetime degrees of freedom (of a vector field with polarization vector $\xi_m$) it takes the following form:\footnote{The expansion in \cite{Schlotterer:2011psa} is more general including also fermionic spacetime degrees of freedom. Moreover, the momenta in \eqref{expansion} are real, i.e$.$ they differ from the corresponding momenta of \cite{Schlotterer:2011psa} by a factor of $i$.}
\begin{IEEEeqnarray}l
	\begin{IEEEeqnarraybox}[][c]{rCl}
	A_{\alpha}(X,\theta)&=&e^{ik\cdot X}\left\{\frac{\xi_m}{2}(\gamma^m\theta)_\alpha-\frac{1}{16}(\gamma_p\theta)_\alpha(\theta\gamma^{mnp}\theta)i k_{[m}\xi_{n]} +\mathcal O(\theta^5)\right\}\ ,\\
	A_m(X,\theta)&=&e^{ik\cdot X}\left\{\xi_m-\frac14 ik_p(\theta\gamma_m^{\hphantom{m}pq}\theta)\xi_q +\mathcal O(\theta^4)\right\} \ , \\
	W^\alpha(X,\theta)&=&e^{ik\cdot X}\left\{-\frac12ik_{[m}\xi_{n]}(\gamma^{mn}\theta)^\alpha +\mathcal O(\theta^3)\right\}\ ,\\
	\mathcal{F}_{mn}(X,\theta)&=&e^{ik\cdot X}\left\{2ik_{[m}\xi_{n]}-\frac12ik_{[p}\xi_{q]}ik_{[m}(\theta\gamma_{n]}^{\hphantom{n]}pq}\theta) +\mathcal O(\theta^4)\right\}\ .
	\IEEEstrut
	\end{IEEEeqnarraybox}
	\label{expansion}
\end{IEEEeqnarray}
It is sufficient to display the expansions only up to a certain order in $\theta$ that is relevant for the computation of a scattering amplitude. All higher orders will not contribute, because they drop out due to the zero mode prescription of the pure spinor formalism, cf$.$ section \ref{sec::zmp}. \par
We have separated $X^m(z, \overline z)$ into left- and right-movers, i.e$.$ $X^m(z, \overline z) = X^m(z)  + \overline X^m(\overline z)$ such that the plane wave factor of the superfields in \eqref{expansion} depend holomorphically on $z$, i.e$.$ $X^m = X^m(z)$. Nevertheless, the full closed string vertex operator \eqref{closed_vops} contains a factor $e^{i k \cdot X (z, \overline z)}$,
because $\overline V \, \! ^{(b)}$ in \eqref{closed_vops} is obtained from \eqref{V0}, \eqref{V1} and \eqref{expansion} by replacing the holomorphic fields $X(z), \theta(z), \lambda^\alpha (z)$ with their antiholomorphic counterparts $\overline X(\overline z), \overline \theta(\overline z), \overline \lambda^\alpha (\overline z)$ and by exchanging the polarisation vector $\xi_m$ with $\overline \xi_m$. Hence, the plane wave factors of the holomorphic part $V \, \! ^{(a)}$ and the antiholomorphic part $\overline V \, \! ^{(b)}$ of a closed string vertex operator combine into $e^{i k \cdot X (z, \overline z)}$. Moreover, the polarization tensor is given by $\epsilon_{mn} = \xi_m \otimes \overline \xi_n$.\par
For closed string amplitudes on the sphere the corresponding conformal Killing group $PSL(2,\mathbb{C})$ can be used to fix three closed string vertex operators completely leaving all others integrated. Thus, in that case the choice $a=b$ is possible and is always made in the literature. But when calculating scattering amplitudes of closed strings on the disk we allow $a\neq b$ in \eqref{closed_vops}, because the conformal Killing group $PSL(2,\mathbb R)$ of the disk does not allow to fix the positions of two closed string vertex operators completely. Therefore, for a disk amplitude involving only closed strings we have to allow the possibility $a \neq b$ \cite{2pt}. This possibility was also discussed in \cite{Alencar:2011tk,Grassi:2004ih}.

Due to the boundary of the disk the left- and right-moving part of a closed string vertex operator are not independent any more, i.e$.$  the boundary of the disk imposes an interaction between the holomorphic and antiholomorphic fields. For the computation of the three-point function of closed strings on the disk we can use the doubling trick to rewrite the right-moving part of the vertex operator \eqref{closed_vops} in order to allow for a unified treatment of the left- and right-movers.
Concretely, in the following we consider type IIB theory in a flat ten dimensional spacetime, which contains a D$p$-brane that is spanned in the $X_1\times X_2\times\ldots\times X_p$ plane. As usual, we use the fact that the D-brane is infinitely heavy in the small coupling regime, i.e$.$ it can absorb an arbitrary amount of momentum in the $X_{p+1},\ldots,X_9$ directions transversal to the D-brane. Thus, momentum is effectively only conserved along the D-brane in the perturbative regime that we are working in. 

Left- and right-movers separately have the standard correlators on the upper half plane 
\begin{IEEEeqnarray}{l}
	\begin{IEEEeqnarraybox}[][c]{rCl}
		\IEEEstrut
		\langle X^m(z)X^n(w)\rangle&=&-\eta^{mn}\ln(z-w)\ ,\\
		\langle p_{\alpha}(z)\theta^\beta(w)\rangle&=&\frac{\delta_\alpha^{\hphantom{\alpha}\beta}}{z-w}\ ,\\
		\langle w_{\alpha}(z)\lambda^\beta(w)\rangle&=&\frac{\delta_\alpha^{\hphantom{\alpha}\beta}}{z-w}\ ,
		\IEEEstrut
	\end{IEEEeqnarraybox}
	\label{eq::correlators}
\end{IEEEeqnarray}
where the antiholomorphic part is analogous. At the boundary of $\mathbb H_+$, i.e$.$ at the real axis, the first $p+1$ components of the world-sheet fields satisfy Neumann  boundary conditions and the remaining $9-p$ components Dirichlet boundary conditions. Both of these boundary conditions impose non-vanishing correlators between the holomorphic and antiholomorphic parts of the fields. We can simplify the calculations by employing the doubling trick, i.e$.$ we replace the right moving spacetime vectors and spacetime spinors by
\begin{IEEEeqnarray}{l}
	\begin{IEEEeqnarraybox}{rl}
	\IEEEstrut
	\text{vectors: }&\overline X^m(\overline z)=D^m_{\hphantom mn}X^n(\overline z)\ ,\\ 
	\text{spinors: }&\overline\Psi^\alpha(\overline z) = M^\alpha_{\hphantom{\alpha}\beta}\Psi^\beta(\overline z) \quad \text{ or } \quad \overline{\Psi}_\alpha(\overline z) = ((M^T)^{-1}) _{\alpha}^{\hphantom{\alpha}\beta}\Psi_\beta(\overline z)\ ,\IEEEeqnarraynumspace
	\label{eq::replace}
	\IEEEstrut
	\end{IEEEeqnarraybox}
\end{IEEEeqnarray}
with $\Psi^\alpha \in \{ \theta^\alpha, \lambda^\alpha\}$ and $\Psi_\alpha \in \{ p_\alpha, w_\alpha\}$ and constant matrices $D$ and $M$.\footnote{A priori, one would introduce two matrices $M$ and $N$ that account for the boundary conditions of the fermions, one for each chirality. Here we already used the result of \cite{2pt} that $N=(M^T)^{-1}$.} Effectively, this corresponds to extending the world-sheet fields to the entire complex plane. This allows us to use only the correlators in \eqref{eq::correlators}, leading to
\begin{IEEEeqnarray}{rClCrCl}
		\IEEEstrut
		\langle X^m(z)\overline X^n(\overline w)\rangle&=&-D^{mn}\ln(z-\overline w)\ ,\nonumber\\
		\langle p_{\alpha}(z)\overline\theta^\beta(\overline w)\rangle&=&\frac{M^{\beta}_{\hphantom{\beta} \alpha}}{z-\overline w}\ ,&\quad&\langle \overline p_{\alpha}(\overline z)\theta^\beta(w)\rangle&=&\frac{((M^T)^{-1})_\alpha^{\hphantom{\alpha}\beta}}{\overline z-w}\ ,\label{eq::correlator2ptdisk}\\
		\langle w_\alpha(z)\overline \lambda^\beta(\overline w)\rangle&=&\frac{M^{\beta}_{\hphantom{\beta} \alpha}}{z-\overline w}\ ,&\quad&\langle \overline w_\alpha(\overline z)\lambda^\beta(w)\rangle&=&\frac{((M^T)^{-1})_\alpha^{\hphantom\alpha\beta}}{\overline z-w}\ .\nonumber
\end{IEEEeqnarray}
The matrix
\begin{IEEEeqnarray}{l}
	D^{mn}=\left\{\begin{matrix} \eta^{mn} \quad&m,n\in\{0,1,\ldots,p\} \\
		-\eta^{mn}\quad&m,n\in\{p+1,\ldots,9\} \\
		0 \quad& {\rm otherwise} \end{matrix}\right. 
	\label{eq::boundary_matrix}
\end{IEEEeqnarray} 
is the same as in the RNS formalism \cite{Garousi:1996ad,Hashimoto:1996bf}, but the matrix $M$ here  differs from the RNS version due to the different spinor representations of the RNS and the PS formalisms. A detailed discussion of the matrix $M$ and its properties can be found in \cite{2pt} and a similar analysis in the context of the RNS formalism can be found for instance in \cite{Garousi:1996ad}. 
As described above, only the momentum parallel to the brane is conserved, because D-branes are infinitely heavy objects, that can absorb momentum in the transverse direction. We can introduce a parallel and transverse momentum
\begin{IEEEeqnarray}{l}
	{k_i}_\parallel=\frac12(k_i+D{\cdot}k_i)\ ,\qquad	{k_i}_\perp=\frac12(k_i-D{\cdot}k_i)\ ,
\end{IEEEeqnarray}
so that for momentum conservation we have
\begin{IEEEeqnarray}{l}
	\sum_{i=1}^N {k_i}_\parallel=0\ .\label{eq::mom_con}
\end{IEEEeqnarray}
Making the replacements \eqref{eq::replace} in the right-moving parts of the vertex operators \eqref{closed_vops} one can show that \cite{2pt}
\begin{IEEEeqnarray}{rCl}
	\overline V^{(0)}(\overline z)&=&\biggl[ \lambda^\alpha A_\alpha[D{\cdot}\overline\xi,D{\cdot}k](X,\theta)\biggl](\overline z)\ ,
	\label{eq::redefinedvertex1}\\
	\overline V^{(1)}(\overline z)&=&\biggl[ \overline\partial\theta^\alpha A_\alpha[D{\cdot}\overline\xi,D{\cdot}k](X,\theta)+\Pi^mA_m[D{\cdot}\overline\xi,D{\cdot}k](X,\theta)\nonumber\\
	&& + d_\alpha W^\alpha[D{\cdot}\overline\xi,D{\cdot}k](X,\theta)+\frac12N^{mn}{\mathcal F}_{mn}[D{\cdot}\overline\xi,D{\cdot}k](X,\theta)\biggl](\overline z)\ ,
	\label{eq::redefinedvertex2}
\end{IEEEeqnarray}
i.e.\ the doubling trick amounts to substituting every antiholomorphic superfield by its holomorphic counterpart and simultaneously multiplying the polarisation vector and momentum with $D$ to account for the boundary conditions.
To simplify the notation we will drop the explicit dependence of the superfields on the polarisation vector and momentum, but instead introduce the following notation
\begin{IEEEeqnarray}{rCl}
	\mathcal{V}_{\overline \imath}(\overline z)\equiv\overline{\mathcal{V}}_i[\overline\xi,k](\overline X(\overline z),\overline \theta(\overline z))=\mathcal{V}_i[D{\cdot}\overline\xi,D{\cdot}k](X(\overline z),\theta(\overline z)) \ ,
\end{IEEEeqnarray}
where ${\mathcal{V}} \in \{A_\alpha, A_m, W^\alpha, {\mathcal F}_{mn}\}$ and $i$ denotes the label of an external string state. Moreover, we will also use this notation for the vertex operators. To summarize this: An overlined label indicates that the field or vertex operator originates from the right-moving part of a string state after employing the doubling trick \eqref{eq::replace}.


\subsection{Calculating correlators of closed strings on the disk}
\label{sec::zmp}

The prescription to calculate open string amplitudes on the disk is well known and tested in the pure spinor formalism \cite{5pt,6pt,npt_1} and therefore also closed string amplitudes on the sphere are straightforwardly calculated using the KLT relations \cite{KLT}. Both world-sheets do not have any moduli and their conformal Killing vectors (CKVs) allow the fixing of three closed or open vertex operators, respectively. For an $N$-point function of closed strings on the disk we can only gauge fix one and a half vertex operators -- i.e.\ three of the real position coordinates of the vertex operators. This is due to the fact that the disk has only three real CKVs, which allow to gauge fix one real position each. For example, we can take the first two vertex operator insertions on the disk at the points $z_1=x_1+iy_1$ and $z_2=x_2+iy_2$ and fix those to the positions $x_1=0,y_1=1$ and $x_2=0$ and keep the integration over $y_2=y$. As discussed in \cite{2pt} this leads to the following $N$-point prescription
\begin{IEEEeqnarray}{rCl}
	\mathcal{A}_{D_2}^{\rm closed}(1,\ldots,N)&=&2i g_c^N T_p \int_0^1\mathrm{d}y\,\biggl\langle V^{(0,0)}_1(i,-i)V^{(0,1)}_2(iy,-iy)\prod_{j=3}^{N}\int_{\mathbb{H}_+}\mathrm{d}^2z_j\,V_j^{(1,1)}(z_j,\overline{z}_j)\biggl\rangle \nonumber\\
	&=&2i g_c^N T_p \int_0^1\mathrm{d}y\,\biggl\langle V_1(i)\overline V_1(-i)V_2(iy)\overline U_2(-iy)\prod_{j=3}^{N}\int_{\mathbb{H}_+}\mathrm{d}^2z_j\,U_j(z_j)\overline U_j(\overline{z}_j)\biggl\rangle\ ,\nonumber\\\label{eq::treepres}
\end{IEEEeqnarray}
where we switched notation in the second line, cf.\ footnote \ref{VUV0V1} above. We have also used the independence of the localization of the integrated vertex operator, see appendix C of \cite{2pt}. Moreover, the above gauge fixing of the vertex operator position restricts the integration over $z_2$ and $\overline z_2$ to the purely imaginary axis. But we don't integrate over the entire imaginary axis, because we have to limit the integration to the moduli space of a punctured disk (as before we are actually talking about the upper half of the complex plane when referring to the disk). To find the moduli space we consider a disk with two vertex operators at $z_1$ and $z_2$. We define the two transformations 
	\begin{IEEEeqnarray}l
		f_{\pm}(z;z_1,z_2)=
		\frac{(x_2-x_1) y_1 z +\left((x_1-x_2)x_2+(y_1 y_\pm-y_2)y_2\right)y_1}{((x_1-x_2)^2+(y_2-y_1 y_\pm)y_2)z-x_1^3+2x_1^2x_2+x_2y_1^2-x_1 (x_2^2+y_1^2+y_2^2-y_1 y_2 y_\pm)}\nonumber\\
	\end{IEEEeqnarray}
with
	\begin{IEEEeqnarray}l
		y_\pm=\frac{\left((x_1-x_2)^2+y_1^2+y_2^2\right)\pm\sqrt{4(x_1-x_2)^2y_1^2+\left((x_1-x_2)^2-y_1^2+y_2^2\right)^2}}{2y_1 y_2}\ ,\nonumber\\
	\end{IEEEeqnarray}
where $x_i \in ]-\infty,\infty[$ and $y_i \in [0,\infty[$  are the real and imaginary parts of $z_i$. For two different points $z_1$ and $z_2$ in the upper half plane, $f_\pm$ are $PSL(2, \mathbb{R})$ transformations. These transformations map $z_1$ to $i$ and $z_2$ to $iy_\pm$. One now observes that ($i$) $y_- \in [0,1]$ and $y_+ \in [1, \infty[$ and ($ii$) for $x_1 = x_2$ all values in the interval $[0,1[$ (or $]1, \infty[$) are assumed for $y_-$ (or $y_+$) for particular values of $y_1$ and $y_2$ (the limiting value 1 would require $y_1=y_2$, i.e.\ $z_1 = z_2$, which we excluded). Together these two facts imply that the moduli space of a disk with two closed string punctures is contained in the interval $[0,1[$ (or alternatively in $]1, \infty[$). Focusing for concreteness on the case of the interval $[0,1[$, the question remains whether a disk with punctures at $i$ and $i y$ could be mapped via a $PSL(2, \mathbb{R})$ transformation to a disk with punctures at $i$ and $i y'$ with $y \neq y'$ and both $y,y' \in [0,1[$. If that were the case, the two values $y$ and $y'$ would describe the same punctured disk and the moduli space would be smaller than $[0,1[$. However, it can easily be shown that this is not the case. The two points $i$ and $i y$ (with $y \in [0,1[$) can be mapped via a $PSL(2, \mathbb{R})$ transformation to the points $i$ and $i y'$ (with $y' \in [0,1[$) only if $y'=y$. Hence the moduli space is indeed $[0,1[$ (or equivalently $]1, \infty[$). \par  
The bracket $\langle\ldots\rangle$ of the correlator in \eqref{eq::treepres} includes also a zero mode prescription for $\lambda^\alpha$ and $\theta^\alpha$, see below. Moreover, $g_c$ is the closed string coupling and $T_p$ the D$p$-brane tension. 
At tree level only $\lambda^\alpha,\theta^\alpha$ and $X^m$ contain zero modes, because they have conformal dimension zero. All other fields, which have conformal weight $h=1$, have no zero modes on the disk \cite{DHoker:1988pdl}. To evaluate the correlator in \eqref{eq::treepres} we would integrate out the non-zero modes first.
A systematic method to do so is presented in appendix \ref{sec::block}.\par
Finally, the evaluation of the $X^m$ zero modes gives a momentum preserving $\delta$-function. Moreover, we are left with an expression in the pure spinor superspace that contains only the zero modes of $\lambda^\alpha$ and $\theta^\alpha$
\begin{IEEEeqnarray}{l}
	\biggl\langle V_{1}(z_{1})V_1(\overline z_1) V_2(z_2) \overline U_2(\overline z_2)\prod_{i=3}^{N} U_i(z_i) \overline U_i(\overline z_i)\biggl\rangle=\left\langle\lambda^\alpha\lambda^\beta\lambda^\gamma f_{\alpha\beta\gamma}(\theta;z_i,\overline z_i)\right\rangle_0\ ,
	\label{eq::zero_mode_expression}
\end{IEEEeqnarray}
with the gauge fixed vertex operator positions $(z_1,\overline z_1,z_2)=(i,-i,iy)$. The subscript on $\langle \ldots \rangle_0$ indicates that the bracket on the right hand side denotes a zero mode prescription, as all the non-zero modes are already integrated out. The functional $f_{\alpha\beta\gamma}(\theta;z_i,\overline z_i)$ is a composite superfield of the external states; it also contains the kinematic content of those states and is therefore $\alpha'$ dependent. The explicit form of $f_{\alpha\beta\gamma}(\theta;z_i,\overline z_i)$ in terms of the superfields $A^i_\alpha,A^i_m,W^\alpha_i$ and $\mathcal F^i_{mn}$  is determined by the OPE contractions while the dependence on $\alpha'$ follows from the momentum expansion. The argument of $\langle \ldots \rangle_0$ in \eqref{eq::zero_mode_expression} has a finite power series expansion of the enclosed superfields in $\theta^\alpha$ and it was argued in \cite{Berkovits} that only terms involving five powers of $\theta$ and three powers of $\lambda$ contribute: Given that the tensor product of three $\lambda$ and five $\theta$ contains a unique scalar, which is the unique element of the cohomology of the BRST operator in the pure spinor formalism at $\mathcal{O}(\lambda^3\theta^5)$, all terms of this type are proportional to each other \cite{Mafra:2008gkx} and are determined by\footnote{Concerning the normalization we follow the convention used for instance in \cite{Mafra:2014gja}.}
\begin{IEEEeqnarray}{rCl}
	\langle(\lambda\gamma^m\theta)(\lambda\gamma^n\theta)(\lambda\gamma^p\theta)(\theta\gamma_{mnp}\theta)\rangle_0 = 2880\ .\label{eq::zmp}
\end{IEEEeqnarray}
Even though there are only five $\theta^\alpha$ out of 16 present in \eqref{eq::zmp} the zero mode prescription can be shown to be supersymmetric \cite{Berkovits}. Because there is only one unique element \eqref{eq::zmp} of the cohomology of $Q$ at order $\mathcal{O}(\lambda^3\theta^5)$ we can evaluate any zero-mode correlator using symmetry arguments together with the normalization condition in equation \eqref{eq::zmp} \cite{Mafra:2022wml}.\par
In order to set the stage for the upcoming computation of the three-point function of closed strings on the disk, we apply the amplitude prescription \eqref{eq::treepres} for $N=3$ and obtain
\begin{IEEEeqnarray}{rCl}
	\mathcal{A}_{D_2}^{\rm closed}(1,2,3)&=&2i g_c^3T_p \int_0^1\mathrm{d}y\int_{\mathbb{H}_+}\mathrm{d}^2z_3\,\left\langle V_1(i)\overline V_1(-i)V_2(iy)\overline U_2(-iy)U_3(z_3)\overline U_3(\overline{z}_3)\right\rangle\ .\nonumber\\\label{eq::3pt}
\end{IEEEeqnarray}

	\section{Three closed strings as six open strings}\label{sec::integration}

As shown in \cite{ovsc} one can express the scattering of three closed strings on the disk $D_2$ as six open strings interacting on $D_2$. This was done by using the methods originally proposed in \cite{KLT} and for example explicitly applied to mixed open and closed string amplitudes on the disk in \cite{Stieberger:2015vya}. In this section we want to follow these lines to simplify the scattering amplitude of three closed strings such that we can write the complex integrals over the upper half plane as integrals over parts of the real line, which correspond to well known open string integrals that would usually arise from colour ordered scattering of six open strings on the disk.\par
The gauge fixing leaves one with already one completely position fixed vertex operator and one real integral over the world sheet coordinate of another vertex operator as described in section \ref{sec::zmp}. Hence, we only have to decompose the integration in \eqref{eq::3pt} over the coordinates $z_3$ and simultaneously $\overline z_3$ of the third vertex operator into two real integrals by applying the KLT relations. The integration over the upper half plane will then be transformed to an integration over 15 integration regions along the real axis. These 15 integration regions are not independent and can be reduced by using monodromy relations as shown in \cite{ovsc}. Moreover, we want to perform a $PSL(2,\mathbb R)$ transformation to change our choice of gauge fixing, which will make the correspondence between the scattering of closed and open strings explicit and simplify the computation in appendix \ref{sec::correlator}, because we can apply the methods proposed in \cite{npt_1}. After the transformation the resulting integrals can be identified as open string partial amplitudes with a certain colour ordering of the open string vertex operators.

\subsection{Analytic continuation and monodromy relations for closed strings}\label{sec::analytic}
Explicitly, the closed string three-point function with the choice of vertex operator positions $(z_1,z_2,z_3)=(i,iy,z)$ and corresponding values for the antiholomorphic world sheet coordinates is given by
\begin{IEEEeqnarray}{rCl}
	\mathcal{A} & = & 2ig_c^3T_p\int_0^1\mathrm dy\int_{\mathbb H_+}\mathrm d^2z\,\left\langle V_1(i)V_{\overline1}(-i)V_2(iy)U_{\overline2}(-iy)U_3(z)U_{\overline3}(\overline z)\right\rangle\ .\label{eq::1.1}
\end{IEEEeqnarray}
To perform the analytic continuation of the amplitude in \eqref{eq::1.1} we write the integral over the upper half plane as an integral over two real variables. After splitting $z$ in real and imaginary part $z=z_1+iz_2$ the integrand in \eqref{eq::1.1} becomes an analytic function in $z_1$ with branch points at $\pm i(1-z_2),\pm i(1+z_2),\pm i(y-z_2),\pm i(y+z_2)$. We would now like to deform the integration contour of the $z_1$ integration from the real to the imaginary axis. More precisely, we deform the $z_1$ integral along the real axis at $\Im(z_1)=0$ to the purely imaginary axis $\Re(z_1)=0$, which is illustrated in figure \ref{fig::z2}. The contributions of both arcs vanish, see also the discussion concerning figure \ref{fig::int} below.
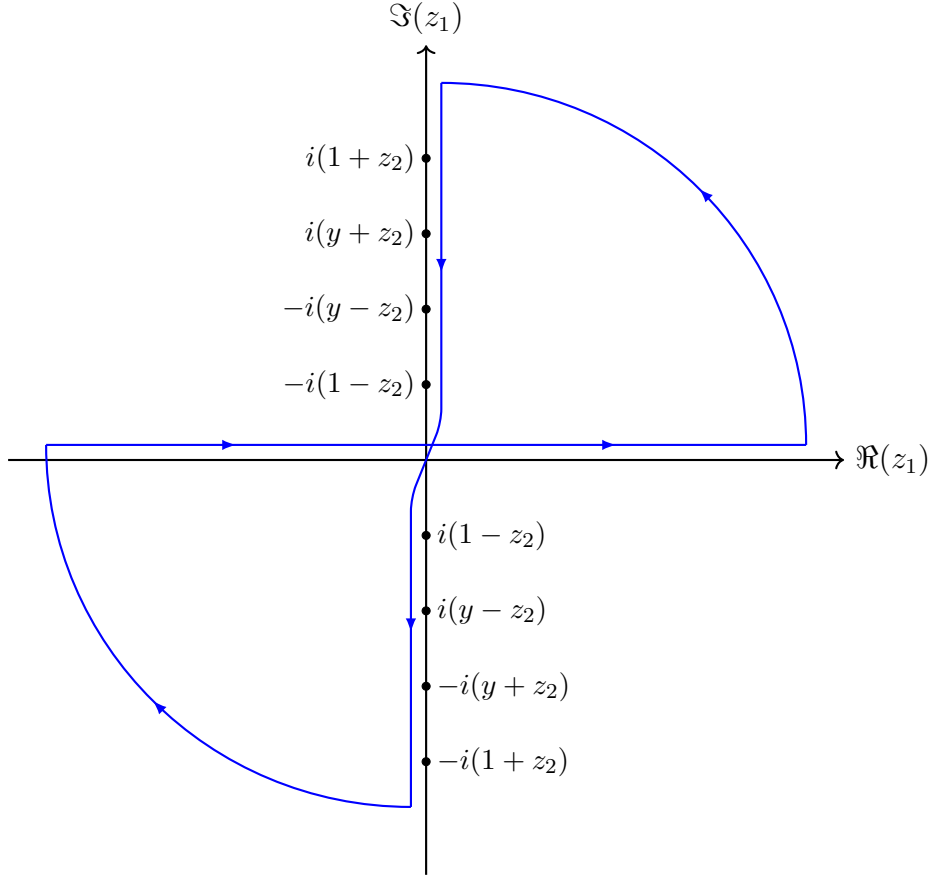
\begin{figure}[h]
	\begin{center}
		\begin{tikzpicture}[decoration={
				markings,
				mark=at position 0.5 with {\arrow{latex}}}]
		\draw[thick,->] (-5.5, 0) -- (5.5, 0) node[right] {$\Re(z_1)$};
		\draw[thick,->] (0, -5.5) -- (0, 5.5) node[above] {$\Im(z_1)$};

		\filldraw[black] (0,4) circle (1.5pt) node[left,font=\small]{$i(1+z_2)$};
		\filldraw[black] (0,3) circle (1.5pt) node[left,font=\small]{$i(y+z_2)$};
		\filldraw[black] (0,1) circle (1.5pt) node[left,font=\small]{$-i(1-z_2)$};
		\filldraw[black] (0,2) circle (1.5pt) node[left,font=\small]{$-i(y-z_2)$};
		
		\filldraw[black] (0,-1) circle (1.5pt) node[right,font=\small]{$i(1-z_2)$};
		\filldraw[black] (0,-2) circle (1.5pt) node[right,font=\small]{$i(y-z_2)$};
		\filldraw[black] (0,-3) circle (1.5pt) node[right,font=\small]{$-i(y+z_2)$};
		\filldraw[black] (0,-4) circle (1.5pt) node[right,font=\small]{$-i(1+z_2)$};

		\draw[blue,thick, postaction={decorate}] (0, 0.2) -- (5, 0.2);
		\draw[blue,thick, postaction={decorate}] (-5, 0.2) -- (0, 0.2);
		\draw[blue,thick,rounded corners=5pt, postaction={decorate}] (0,0)  -- (-0.2, -0.5) -- (-0.2,-4.6);
		\draw[blue,thick,rounded corners=5pt, postaction={decorate}] (0.2,5) -- (0.2,0.5) -- (0, 0);

		\draw[thick,blue, postaction={decorate}] (5,0.2) arc (0:90:4.8);
		\draw[thick,blue, postaction={decorate}] (-0.2,-4.6) arc (-90:-180:4.8);
		\end{tikzpicture}
	\end{center}
	\caption{Branch point structure and contour deformation in the complex $z_1$-plane for $z_2>1$ (as an example).}\label{fig::z2}
\end{figure}	
With the Jacobian $\left|\det\frac{\partial(z,\overline z)}{\partial(z_1,z_2)}\right|=2$ we can write the amplitude after the contour deformation as
\begin{IEEEeqnarray}{rCl}
	\mathcal{A} & = & -4ig_c^3T_p\int_0^1\mathrm dy\int_{-i\infty}^{i\infty}\mathrm dz_1\int_{0}^{\infty}\mathrm dz_2\,\left\langle V_1(i)V_{\overline1}(-i)V_2(iy)U_{\overline2}(-iy)U_3(z_1+iz_2)U_{\overline3}(z_1-iz_2)\right\rangle\nonumber\\
	& = & 4g_c^3T_p\int_0^1\mathrm dy\int_{-\infty}^{\infty}\mathrm dz_1\int_{0}^{\infty}\mathrm dz_2\,\left\langle V_1(i)V_{\overline1}(-i)V_2(iy)U_{\overline2}(-iy)U_3(i(z_1+z_2))U_{\overline3}(i(z_1-z_2))\right\rangle \ .\nonumber\\
\end{IEEEeqnarray}
Then, we define the real variables
\begin{equation}
	\xi=z_1+z_2\ ,\qquad \eta=z_1-z_2\ ,
\end{equation}
which have to fulfil $\xi-\eta\geq0$, which ensures that we still integrate over the upper half plane, i.e$.$ that $z_2\geq0$. We can then change the integration variables and perform the substitution $(z_1,z_2)\rightarrow(\xi,\eta)$, which gives
\begin{IEEEeqnarray}{rCl}
	\mathcal{A}&=&2g_c^3T_p\int_0^1\mathrm dy\int_{-\infty}^{\infty}\mathrm d\xi\int^{\xi}_{-\infty}\mathrm d\eta\,\Pi(y,\xi,\eta)\left\langle V_1(i)V_{\overline1}(-i)V_2(iy)U_{\overline2}(-iy)U_3(i\xi)U_{\overline3}(i\eta)\right\rangle\ ,\nonumber\\\label{eq::q3.1}
\end{IEEEeqnarray}
where it was used that the Jacobian $\left|\det\frac{\partial(z_1,z_2)}{\partial(\xi,\eta)}\right|=\frac12$ cancels against the Jacobian determinant $\left|\det\frac{\partial(z,\overline z)}{\partial(z_1,z_2)}\right|=2$ of the transformation $(z,\overline z)\to(z_1,z_2)$, which we have preformed previously.\par
The correlator in \eqref{eq::q3.1} is invariant under dilatations and rotations, which are the transformations generated by $L_0$, see \eqref{eq::trafo_generator} and also \cite{BLT}. According to \eqref{eq::trafo_generator} for each vertex operator we find that $V^{(a)}(iz)=i^{-h}V^{(a)}(z)$ (using the notation $V^{(0)}$ for $V$ and $V^{(1)}$ for $U$, cf.\ footnote \ref{VUV0V1}). Hence, we can pull out the factor of $i$ in each vertex operator by using that the conformal dimension of the integrated vertex operators and the unintegrated vertex operators is given by $h=1$ and $h=0$, respectively. Doing so, we find
\begin{IEEEeqnarray}{rCl}
	\mathcal{A}&=&2ig_c^3T_p\int_0^1\mathrm dy\int_{-\infty}^{\infty}\mathrm d\xi\int^{\xi}_{-\infty}\mathrm d\eta\,\Pi(y,\xi,\eta)\left\langle V_1(1)V_{\overline1}(-1)V_2(y)U_{\overline2}(-y)U_3(\xi)U_{\overline3}(\eta)\right\rangle\ .\label{eq::q3}
\end{IEEEeqnarray}
For more details on the analytic continuation see appendix \ref{sec::phase}.\par
For splitting the integral over the complex upper half plane into the integration over two real variables $\eta$ and $\xi$, we have introduced the phase factor $\Pi(y,\xi,\eta)$. This phase factor is independent of the kinematical structure of the correlator and it accounts for the correct branch of the integrand to ensure that it is well defined. Moreover, for each integration region, i.e$.$ for each subamplitude,  separately this phase factor becomes independent of the particular value of the world sheet variables, but depends on the ordering of $\eta$ and $\xi$ with respect to the other four vertex operator positions. Explicitly, the phase $\Pi(y,\xi,\eta)$ can be written as
\begin{IEEEeqnarray}{rCl}
	\Pi(y,\xi,\eta)&=&e^{i\pi s_{13}\Theta(-(1-\xi)(1+\eta))}e^{i\pi s_{1\overline3}\Theta(-(1+\xi)(1-\eta))} e^{i\pi s_{23}\Theta(-(y-\xi)(y+\eta))}\nonumber\\
	&&\times e^{i\pi  s_{2\overline3}\Theta(-(y+\xi)(y-\eta))} e^{i\pi  s_{3\overline3}\Theta(-(\xi-\eta))}\ ,\label{eq::phase}
\end{IEEEeqnarray}
where $\Theta$ is the Heaviside step function and the kinematic invariants are defined as\footnote{We made $\ap$ explicit here, but otherwise use $\ap=2$.}
\begin{IEEEeqnarray}l
	s_{ij}=\frac{\alpha'}{4}(k_i+k_j)^2=\frac{\alpha'}{2}k_i{\cdot}k_j\ ,\qquad s_{i\overline\jmath}=\frac{\alpha'}{4}(k_i+D{\cdot}k_j)^2=\frac{\alpha'}{2}k_i{\cdot}D{\cdot}k_j\ .
\end{IEEEeqnarray}
Note, that these are not independent and  momentum conservation \eqref{eq::mom_con} leads to the following kinematical relations:
\begin{align}
	s_{1\bar 1}&=-s_{12}-s_{1\bar2}-s_{13}-s_{1\bar3}\ ,\label{kin1}\\
	s_{2\bar 2}&=-s_{12}-s_{1\bar2}-s_{23}-s_{2\bar3}\ ,\label{kin2}\\
	s_{3\bar 3}&=-s_{13}-s_{1\bar3}-s_{23}-s_{2\bar3}\ .\label{kin3}
\end{align}
Consequently, there are six independent kinematic invariants for  scattering three closed strings in the presence of D--branes \cite{ovsc}.\par
The phase factor \eqref{eq::phase} can be derived following \cite{ovsc, Stieberger:2015vya}, cf.\ appendix \ref{sec::phase}.   We have also added $e^{i\pi  s_{3\overline3}\Theta(-(\xi-\eta))}$ for completeness although above we are only integrating over $\eta<\xi$ and hence the contribution of this term is always one in \eqref{eq::q3}. But later we will use dual Ward identities and thereby we will encounter also integration regions with $\xi<\eta$ such that it will be important to include $e^{i\pi  s_{3\overline3}\Theta(-(\xi-\eta))}$ in the phase \eqref{eq::phase}.\par
Since the world sheet coordinates appear in the phase factor only inside the $\Theta$-function, we immediately want to perform the integration over those $\Theta$-functions. Evaluating the phase factor \eqref{eq::phase} is achieved by dividing the integration region into smaller patches, where in each patch the $\Theta$-functions yield a constant phase $\Pi(y,\eta,\xi)$, which does not depend on the world sheet coordinates any more. We find that the integration region in the $(\xi,\eta)$-plane in \eqref{eq::q3} can be split into 15 smaller regions of integration. All integration regions for $(\xi,\eta)$  are listed in table \ref{tab::3}.
\begin{table}[h]
	\begin{center}
		\resizebox{\textwidth}{!}{%
			\begin{tabular}{|c |c |c |c |c |c |}
				\hline
				     \rule[-1ex]{0pt}{3.5ex}       & $\eta<\xi$                                                                       & \multicolumn{1}{c}{}                      & \multicolumn{1}{c}{}             & \multicolumn{1}{c}{}                                     &                                                                                  \\ \hline
				 \rule[-1ex]{0pt}{3.5ex}$\xi<-1$   & $e^{i\pi s_{13}}e^{i\pi s_{1\overline3}}e^{i\pi s_{23}}e^{i\pi s_{2\overline3}}$ & \multicolumn{1}{c}{}                      & \multicolumn{1}{c}{}             & \multicolumn{1}{c}{}                                     &                                                                                  \\ \hline
				       \multicolumn{1}{c}{}        & \multicolumn{1}{c}{}                                                             & \multicolumn{1}{c}{}                      & \multicolumn{1}{c}{}             & \multicolumn{1}{c}{}                                     & \multicolumn{1}{c}{}                                                             \\ \hline
				     \rule[-1ex]{0pt}{3.5ex}       & $\eta<-1$                                                                        & $-1<\eta<\xi$                             & \multicolumn{1}{c}{}             & \multicolumn{1}{c}{}                                     &                                                                                  \\ \hline
				\rule[-1ex]{0pt}{3.5ex}$-1<\xi<-y$ & $e^{i\pi s_{13}}e^{i\pi s_{23}}e^{i\pi s_{2\overline3}}$                         & $e^{i\pi s_{23}}e^{i\pi s_{2\overline3}}$ & \multicolumn{1}{c}{}             & \multicolumn{1}{c}{}                                     &                                                                                  \\ \hline
				       \multicolumn{1}{c}{}        & \multicolumn{1}{c}{}                                                             & \multicolumn{1}{c}{}                      & \multicolumn{1}{c}{}             & \multicolumn{1}{c}{}                                     & \multicolumn{1}{c}{}                                                             \\ \hline
				     \rule[-1ex]{0pt}{3.5ex}       & $\eta<-1$                                                                        & $-1<\eta<-y$                              & $-y<\eta<\xi$                    & \multicolumn{1}{c}{}                                     &                                                                                  \\ \hline
				\rule[-1ex]{0pt}{3.5ex}$-y<\xi<y$  & $e^{i\pi s_{13}}e^{i\pi s_{23}}$                                                 & $e^{i\pi s_{23}}$                         & $1$                              & \multicolumn{1}{c}{}                                     &                                                                                  \\ \hline
				       \multicolumn{1}{c}{}        & \multicolumn{1}{c}{}                                                             & \multicolumn{1}{c}{}                      & \multicolumn{1}{c}{}             & \multicolumn{1}{c}{}                                     & \multicolumn{1}{c}{}                                                             \\ \hline
				     \rule[-1ex]{0pt}{3.5ex}       & $\eta<-1$                                                                        & $-1<\eta<-y$                              & $-y<\eta<y$                      & $y<\eta<\xi$                                             &                                                                                  \\ \hline
				\rule[-1ex]{0pt}{3.5ex}	$y<\xi<1$  & $e^{i\pi s_{13}}$                                                                & $1$                                       & $e^{i\pi s_{23}}$                & $e^{i\pi s_{23}}e^{i\pi s_{2\overline3}}$                &                                                                                  \\ \hline
				       \multicolumn{1}{c}{}        & \multicolumn{1}{c}{}                                                             & \multicolumn{1}{c}{}                      & \multicolumn{1}{c}{}             & \multicolumn{1}{c}{}                                     & \multicolumn{1}{c}{}                                                             \\ \hline
				     \rule[-1ex]{0pt}{3.5ex}       & $\eta<-1$                                                                        & $-1<\eta<-y$                              & $-y<\eta<y$                      & $y<\eta<1$                                               & $1<\eta<\xi$                                                                     \\ \hline
				  \rule[-1ex]{0pt}{3.5ex}$1<\xi$   & $1$                                                                              & $e^{i\pi s_{13}}$                         & $e^{i\pi s_{13}}e^{i\pi s_{23}}$ & $e^{i\pi s_{13}}e^{i\pi s_{23}}e^{i\pi s_{2\overline3}}$ & $e^{i\pi s_{13}}e^{i\pi s_{1\overline3}}e^{i\pi s_{23}}e^{i\pi s_{2\overline3}}$ \\ \hline
			\end{tabular}}
		\caption{$\Pi(y,\xi,\eta)$ for each integration region in the $(\xi,\eta)$-plane.}\label{tab::3}
	\end{center}
\end{table}\par
In string theory we find an analogue to the dual Ward identity in field theory\footnote{Compared to for example \cite{ovsc} we are using a slightly different monodromy relation in \eqref{eq::monodromy}, where the contributions $\mathbb A(i_1,i_2,\ldots,i_{N-1},i_N)$ and $\mathbb A(i_2,\ldots,i_{N-1},i_N,i_1)$ are not equivalent, because for a closed string amplitude there is no vertex operator position $z\to\infty$ after gauge fixing. Moreover, the integration regions corresponding to $\mathbb A(i_1,i_2,\ldots,i_{N-1},i_N)$ and $\mathbb A(i_2,\ldots,i_{N-1},i_N,i_1)$ appear with the same phase in \eqref{eq::monodromy}, i.e$.$ we don't pick up a phase when jumping from $+\infty$ to $-\infty$, because there is no vertex operator localized at infinity. Hence, they can be combined to become a proper open string subamplitude.} 
\begin{IEEEeqnarray}{rCl}
	0&=&\mathbb A(i_1,i_2,\ldots,i_{N-1},i_N)+e^{i\pi s_{i_1i_2}}\mathbb A(i_2,i_1,i_3,\ldots,i_{N-1},i_N)\nonumber\\
	&&+e^{i\pi (s_{i_1i_2}+s_{i_1i_3})}\mathbb A(i_2,i_3,i_1,\ldots,i_{N-1},i_N)+\ldots\nonumber\\
	&&+e^{i\pi(s_{i_1i_2}+s_{i_1i_3}+\ldots+s_{i_1i_{N-1}})}\mathbb A(i_2,i_3,\ldots,i_{N-1},i_1,i_N)+\mathbb A(i_2,\ldots,i_{N-1},i_N,i_1)\ ,\label{eq::monodromy}
\end{IEEEeqnarray}
where $i_j\in\{1,\overline1,2,\overline2,\ldots,M,\overline M\}$ such that $M+\overline M=N$, i.e.\ $j\in\{1,2,\ldots,2M\}$. Using these relations for $M=3$ we want to decompose the disk amplitude of three closed strings into a sum over colour ordered partial subamplitudes of six open strings, but we have to be careful: The integration regions beginning or ending at $\pm\infty$ are no open string subamplitudes.\footnote{This can be seen more clearly, when transforming the integration region under the $PSL(2,\mathbb{R})$ transformation \eqref{eq::q5}. Under this transformation $\pm\infty$ is mapped onto $-\sqrt{x}$. Hence, the upper or lower boundary of the $\xi$ or $\eta$ integration is $-\sqrt{x}$, which is not a world sheet position of any vertex operator, which implies that this cannot be an open string subamplitude, see also section \ref{sec::psl2r_trafo}.} Hence, not all $\mathbb A(i_1,i_2,\ldots,i_{N-1},i_N)$ correspond to open string subamplitudes with different colour orderings of $N$ open strings immediately, but it is possible to combine and rewrite the $\mathbb A(i_1,i_2,\ldots,i_{N-1},i_N)$ so that they become open string subamplitudes. To discriminate between closed string contributions \eqref{eq::open} and open string subamplitudes \eqref{eq::open_2} we use $\mathbb{A}(\ldots)$ and $A(\ldots)$, respectively.\par
To derive the relation \eqref{eq::monodromy} we consider the more general form of an amplitude of $M$ closed strings after analytic continuation
\begin{IEEEeqnarray}{rCl}
	\mathbb A(i_1,\ldots,i_{N})&\sim&\int_0^1\mathrm dz_{i_d}\,\smashoperator{\prod_{\substack{p=1\\p\neq a,b,c,d}}^{N}}\,\int_{-\infty\leq z_{i_1}\leq z_{i_2}\leq\ldots\leq z_{i_{N}}\leq\infty}\mathrm{d}z_{i_p}
	 \sum_I \prod_{j<k}|z_{i_j}-z_{i_k}|^{s_{i_{j}i_{k}}} (z_{i_j}-z_{i_k})^{n^I_{i_{j}i_{k}}}
	\mathcal{K}^I\ ,\nonumber\\\label{eq::open}
\end{IEEEeqnarray}
where the vertex operators $(i_a,i_b,i_c)$ are position fixed and we don't integrate over these world sheet positions. We cannot choose arbitrary vertex operators, but have to follow the gauge fixing of section \ref{sec::zmp}, which implies that before analytic continuation two of $(z_{i_a},z_{i_b},z_{i_c})$ have to satisfy $z_{i_b}=\overline z_{i_a}$ with $z_{i_a}=i$ and a third one $z_{i_c}=\overline z_{i_d}$, which is the complex conjugate to the vertex operator position $z_{i_d}=-iy$ and $y$ is integrated from $0$ to $1$, because we can only fix one and a half vertex operator positions\footnote{After analytic continuation we find $(z_{i_a},z_{i_b})\to(1,-1)$ and $(z_{i_c},z_{i_d})\to(y,-y)$, see \eqref{eq::q3} for three closed strings.}. The integers $n_{i_{j}i_{k}}^I$ are specific for a given kinematic factor $\mathcal K^I$ and originate from the evaluation of the correlator of the $M$-point amplitude. Without loss of generality we can assume that $i_1 \notin \{ i_a, i_b, i_c, i_d\}$. Then we can analytically continue the $z_{i_1}$-dependence of the integrand to the whole complex plane and integrate $z_{i_1}$ along the contour depicted in figure \ref{fig::int} rather than over $(-\infty,z_{i_2})$. The semicircle at infinity in the complex plane does not contribute, because the integrand in \eqref{eq::open} behaves like $z_{i_1}^{-2h_{i_1}}$ as $|z_{i_1}|\to\infty$, where $h_{i_1}$ is the conformal weight of the vertex operator $U_{i_1}(z_{i_1})$ and in the pure spinor formalism an integrated vertex operator has a conformal weight of one. Thus, $z_{i_1}^{-2}\to0$ for $|z_{i_1}|\to\infty$ and we have no contribution from the semicircle. Moreover, by performing the $z_{i_1}$ integration along the real axis each time when we encircle one of the vertex operator positions $z_{i_j}$ for $j=2,3,\ldots,N$ we pick up a phase factor. This arises from using \eqref{eq::sign_z} when expressing the integrand of $\mathbb{A}(i_1,i_2,\ldots,i_N)$ in terms of the integrand of $\mathbb{A}(i_2,\ldots,i_j,i_1,i_{j+1},\ldots,i_N)$. For instance $\mathbb{A}(i_1,i_2,\ldots,i_N)$ contains a factor $(z_{i_2}-z_{i_1})^{s_{i_1 i_2}}$ while $\mathbb{A}(i_2,i_1,\ldots,i_N)$ contains a factor $(z_{i_1}-z_{i_2})^{s_{i_1 i_2}}$ \cite{Plahte:1970wy}. By applying Cauchy's theorem the integral along the closed contour in figure \ref{fig::int} vanishes and we end up with \eqref{eq::monodromy}. \par
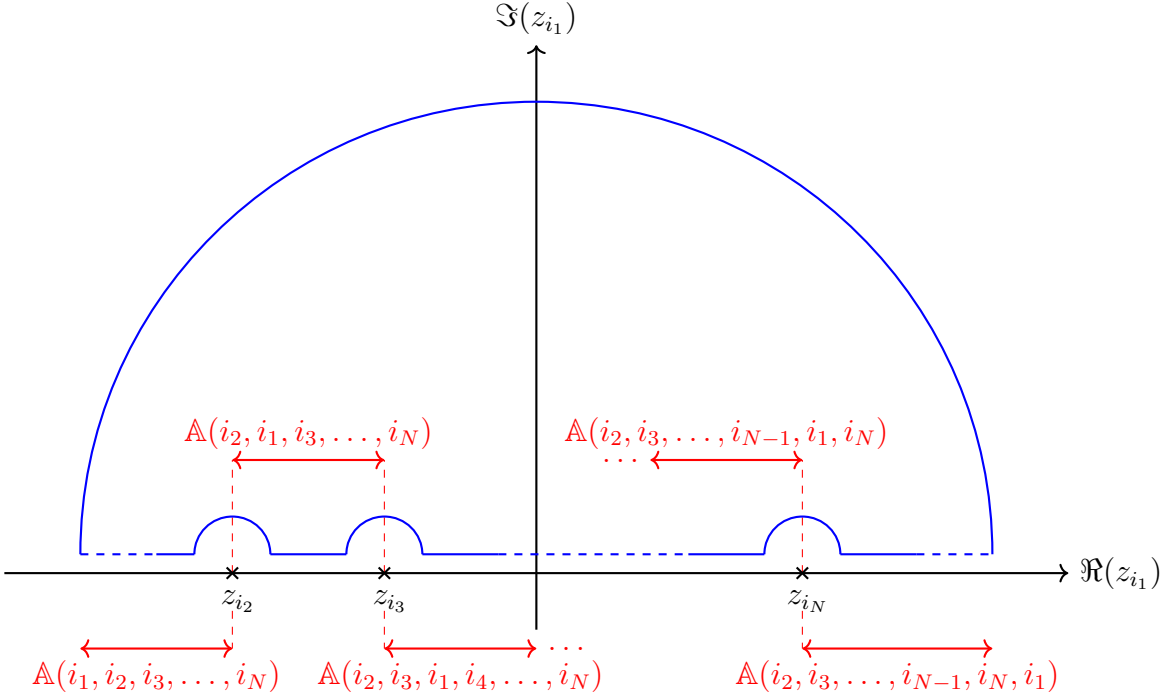
\begin{figure}
	\begin{center}
		\begin{tikzpicture}
			\draw[red, dashed] (-2,0)--(-2,1.5);
			\draw[red, dashed] (-4,0)--(-4,1.5);
			\draw[red, dashed] (3.5,0)--(3.5,1.5);
			\draw[red, dashed] (-2,-0.5)--(-2,-1);
			\draw[red, dashed] (-4,-0.5)--(-4,-1);
			\draw[red, dashed] (3.5,-0.5)--(3.5,-1);
			\draw[<->, thick, red] (-6,-1) -- (-5,-1) node[below]{$\mathbb A(i_1,i_2,i_3,\ldots,i_N)$} -- (-4,-1);
			\draw[<->, thick, red] (-2,1.5) -- (-3,1.5) node[above]{$\mathbb A(i_2,i_1,i_3,\ldots,i_N)$} -- (-4,1.5);
			\draw[<->, thick, red] (-2,-1) -- (-1,-1) node[below]{$\mathbb A(i_2,i_3,i_1,i_4,\ldots,i_N)$} -- (0,-1) node[right]{$\ldots$};
			\draw[<->, thick, red] (1.5,1.5) node[left]{$\ldots$} -- (2.5,1.5) node[above]{$\mathbb A(i_2,i_3,\ldots,i_{N-1},i_1,i_N)$} -- (3.5,1.5);
			\draw[<->, thick, red] (3.5,-1) -- (4.75,-1) node[below]{$\mathbb A(i_2,i_3,\ldots,i_{N-1},i_N,i_1)$} -- (6,-1);
			\draw[->, thick] (-7, 0) -- (7, 0) node[right] {$\Re(z_{i_1})$};
			\draw[->, thick] (0, -0.75) -- (0, 7) node[above] {$\Im(z_{i_1})$};
			
			\draw[scale=1, domain=-6:-5, dashed, variable=\x, thick, blue] plot ({\x}, {0.25});
			\draw[scale=1, domain=-5:-4.5, smooth, variable=\x, thick, blue] plot ({\x}, {0.25});
			\draw[scale=1, domain=-3.5:-2.5, smooth, variable=\x, thick, blue] plot ({\x}, {0.25});
			\draw[scale=1, domain=-1.5:-0.5, smooth, variable=\x, thick, blue] plot ({\x}, {0.25});
			\draw[scale=1, domain=-0.5:2, dashed, variable=\x, thick, blue] plot ({\x}, {0.25});
			\draw[scale=1, domain=2:3, smooth, variable=\x, thick, blue] plot ({\x}, {0.25});
			\draw[scale=1, domain=4:5, smooth, variable=\x, thick, blue] plot ({\x}, {0.25});
			\draw[scale=1, domain=5:6, dashed, variable=\x, thick, blue] plot ({\x}, {0.25});
			\draw[thick,blue] (6,0.25) arc (0:180:6);
			\draw[thick,blue] (-2.5,0.25) arc (180:0:0.5);
			\draw[thick,blue] (-4.5,0.25) arc (180:0:0.5);
			\draw[thick,blue] (3,0.25) arc (180:0:0.5);
			\draw[thick] (3.5-0.075,0.075)--(3.5+0.075,-0.075) node[below]{$z_{i_N}$};
			\draw[thick] (3.5+0.075,0.075)--(3.5-0.075,-0.075);
			\draw[thick] (-4-0.075,0.075)--(-4+0.075,-0.075) node[below]{$z_{i_2}$};
			\draw[thick] (-4+0.075,0.075)--(-4-0.075,-0.075);
			\draw[thick] (-2-0.075,0.075)--(-2+0.075,-0.075) node[below]{$z_{i_3}$};
			\draw[thick] (-2+0.075,0.075)--(-2-0.075,-0.075);
		\end{tikzpicture}
		\caption{Contour integral in the complex $z_{i_1}$-plane.} \label{fig::int}
	\end{center}
\end{figure}
Next, we want to write the amplitude \eqref{eq::q3} in terms of $\mathbb A(i_1,i_2,i_3,i_4,i_5,i_6)$: Taking all phases in table \ref{tab::3} and the corresponding integration region we can write \eqref{eq::q3} as 
\begin{IEEEeqnarray*}{rCl}
	\mathcal{A} & = & e^{i\pi s_{13}}e^{i\pi s_{1\overline3}}e^{i\pi s_{23}}e^{i\pi s_{2\overline3}}\mathbb A(\overline3,3,\overline1,\overline2,2,1)+e^{i\pi s_{13}}e^{i\pi s_{23}}e^{i\pi s_{2\overline3}}\mathbb A(\overline3,\overline1,3,\overline2,2,1)\\
	&&+e^{i\pi s_{23}}e^{i\pi s_{2\overline3}}\mathbb A(\overline1,\overline3,3,\overline2,2,1)+e^{i\pi s_{13}}e^{i\pi s_{23}}\mathbb A(\overline3,\overline1,\overline2,3,2,1)+e^{i\pi s_{23}}\mathbb A(\overline1,\overline3,\overline2,3,2,1)\\
	&&+\mathbb A(\overline1,\overline2,\overline3,3,2,1)+e^{i\pi s_{13}}\mathbb A(\overline3,\overline1,\overline2,2,3,1)+\mathbb A(\overline1,\overline3,\overline2,2,3,1)+e^{i\pi s_{23}}\mathbb A(\overline1,\overline2,\overline3,2,3,1)\\
	&&+e^{i\pi s_{23}}e^{i\pi s_{2\overline3}}\mathbb A(\overline1,\overline2,2,\overline3,3,1)+\mathbb A(\overline3,\overline1,\overline2,2,1,3)+e^{i\pi s_{13}}\mathbb A(\overline1,\overline3,\overline2,2,1,3)\\
	&&+e^{i\pi s_{13}}e^{i\pi s_{23}}\mathbb A(\overline1,\overline2,\overline3,2,1,3)+e^{i\pi s_{13}}e^{i\pi s_{23}}e^{i\pi s_{2\overline3}}\mathbb A(\overline1,\overline2,2,\overline3,1,3)\\
	&&+e^{i\pi s_{13}}e^{i\pi s_{1\overline3}}e^{i\pi s_{23}}e^{i\pi s_{2\overline3}}\mathbb A(\overline1,\overline2,2,1,\overline3,3)\ .\IEEEyesnumber\label{eq::amp_mon_1}
\end{IEEEeqnarray*}
From the general monodromy relation in \eqref{eq::monodromy} we obtain by permutation

\begin{IEEEeqnarray}{rCl}
	0&=&\mathbb A(\overline3,\overline1,\overline2,2,1,3)+e^{i\pi s_{13}}\mathbb A(\overline1,\overline3,\overline2,2,1,3)+e^{i\pi s_{13}}e^{i\pi s_{23}}\mathbb A(\overline1,\overline2,\overline3,2,1,3)\nonumber\\
	&&+e^{i\pi s_{13}}e^{i\pi s_{23}}e^{i\pi s_{2\overline3}}\mathbb A(\overline1,\overline2,2,\overline3,1,3)+e^{i\pi s_{1\overline3}}e^{i\pi s_{2\overline3}}e^{i\pi s_{23}}e^{i\pi s_{13}}\mathbb A(\overline1,\overline2,2,1,\overline3,3)\nonumber\\
	&&+\mathbb A(\overline1,\overline2,2,1,3,\overline3)\ ,\label{eq::mon_eq_1}\\
	0&=&\mathbb A(\overline3,\overline1,\overline2,2,1,3)+e^{i\pi s_{13}}\mathbb A(\overline3,\overline1,\overline2,2,3,1)+e^{i\pi s_{13}}e^{i\pi s_{23}}\mathbb A(\overline3,\overline1,\overline2,3,2,1)\nonumber\\
	&&+e^{i\pi s_{13}}e^{i\pi s_{23}}e^{i\pi s_{2\overline3}}\mathbb A(\overline3,\overline1,3,\overline2,2,1)+e^{i\pi s_{13}}e^{i\pi s_{1\overline3}}e^{i\pi s_{23}}e^{i\pi s_{2\overline3}}\mathbb A(\overline3,3,\overline1,\overline2,2,1)\nonumber\\
	&&+\mathbb A(3,\overline3,\overline1,\overline2,2,1)\label{eq::mon_eq_2}
\end{IEEEeqnarray}
to reduce the number of integration regions in table \ref{tab::3} and simplify equation \eqref{eq::amp_mon_1}. Note, that one actually obtains \eqref{eq::mon_eq_2} by the complex conjugation of \eqref{eq::monodromy} and multiplication by $(-1)$ to take into account the reversal of the contour
\begin{IEEEeqnarray}{rCl}
	0&=&-\overline{\mathbb A}(3,\overline3,\overline1,\overline2,2,1)-e^{-i\pi s_{3\overline3}}\overline{\mathbb A}(\overline3,3,\overline1,\overline2,2,1)-e^{-i\pi s_{3\overline3}}e^{-i\pi s_{1\overline3}}
	\overline{\mathbb A}(\overline3,\overline1,3,\overline2,2,1)\nonumber\\
	&&-e^{-i\pi s_{3\overline3}}e^{-i\pi s_{1\overline3}}e^{-i\pi s_{2\overline3}}\overline{\mathbb A}(\overline3,\overline1,\overline2,3,2,1)-e^{-i\pi s_{3\overline3}}e^{-i\pi s_{1\overline3}}e^{-i\pi s_{2\overline3}}e^{-i\pi s_{23}}\overline{\mathbb A}(\overline3,\overline1,\overline2,2,3,1)\nonumber\\
	&&-\overline{\mathbb A}(\overline3,\overline1,\overline2,2,1,3)\ .
\end{IEEEeqnarray}
The partial amplitudes are purely imaginary -- the amplitude in \eqref{eq::open} is real, but the overall amplitude contains an additional factor of $i$, which is obtained from gauge fixing, see \eqref{eq::q3} -- hence, $\overline{\mathbb A}(i_1,i_2,i_3,i_4,i_5,i_6)=-\mathbb A(i_1,i_2,i_3,i_4,i_5,i_6)$ and we find
\begin{IEEEeqnarray}{rCl}
	0&=&\mathbb A(3,\overline3,\overline1,\overline2,2,1)+e^{-i\pi s_{3\overline3}}\mathbb A(\overline3,3,\overline1,\overline2,2,1)+e^{-i\pi s_{3\overline3}}e^{-i\pi s_{1\overline3}}
	\mathbb A(\overline3,\overline1,3,\overline2,2,1)\nonumber\\
	&&+e^{-i\pi s_{3\overline3}}e^{-i\pi s_{1\overline3}}e^{-i\pi s_{2\overline3}}\mathbb A(\overline3,\overline1,\overline2,3,2,1)+e^{-i\pi s_{3\overline3}}e^{-i\pi s_{1\overline3}}e^{-i\pi s_{2\overline3}}e^{-i\pi s_{23}}\mathbb A(\overline3,\overline1,\overline2,2,3,1)\nonumber\\
	&&+\mathbb A(\overline3,\overline1,\overline2,2,1,3)\ .
\end{IEEEeqnarray}
Finally, with momentum conservation we arrive at equation \eqref{eq::mon_eq_2}
\begin{IEEEeqnarray}{rCl}
	0&=&\mathbb A(3,\overline3,\overline1,\overline2,2,1)+e^{i\pi s_{13}}e^{i\pi s_{1\overline3}}e^{i\pi s_{23}}e^{i\pi s_{2\overline3}}\mathbb A(\overline3,3,\overline1,\overline2,2,1)+e^{i\pi s_{13}}e^{i\pi s_{23}}e^{i\pi s_{2\overline3}}
	\mathbb A(\overline3,\overline1,3,\overline2,2,1)\nonumber\\
	&&+e^{i\pi s_{13}}e^{i\pi s_{23}}\mathbb A(\overline3,\overline1,\overline2,3,2,1)+e^{i\pi s_{13}}\mathbb A(\overline3,\overline1,\overline2,2,3,1)+\mathbb A(\overline3,\overline1,\overline2,2,1,3)\ .
\end{IEEEeqnarray}
We use the first equation \eqref{eq::mon_eq_1} and see that the sum of the terms corresponding to the bottom row $(1<\xi)$ in table \ref{tab::3} can be written in terms of only $\mathbb A(\overline1,\overline2,2,1,3,\overline3)$, because these integration regions appear with the correct phases such that they build a closed contour in the complex  $\eta$-plane and only $\mathbb A(\overline1,\overline2,2,1,3,\overline3)$ is missing in this row. In a similar manner we can reduce the sum of the terms corresponding to the left column $(\eta<-1)$ with the second equation \eqref{eq::mon_eq_2} to two integration regions $\mathbb A(\overline3,\overline1,\overline2,2,1,3)$ and $\mathbb A(3,\overline3,\overline1,\overline2,2,1)$. Hence, we have written \eqref{eq::amp_mon_1} as
\begin{IEEEeqnarray*}{rCl}
	\mathcal{A} & = & -\mathbb A(3,\overline3,\overline1,\overline2,2,1)+e^{i\pi s_{23}}e^{i\pi s_{2\overline3}}\mathbb A(\overline1,\overline3,3,\overline2,2,1)+e^{i\pi s_{23}}\mathbb A(\overline1,\overline3,\overline2,3,2,1)\\
	&&+\mathbb A(\overline1,\overline2,\overline3,3,2,1)+\mathbb A(\overline1,\overline3,\overline2,2,3,1)+e^{i\pi s_{23}}\mathbb A(\overline1,\overline2,\overline3,2,3,1)\\
	&&+e^{i\pi s_{23}}e^{i\pi s_{2\overline3}}\mathbb A(\overline1,\overline2,2,\overline3,3,1)-\mathbb A(\overline3,\overline1,\overline2,2,1,3)-\mathbb A(\overline1,\overline2,2,1,3,\overline3)\ ,\IEEEyesnumber\label{eq::amp_mon_2}
\end{IEEEeqnarray*} 
where the integration regions can also be found in table \ref{tab::4}. As we will show \eqref{eq::amp_mon_2} can be written in terms of open string partial amplitudes, but to see that it is necessary to perform an appropriate $PSL(2,\mathbb R)$ transformation.
\begin{table}
	\begin{center}
		\begin{tabular}{|c |c |c |c |c |c | c|}
			\hline
			      \rule[-1ex]{0pt}{3.5ex}       & $\xi<\eta<-1$        & \multicolumn{1}{c}{}                      & \multicolumn{1}{c}{} & \multicolumn{1}{c}{}                      & \multicolumn{1}{c}{} &                      \\ \hline
			 \rule[-1ex]{0pt}{3.5ex} $\xi<-1$   & $-1$                 & \multicolumn{1}{c}{}                      & \multicolumn{1}{c}{} & \multicolumn{1}{c}{}                      & \multicolumn{1}{c}{} &                      \\ \hline
			       \multicolumn{1}{c}{}         & \multicolumn{1}{c}{} & \multicolumn{1}{c}{}                      & \multicolumn{1}{c}{} & \multicolumn{1}{c}{}                      & \multicolumn{1}{c}{} & \multicolumn{1}{c}{} \\ \hline
			      \rule[-1ex]{0pt}{3.5ex}       &                      & $-1<\eta<\xi$                             & \multicolumn{1}{c}{} & \multicolumn{1}{c}{}                      & \multicolumn{1}{c}{} &                      \\ \hline
			\rule[-1ex]{0pt}{3.5ex} $-1<\xi<-y$ &                      & $e^{i\pi s_{23}}e^{i\pi s_{2\overline3}}$ & \multicolumn{1}{c}{} & \multicolumn{1}{c}{}                      & \multicolumn{1}{c}{} &                      \\ \hline
			       \multicolumn{1}{c}{}         & \multicolumn{1}{c}{} & \multicolumn{1}{c}{}                      & \multicolumn{1}{c}{} & \multicolumn{1}{c}{}                      & \multicolumn{1}{c}{} & \multicolumn{1}{c}{} \\ \hline
			      \rule[-1ex]{0pt}{3.5ex}       &                      & $-1<\eta<-y$                              & $-y<\eta<\xi$        & \multicolumn{1}{c}{}                      & \multicolumn{1}{c}{} &                      \\ \hline
			 \rule[-1ex]{0pt}{3.5ex}$-y<\xi<y$  &                      & $e^{i\pi s_{23}}$                         & $1$                  & \multicolumn{1}{c}{}                      & \multicolumn{1}{c}{} &                      \\ \hline
			       \multicolumn{1}{c}{}         & \multicolumn{1}{c}{} & \multicolumn{1}{c}{}                      & \multicolumn{1}{c}{} & \multicolumn{1}{c}{}                      & \multicolumn{1}{c}{} & \multicolumn{1}{c}{} \\ \hline
			      \rule[-1ex]{0pt}{3.5ex}       &                      & $-1<\eta<-y$                              & $-y<\eta<y$          & $y<\eta<\xi$                              & \multicolumn{1}{c}{} &                      \\ \hline
			 \rule[-1ex]{0pt}{3.5ex} $y<\xi<1$  &                      & $1$                                       & $e^{i\pi s_{23}}$    & $e^{i\pi s_{23}}e^{i\pi s_{2\overline3}}$ & \multicolumn{1}{c}{} &                      \\ \hline
			       \multicolumn{1}{c}{}         & \multicolumn{1}{c}{} & \multicolumn{1}{c}{}                      & \multicolumn{1}{c}{} & \multicolumn{1}{c}{}                      & \multicolumn{1}{c}{} & \multicolumn{1}{c}{} \\ \hline
			      \rule[-1ex]{0pt}{3.5ex}       & $\eta<-1$            & \multicolumn{1}{c}{}                      & \multicolumn{1}{c}{} & \multicolumn{1}{c}{}                      &                      & $\xi<\eta$           \\ \hline
			  \rule[-1ex]{0pt}{3.5ex} $1<\xi$   & $-1$                 & \multicolumn{1}{c}{}                      & \multicolumn{1}{c}{} & \multicolumn{1}{c}{}                      &                      & $-1$                 \\ \hline
		\end{tabular}
	\caption{$\Pi(y,\xi,\eta)$ for each integration region in the $(\xi,\eta)$-plane after applying monodromy relations.}\label{tab::4}
	\end{center}
\end{table}

\subsection[$PSL(2,\mathbb R)$ transformation and monodromy relations for open strings]{\boldmath{$PSL(2,\mathbb R)$} transformation and monodromy relations for open strings}\label{sec::psl2r_trafo}

To write \eqref{eq::q3} as partial open string amplitudes it is very convenient to change the vertex operator position fixing by performing a $PSL(2,\mathbb R)$ transformation similar to the scattering of two closed strings on the disk as already described in \cite{Garousi:1996ad,Hashimoto:1996bf,2pt}. For two closed strings one has only one real world sheet variable, which is integrated over. In equation $(3.9)$ of \cite{2pt} we can find the corresponding transformation for the scattering of two closed strings. If we compare the amplitude of two closed strings with the amplitude of four open strings on the disk, we can conclude that the transformation maps the vertex operator position fixing from $(-1,y,1)$ to $(0,1,\infty)$. In the following we want to generalise the $PSL(2,\mathbb R)$ transformation of \cite{2pt} for a correlator of three closed strings on the disk.\par
We can take a general linear fractional transformation $z':=f(z)=\frac{az+b}{cz+d}$ with $ad-bc=1$ and find the parameters $a,b,c$ and $d$ for this transformation by solving $f(z)=z'$ for $z\in\{-1,y,1\}$ and $z'\in\{0,1,\infty\}$. We then end up with the following $PSL(2,\mathbb R)$ transformation
\begin{equation}
	\frac{1}{\sqrt{2(1-y^2)}}
	\begin{pmatrix}
		1-y&1-y\\
		-(1+y)&1+y
	\end{pmatrix}\in PSL(2,\mathbb{R})\ .\label{eq::psl}
\end{equation}
For the transformation of the coordinates of the vertex operators we have to use the fractional linear transformation of the above $PSL(2,\mathbb R)$ transformation
\begin{equation}
	z':=f(z)=\frac{(1-y)(1+z)}{(1+y)(1-z)}\ .\label{eq::q5}
\end{equation}
We can use the fractional linear transformation to map the amplitude from the vertex operator position fixing $(-1,y,1)$ to $(0,1,\infty)$. For this purpose we define the new variables
\begin{IEEEeqnarray}{l}
	\begin{IEEEeqnarraybox}[][c]{rCl}
		\IEEEstrut
		x & := & f(-y) =\frac{(1-y)^2}{(1+y)^2}\ ,            \\
		\tilde\xi & := & f(\xi) = \frac{(1-y)(1+\xi)}{(1+y)(1-\xi)}\ ,               \\
		\tilde\eta & := & f(\eta) = \frac{(1-y)(1+\eta)}{(1+y)(1-\eta)}\ .
		\IEEEstrut
	\end{IEEEeqnarraybox}\label{eq::6}
\end{IEEEeqnarray}
Because we want to map from the old coordinates $(y,\xi,\eta)$ to the knew ones $(x,\tilde\xi,\tilde\eta)$, it is useful to also have the inverse transformations of \eqref{eq::6}, which are given by\footnote{Note, that we have the same transformation for $y$ here as in $(3.9)$ of \cite{2pt} for the scattering of two closed strings.}
\begin{IEEEeqnarray}{rCl}
	\begin{IEEEeqnarraybox}[][c]{rCl}
		\IEEEstrut
		y & = & \frac{1-\sqrt x}{1+\sqrt x}\ ,                       \\
		\xi & = & \frac{\tilde\xi-\sqrt{x}}{\tilde\xi+\sqrt{x}}\ ,          \\
		\eta & = & \frac{\tilde\eta-\sqrt{x}}{\tilde\eta+\sqrt{x}}\ .
		\IEEEstrut
	\end{IEEEeqnarraybox}\label{eq::8}
\end{IEEEeqnarray}
From \eqref{eq::6} one could also find a second solution $y=\frac{1+\sqrt x}{1-\sqrt x}$ for $x=f(-y)$. For this other solution we would find for $x\in[0,1]$ that $y\in[1,\infty[$. Hence, we ignore this solution.\par
The vertex operators transform under a global conformal transformation as
\begin{IEEEeqnarray}l
	V(z)\to V'(z')\ ,\qquad U(z)\to \left(\frac{\partial z}{\partial z'}\right)^{-1}U'(z')\ ,
\end{IEEEeqnarray}
where we have used that an unintegrated vertex operator has conformal weight $h=0$ and an integrated vertex operator has conformal weight $h=1$. Hence, the complete amplitude transforms as follows
\begin{IEEEeqnarray}{l}
	\int\mathrm dy\,\mathrm d\xi\,\mathrm d\eta\,\left\langle V_1(1)V_{\overline1}(-1)V_2(y)U_{\overline2}(-y)U_3(\xi)U_{\overline3}(\eta)\right\rangle=\nonumber\\
	\to\frac12\int\mathrm dx\,\mathrm d\tilde\xi\,\mathrm d\tilde\eta\,\frac{\partial(y,\xi,\eta)}{\partial(x,\tilde\xi,\tilde\eta)}\left\langle V'_1(\infty)V'_{\overline1}(0)V'_2(1)\left(\frac{\partial y}{\partial x}\right)^{-1}U'_{\overline2}(x)\left(\frac{\partial\xi}{\partial \tilde\xi}\right)^{-1}U'_3(\tilde\xi)\left(\frac{\partial\eta}{\partial\tilde \eta}\right)^{-1}U'_{\overline3}(\tilde\eta)\right\rangle\nonumber\\
	=\frac12\int\mathrm dx\,\mathrm d\tilde\xi\,\mathrm d\tilde\eta\,\left\langle V_1(\infty)V_{\overline1}(0)V_2(1)U_{\overline2}(x)U_3(\tilde\xi)U_{\overline3}(\tilde\eta)\right\rangle\ ,\label{eq::psl_amplitude}
\end{IEEEeqnarray}
where in the last line we have used that a correlator is invariant under global conformal transformations, cf.\ appendix \ref{sec::correlator1} and especially equation \eqref{eq::inv_disk}. The factor of $1/2$ can be understood from the fact that having the vertex operators $2$ and $\overline 2$ at $y$ and $-y$ (in contrast to $1$ and $x$) leads to an extra factor of $2$ due to the choice which one to consider as fixed and which one as integrated, cf.\ formulas (3.4) and (3.5) in \cite{2pt}. Furthermore, the Jacobian of the $PSL(2,\mathbb{R})$ transformation cancels against the derivatives coming from the transformation of the integrated vertex operators. Hence, we can conclude that the integrand of each of the different integration regions in table \ref{tab::4} is mapped correctly.\footnote{We have explicitly checked this and the details can be found in appendix \ref{sec::correlator1}.}\par
So far, we have only discussed the transformation behaviour of the integrand of the three-point function, but under the transformation \eqref{eq::q5} also the integration regions in table \ref{tab::4} change. The integration over $x$ is for all integration regions the same and has identical boundaries as for $y$ before the transformation \eqref{eq::q5} and so we integrate $x$ from $0$ to $1$. The integration regions of the world sheet coordinates of the third vertex operator change. The boundaries of the integration should be determined by the position of the other vertex operators. However, this is not the case for all intervals given in table \ref{tab::4}: For example the interval $[1,\infty[$ is mapped under \eqref{eq::q5} onto $]-\infty,-\sqrt{x}]$, which doesn't resemble the integration region of an open string subamplitude, because $-\sqrt{x}$ is not a world sheet position of any of the vertex operators in \eqref{eq::psl_amplitude}. Fortunately, this is not a problem, because with the transformation \eqref{eq::q5} we can combine\footnote{This can be seen in more detail as follows
\begin{IEEEeqnarray}{rCl}
	\IEEEeqnarraymulticol{3}{l}{\int_{-\infty}^{-1}\mathrm{d}\xi\int_{\xi}^{-1}\mathrm{d}\eta+\int_{1}^{\infty}\mathrm{d}\xi\int_{\xi}^{\infty}\mathrm{d}\eta+\int_{1}^{\infty}\mathrm{d}\xi\int_{-\infty}^{-1}\mathrm{d}\eta}\nonumber\\
	&\overset{\eqref{eq::q5}}{\sim}&\int_{-\sqrt{x}}^{0}\mathrm{d}\tilde\xi\int_{\tilde\xi}^{0}\mathrm{d}\tilde\eta+\int_{-\infty}^{-\sqrt{x}}\mathrm{d}\tilde\xi\int_{\tilde\xi}^{-\sqrt{x}}\mathrm{d}\tilde\eta+\int_{-\infty}^{-\sqrt{x}}\mathrm{d}\tilde\xi\int_{-\sqrt{x}}^{0}\mathrm{d}\tilde\eta\nonumber\\
	&=&\int_{-\infty}^0\mathrm{d}\tilde\xi\int_{\tilde\xi}^0\mathrm{d}\tilde\eta\ .
\end{IEEEeqnarray}}
\begin{IEEEeqnarray}l
	\mathbb A(3,\overline3,\overline1,\overline2,2,1)+\mathbb A(\overline1,\overline2,2,1,3,\overline3)+\mathbb A(\overline3,\overline1,\overline2,2,1,3)\xrightarrow[]{PSL(2,\mathbb{R})}A(3,4,1,2,5,6) \label{eq::combine}
\end{IEEEeqnarray}
to one open string integration region, where after the $PSL(2,\mathbb{R})$ transformation the vertex operator positions for the six open strings of each subamplitude are given by
\begin{IEEEeqnarray}l
	z_1=0,\qquad z_2=x,\qquad z_3=\tilde\xi,\qquad z_4=\tilde\eta,\qquad z_5=1,\qquad z_6=\infty\label{eq::open_pos}
\end{IEEEeqnarray} 
and we have identified the closed strings with the open strings as follows
\begin{IEEEeqnarray}l
	\overline 1\leftrightarrow1,\qquad\overline2\leftrightarrow 2,\qquad3\leftrightarrow3,\qquad \overline 3\leftrightarrow4,\qquad2\leftrightarrow5,\qquad 1\leftrightarrow 6.\label{eq::identification}
\end{IEEEeqnarray}
With this identification we can map the momenta of the closed strings onto the corresponding open string momenta:
\begin{IEEEeqnarray}l
	D{\cdot}k_1\leftrightarrow p_1,\quad D{\cdot}k_2\leftrightarrow p_2,\quad k_3\leftrightarrow p_3,\quad D{\cdot}k_3\leftrightarrow p_4,\quad k_2\leftrightarrow p_5,\quad k_1\leftrightarrow p_6,\label{eq::ident_mom}
\end{IEEEeqnarray}
which lets us define the open string Mandelstam variables
\begin{IEEEeqnarray}l
	\hat s_{ij}=\frac{\alpha'}{4}(p_i+p_j)^2=\frac{\alpha'}{2}p_i{\cdot}p_j.
\end{IEEEeqnarray}
As already used in section \ref{sec::psf} to split the vertex operators (from the NSNS sector) in a holomorphic and antiholomorphic part, we split the polarization tensor
\begin{IEEEeqnarray}l
	\epsilon^i_{mp}D^p_n\to\xi^i_m\otimes\overline\xi^i_pD^p_n\ ,
\end{IEEEeqnarray} 
where both $\xi^i_m$ and $\overline\xi^i_p D^p_n$ can be identified with open string polarization vectors. Hence, we can define the corresponding open string polarization vectors $\zeta^i$ as
\begin{IEEEeqnarray}{rCl}
	D{\cdot}\overline\xi^1\leftrightarrow \zeta^1,\quad D{\cdot}\overline\xi^2\leftrightarrow \zeta^2,\quad \xi^3\leftrightarrow \zeta^3,\quad D{\cdot}\overline\xi^3\leftrightarrow \zeta^4,\quad \xi^2\leftrightarrow \zeta^5,\quad \xi^1\leftrightarrow \zeta^6. \label{eq::ident_pol}
\end{IEEEeqnarray}
In general, an open string partial amplitude in terms of the closed string vertex operators is defined similarly as in \eqref{eq::open}
\begin{IEEEeqnarray}l
	A(j_1,\ldots,j_N)\sim\int_0^1\mathrm dz_{j_d}\,\smashoperator{\prod_{\substack{p=1\\p\neq a,b,c,d}}^{N}}\,\int_{z_{j_1}\leq z_{j_2}\leq\ldots\leq z_{j_N}}\mathrm{d}z_{j_p} \sum_I \prod_{i<k}^{N}|z_{j_i}-z_{j_k}|^{2\hat s_{j_ij_{k}}}(z_{j_i}-z_{j_k})^{n^I_{j_ij_{k}}}
	\mathcal{K}^I, \nonumber\\\label{eq::open_2}
\end{IEEEeqnarray}
but here the world sheet coordinates we are not integrating over are fixed to $(z_{j_a},z_{j_b},z_{j_c})=(\infty,0,1)$ and $j_i\in\{1,\ldots,N\}$. The general open string partial amplitude in \eqref{eq::open_2} is related to \eqref{eq::open} by the $PSL(2,\mathbb{R})$ transformation in \eqref{eq::q5} such that $(z_{i_a},z_{i_b},z_{i_c})\to(z_{j_a},z_{j_b},z_{j_c})$ and $z_{i_d}$ in \eqref{eq::open} is mapped to $z_{j_d}$.\footnote{This holds up to the subtlety discussed above that some of the expressions of the form given in \eqref{eq::open} have to be combined in order to yield an open string subamplitude of \eqref{eq::open_2}, cf.\ \eqref{eq::combine}.} This allows us to write \eqref{eq::amp_mon_2} in terms of only open string subamplitudes
\begin{IEEEeqnarray*}{rCl}
	\mathcal{A} & = & -A(3,4,1,2,5,6)+e^{i\pi \hat s_{35}}e^{i\pi \hat s_{45}}A(1,4,3,2,5,6)+e^{i\pi \hat s_{35}}A(1,4,2,3,5,6)\\
	&&+A(1,2,4,3,5,6)+A(1,4,2,5,3,6)+e^{i\pi \hat s_{35}}A(1,2,4,5,3,6)\\
	&&+e^{i\pi \hat s_{35}}e^{i\pi \hat s_{45}}A(1,2,5,4,3,6),\IEEEyesnumber\label{eq::amp_mon_3}
\end{IEEEeqnarray*}
which are also given in table \ref{tab::5}.
Note, that while the representation of the amplitude \req{eq::amp_mon_3} explicitly exhibits the 
symmetry under permutations $3\leftrightarrow5$ and $2\leftrightarrow 4$ the symmetries under 
$1\leftrightarrow2,5\leftrightarrow 6$ and $1\leftrightarrow4,3\leftrightarrow 6$ can only be seen after 
applying subamplitude relations. 

\begin{table}
	\begin{center}
		\begin{tabular}{|c |c |c |c |c|}
			\hline
			        \rule[-1ex]{0pt}{3.5ex}          & $\tilde\xi<\tilde\eta<0$ & \multicolumn{1}{c}{}                      & \multicolumn{1}{c}{}     &                                           \\ \hline
			 \rule[-1ex]{0pt}{3.5ex}  $\tilde\xi<0$  & $-1$                     & \multicolumn{1}{c}{}                      & \multicolumn{1}{c}{}     &                                           \\ \hline
			          \multicolumn{1}{c}{}           & \multicolumn{1}{c}{}     & \multicolumn{1}{c}{}                      & \multicolumn{1}{c}{}     & \multicolumn{1}{c}{}                      \\ \hline
			        \rule[-1ex]{0pt}{3.5ex}          &                          & $0<\tilde\eta<\tilde\xi$                  & \multicolumn{1}{c}{}     &                                           \\ \hline
			\rule[-1ex]{0pt}{3.5ex} $0<\tilde\xi<x$  &                          & $e^{i\pi \hat s_{35}}e^{i\pi \hat s_{45}}$ & \multicolumn{1}{c}{}     &                                           \\ \hline
			          \multicolumn{1}{c}{}           & \multicolumn{1}{c}{}     & \multicolumn{1}{c}{}                      & \multicolumn{1}{c}{}     & \multicolumn{1}{c}{}                      \\ \hline
			        \rule[-1ex]{0pt}{3.5ex}          &                          & $0<\tilde\eta<x$                          & $x<\tilde\eta<\tilde\xi$ &                                           \\ \hline
			\rule[-1ex]{0pt}{3.5ex}  $x<\tilde\xi<1$ &                          & $e^{i\pi \hat s_{35}}$                         & $1$                      &                                           \\ \hline
			          \multicolumn{1}{c}{}           & \multicolumn{1}{c}{}     & \multicolumn{1}{c}{}                      & \multicolumn{1}{c}{}     & \multicolumn{1}{c}{}                      \\ \hline
			        \rule[-1ex]{0pt}{3.5ex}          &                          & $0<\tilde\eta<x$                          & $x<\tilde\eta<1$         & $1<\tilde\eta<\tilde\xi$                  \\ \hline
			 \rule[-1ex]{0pt}{3.5ex} $1<\tilde\xi$   &                          & $1$                                       & $e^{i\pi \hat s_{35}}$        & $e^{i\pi \hat s_{35}}e^{i\pi \hat s_{45}}$ \\ \hline
		\end{tabular}
	\caption{$\Pi(x,\tilde\xi,\tilde\eta)$ for each integration region in the $(\tilde\xi,\tilde\eta)$-plane.}\label{tab::5}
	\end{center}
\end{table}

\def\s{\hat s}

For six open strings there are 120 open string subamplitudes\footnote{For an $N$-point function of open strings there are $(N-1)!$ different colour ordered subamplitudes.} with different colour orderings in total, but in table \ref{tab::5} we have only 7 of the 120 subamplitudes. However, these 120 subamplitudes are not independent and we can apply the reflection and parity symmetry
\begin{IEEEeqnarray}{rCl}
	A(1,2,\ldots,N-1,N)=(-1)^NA(N,N-1,\ldots,2,1)\ ,\label{eq::reflection}
\end{IEEEeqnarray}
which follows from the properties of the string world sheet, to reduce the number of independent amplitudes from $(N-1)!$ down to $\frac12(N-1)!$.\par
Furthermore, there is also an analogue of the dual Ward identity for the open string partial amplitudes of the form \eqref{eq::open_2} \cite{ovsc,Bjerrum-Bohr:2009ulz}
\begin{IEEEeqnarray}{rCl}
	0&=&A(j_1,j_2,\ldots,j_{N-1},j_N)+e^{i\pi \hat s_{12}}A(j_2,j_1,j_3,\ldots,j_{N-1},j_N)\nonumber\\
	&&+e^{i\pi (\hat s_{12}+\hat s_{13})}A(j_2,j_3,j_1,\ldots,j_{N-1},j_N)+\ldots+e^{i\pi(\hat s_{12}+\ldots+\hat s_{1N-1})}A(j_2,j_3,\ldots,j_{N-1},j_1,j_N)\ ,\nonumber\\\label{eq::monodromy_2}
\end{IEEEeqnarray}
which can be derived similarly as \eqref{eq::monodromy}. For more details on the dual Ward identities of open strings, see reference \cite{ovsc}. With these relations it is possible to reduce the $(N-1)!$ subamplitudes of an open string $N$-point function down to $(N-3)!$ independent partial amplitudes: After using \eqref{eq::reflection} for six open strings we would find $\frac12(6-1)!=60$ subamplitudes. Eventually, the monodromy relations in \eqref{eq::monodromy_2} allow us to express these remaining $60$ subamplitudes in terms of only $(N-3)!=6$ \cite{ovsc,Bjerrum-Bohr:2009ulz}. 
In fact, combining specific relations of the sort \req{eq::monodromy_2} allows  to express the first subamplitude 
of \req{eq::amp_mon_3} in terms of the remaining six as:
\begin{align}
A(1,2,5,6,3,4)&=e^{i\pi(\s_{36}-\s_{14})}\;A(1,4,2,5,3,6)+e^{i\pi(\s_{36}-\s_{14}-\s_{24})}\; A(1,2,4,5,3,6)
\nonumber\\
&+e^{i\pi(\s_{36}+\s_{34}+\s_{46})}\; A(1,2,5,4,3,6)+e^{i\pi(\s_{12}-\s_{56})}\;A(1,2,4,3,5,6)\nonumber\\
&+e^{i\pi(\s_{12}+\s_{24}-\s_{56})}\;A(1,4,2,3,5,6)+e^{-i\pi(\s_{25}+\s_{26}+\s_{56})}\;A(1,4,3,2,5,6)\ .\label{FindE}
\end{align}
More precisely,  by considering the following three monodromy relations
\begin{align}
X_1&:=A(1,2,5,6,3,4)+e^{i\pi \s_{36}}\;A(1,2,5,3,6,4)+e^{i\pi(\s_{36}+\s_{46})}\;A(1,2,5,3,4,6)\nonumber\\
&+e^{i\pi(\s_{36}+\s_{46}+\s_{16})}\;A(1,6,2,5,3,4)+e^{-i\pi \s_{56}}\;A(1,2,6,5,3,4)=0\ ,\\
X_2&:=A(1,2,5,3,6,4)+e^{i\pi \s_{14}}\;A(1,4,2,5,3,6)+e^{i\pi(\s_{14}+\s_{24})}\;A(1,2,4,5,3,6)\nonumber\\
&+e^{i\pi(\s_{14}+\s_{24}+\s_{45})}\;A(1,2,5,4,3,6)+e^{-i\pi \s_{46}}\;A(1,2,5,3,4,6)=0\ ,\\
X_3&:=A(1,4,3,5,2,6)+e^{i\pi \s_{26}}\;A(1,4,3,5,6,2)+e^{i\pi(\s_{12}+\s_{26})}\;A(1,2,4,3,5,6)\nonumber\\
&+e^{i\pi(\s_{12}+\s_{24}+\s_{26})}\;A(1,4,2,3,5,6)+e^{-i\pi\s_{25}}\;A(1,4,3,2,5,6)=0\ ,
\end{align}
and computing $X_1-e^{i\pi \s_{36}} \overline{X}_2-e^{i\pi(\s_{36}+\s_{46}+\s_{61})}X_3$  yields \req{FindE} subject to cyclic symmetries and reflection symmetry \eqref{eq::reflection} of the partial amplitudes.
Eventually, inserting \req{FindE} into \req{eq::amp_mon_3} gives:
\begin{align}\label{eq::amp_mon_4}
\mathcal{A}&=2i\sin(\pi  \hat s_{35})\;A(1,2,4,5,3,6)+2i\sin[\pi (\hat s_{35}+\hat s_{45})]\;A(1,2,5,4,3,6)\\
&\equiv 2i\sin(\pi  s_{23})\;A(\overline1,\overline2,\overline3,2,3,1)+2i\sin[\pi (s_{23}+s_{2\overline3})]\;A(\overline1,\overline2,2,\overline3,3,1)\ .\label{eq::amp_mon_4a}
\end{align}
Interestingly, the lowest order in $\ap$ of \req{eq::amp_mon_4} vanishes. Consequently the $\ap$ expansion of \req{eq::amp_mon_4} only starts at the subleading order in $\ap$, cf. also section \ref{LowEXP}.

After performing the $PSL(2,\mathbb R)$ transformation and using the monodromy relations for open strings the integration regions in the $(\tilde\xi,\tilde\eta)$-plane for each of the partial amplitudes are given by
\begin{IEEEeqnarray}l
	\begin{IEEEeqnarraybox}[][c]{rlrClCrCl}
		\IEEEstrut
		A(\overline 1,\overline2,\overline3,2,3,1)&:\qquad \mathcal I_1=\left\{(\tilde\xi,\tilde\eta)\in\mathbb R^2\,|\,1<\tilde\xi<\infty,x<\tilde\eta<1\right\}\ ,\\
		A(\overline 1,\overline2,2,\overline3,3,1)&:\qquad \mathcal I_2=\left\{(\tilde\xi,\tilde\eta)\in\mathbb R^2\,|\,1<\tilde\xi<\infty,1<\tilde\eta<\tilde\xi\right\}
		\IEEEstrut
	\end{IEEEeqnarraybox}\label{eq::int_domain}
\end{IEEEeqnarray}
and $x\in\mathbb R$ is always $0<x<1$. We have written the scattering of three closed strings on the disk in terms of a scattering process of six open strings on the disk. Moreover, we could express the scattering of three closed strings in terms of {\it only two} independent partial open string amplitudes instead of six as originally anticipated in \cite{ovsc}. Further, we have explicitly shown the connection between open and closed strings by changing the vertex operator position fixing from $(-1,y,1)$ to $(0,1,\infty)$, which is more common for open string scattering amplitudes and allows us to apply the methods presented in \cite{npt_1, npt_2}. Hence, for the scattering of 3 NSNS states, we can immediately express the partial amplitudes in \eqref{eq::amp_mon_4a} by using equation $(2.5)$ in \cite{npt_2} (together with the identification \eqref{eq::identification} and including some overall constants appropriate for our definition of the closed string amplitude) as
	\begin{IEEEeqnarray}{rCl}
		A(\overline1,\overline2,\overline3,2,3,1)&=& i g_c^3T_p\sum_{\sigma\in S_{3}}A_{\rm YM}(\overline1,\overline2_\sigma,3_\sigma,\overline3_\sigma,2,1)F^{(\overline2_\sigma3_\sigma\overline3_\sigma)}_{\mathcal I_1}\ ,\label{eq::partial_1}\\
		A(\overline1,\overline2,2,\overline3,3,1)&=& i g_c^3T_p\sum_{\sigma\in S_{3}}A_{\rm YM}(\overline1,\overline2_\sigma,3_\sigma,\overline3_\sigma,2,1)F^{(\overline2_\sigma3_\sigma\overline3_\sigma)}_{\mathcal I_2}\ ,\label{eq::partial_2}
	\end{IEEEeqnarray}
where $\sigma\in S_3$ describes the permutation of the labels $(\overline2,3,\overline3)$, which we are integrating over. 
Plugging this into \eqref{eq::amp_mon_4a} we finally end up with
	\begin{IEEEeqnarray}{rCl}
		\mathcal{A}&=& -2 g_c^3T_p\sum_{\sigma\in S_3}\left\{\sin(\pi s_{23})F^{(\overline2_\sigma3_\sigma\overline3_\sigma)}_{\mathcal{I}_1}+\sin\left[\pi (s_{23}+s_{2\overline3})\right]F^{(\overline2_\sigma3_\sigma\overline3_\sigma)}_{\mathcal{I}_2}\right\}\ A_{\rm YM}(\overline1,\overline2_\sigma,3_\sigma,\overline3_\sigma,2,1)\ .\nonumber\\\label{eq::A3}
	\end{IEEEeqnarray}
The functions $F$ are given in \eqref{Sample1} and $A_{\rm YM}$ in appendix \ref{sec::correlator} below. In that appendix we also give a detailed derivation how one can obtain the result in \eqref{eq::A3} from the CFT correlator of three closed strings on the disk, i.e.\ we explicitly discuss the $PSL(2,\mathbb R)$ transformation of the correlator and compute the contraction of the vertex operators with Wick's theorem. Furthermore, we are going to take the low energy limit of \eqref{eq::amp_mon_4a} in section \ref{LowEXP}. To do so we relate the integrals $F$ to low energy expansions of open string scattering amplitudes given in \cite{npt_2}. 

We would like to stress that in the derivation of \eqref{eq::amp_mon_4a} we did not make any assumptions about the three closed string states except that they are massless. Thus, formula \eqref{eq::amp_mon_4a} holds completely generally and the closed string states could be either bosons or fermions (i.e.\ in the language of the RNS theory, they could be from any of the four sectors, NSNS, RR, RNS or NSR). In the general case the open string amplitudes would then involve both gluons and gluinos. Also formula \eqref{eq::A3} should generalize to that general case when replacing the YM amplitudes by the appropriate SYM amplitudes.

Note, that our result \req{eq::amp_mon_4a} describes the disk amplitude of three closed strings.
In contrast, in \cite{ovsc} the scattering of three closed strings has been considered on the double cover, which incorporates manifest symmetry between left-- and right--movers. The resulting string amplitude can be found in Eq. (3.65) of \cite{ovsc} and reads in terms of closed string labels:
\begin{align}
\mathcal{A}(1,2,3)&=\sin(\pi s_{2\overline3})\ A(\overline 1,\overline 2,3,2,\overline 3,1)
+\sin(\pi s_{23})\ A(\overline 1,2,3,\overline 2,\overline 3,1)\nonumber\\
&+\sin(\pi s_{13})\ \lf[\ A(\overline 1,\overline 2,2,3,1,\overline 3)+A(\overline 1,2,\overline 2,3,1,\overline 3)\ \ri]\nonumber\\
&+\lf(\ \sin\lf[\pi\lf(s_{23}+s_{2\overline3}\ri)\ri]+
\sin\lf[\pi\lf(s_{13}+s_{1\overline3}\ri)\ri]\ \ri)\ \lf[\ A(\overline 1,\overline 2,2,3,\overline 3,1)+A(\overline 1,2,\overline 2,3,\overline 3,1)\ \ri]\ .\label{expression1}
\end{align}
Let us discuss the connection between \req{expression1} and \req{eq::amp_mon_4a} for a specific kinematical configuration of three closed strings.
Imposing (by hand) on \req{eq::amp_mon_4a} the symmetries $3\leftrightarrow\overline3$ (i.e. $3\leftrightarrow4$ at the open string sector) and $2\leftrightarrow\overline2$ (i.e. $2\leftrightarrow5$ at the open string sector) yields the expression:
\begin{align}
\mathcal{A}_{S^2} & := 2i\sin(\pi  s_{23})\;A(1,\overline1,\overline2,\overline3,2,3)+2i\sin[\pi (s_{23}+s_{2\overline3})]\;A(1,\overline1,\overline2,2,\overline3,3)\nonumber\\
&+2i\sin(\pi  s_{2\overline3})\;A(1,\overline1,\overline2,3,2,\overline3)+2i\sin[\pi (s_{23}+s_{2\overline3})]\;A(1,\overline1,\overline2,2,3,\overline3)\nonumber\\
&+2i\sin(\pi  s_{2\overline3})\;A(1,\overline1,2,\overline3,\overline2,3)+2i\sin[\pi (s_{23}+s_{2\overline3})]\;A(1,\overline1,2,\overline2,\overline3,3)\nonumber\\
&+2i\sin(\pi  s_{23})\;A(1,\overline1,2,3,\overline2,\overline3)+2i\sin[\pi (s_{23}+s_{2\overline3})]\;A(1,\overline1,2,\overline2,3,\overline3)\ .\label{expression2}
\end{align}
By applying six--point open string monodromy relations subject to \req{kin1}--\req{kin3} it can be shown that the two expressions \req{expression1} and \req{expression2} agree up to an overall factor, i.e.\  $\mathcal{A}_{S^2} = 2i\; \mathcal{A}(1,2,3)$. This gives for each concrete kinematical factor the link between the world--sheet integrations on the disk and those on its double cover. 

The symmetrization is equivalent to extending the integration regions of $y$ and $\eta$ in \eqref{eq::q3} to $[0,1]$ and $]-\infty, \infty[$, respectively. The resulting integrals miss some poles that are present when working on the disk. We give more details on this in appendix \ref{sec:symmetrization} and we come back to the pole structure of our result in section \ref{InterpretLowEXP}.

\section{Low energy expansion and effective action}\label{LowEXP}
\def\ss{{\small}}
\def\si{\sigma}
\def\t{\hat t}

\subsection[Expansion in inverse string tension $\ap$]{Expansion in inverse string tension \boldmath{$\ap$}}
\label{alphaprime_expand}

In this subsection we shall evaluate the $\alpha'$--expansion of the amplitude \req{eq::amp_mon_4a}, building on the results of \cite{npt_2}. More concretely, the two integrals  in \req{eq::partial_1} and \eqref{eq::partial_2}
\be
F^{(\overline2_\si 3_\si\overline 3_\si)}_{\mathcal
  I_n}=-\int_{\mathcal{I}_n}\mathrm{d}z_{\overline2}\,\mathrm{d}z_3\,\mathrm{d}z_{\overline3}\,\Biggl(\prod_{i<j}|z_{ij}|^{s_{ij}}\Biggl)\frac{s_{\overline1\overline2_\si}}{z_{\overline1\overline2_\si}}\frac{s_{\overline3_\si2}}{z_{\overline3_\si
    2}}\left(\frac{s_{\overline13_\si}}{z_{\overline13_\si}}+\frac{s_{\overline2_\si3_\si}}{z_{\overline2_\si3_\si}}\right)\ ,\quad\si\in S_3,\ n=1,2\ ,\label{Sample1}
\ee
which are integrated over the two domains  \req{eq::int_domain}, need to be expressed in terms of a power series w.r.t.\ small $\ap$. The six permutations $\si\in S_3$ act on the three labels  $\overline2, 3$ and $\overline 3$ as $i_\si:=\si(i)$.
In \req{Sample1} the two domains  \req{eq::int_domain} 
\begin{align}
  \Ic_1:\ z_{\bar 1}<z_{\bar 2}<z_{\bar 3}<z_2<z_3<z_1\ ,\nonumber\\
  \Ic_2:\ z_{\bar 1}<z_{\bar 2}<z_2<z_{\bar 3}<z_3<z_1\ ,\label{Region1}
\end{align}
are subject to the fixing
\be
z_{\bar 1}=0\ ,\ z_2=1\ ,\ z_1=\infty\ ,
\ee
which may be rescinded by introducing the volume of the conformal Killing group:
\be
F^{(\overline2_\si 3_\si\overline 3_\si)}_{\mathcal I_n}=-V_{\rm
  CKG}^{-1}\int_{\mathcal{I}_n}\mathrm{d}z_i\;\Biggl(\prod_{i<j}|z_{ij}|^{s_{ij}}\Biggl)\fc{1}{z_{\bar1
    2}z_{\bar 11}z_{21}}\;\frac{s_{\overline1\overline2_\si}}{z_{\overline1\overline2_\si}}\frac{s_{\overline3_\si2}}{z_{\overline3_\si 2}}\left(\frac{s_{\overline13_\si}}{z_{\overline13_\si}}+\frac{s_{\overline2_\si3_\si}}{z_{\overline2_\si3_\si}}\right)\ ,\quad\si\in S_3\ .\label{Sample2}
\ee
In view of \eqref{Region1}, a more suitable gauge choice would be
\be
z_{\bar 1}=0\ ,\quad z_3=1\ ,\quad z_1=\infty\ ,
\ee
leading to:
\be
F^{(\overline2_\si 3_\si\overline 3_\si)}_{\mathcal I_n}=-\int_{\mathcal{I}_n}\mathrm{d}z_{2}\,\mathrm{d}z_{\overline2}\,\mathrm{d}z_{\bar 3}\ \fc{z_{\bar 13}}{z_{\bar 12}}\;\Biggl(\prod_{i<j}|z_{ij}|^{s_{ij}}\Biggl)\frac{s_{\overline1\overline2_\si}}{z_{\overline1\overline2_\si}}\frac{s_{\overline3_\si2}}{z_{\overline3_\si 2}}\left(\frac{s_{\overline13_\si}}{z_{\overline13_\si}}+\frac{s_{\overline2_\si3_\si}}{z_{\overline2_\si3_\si}}\right)\ ,\quad\si\in S_3\ .\label{Sample3}
\ee
After introducing the maps
\be
\begin{array}{lcl}
\Ic_1,\ \ \varphi_1:\;(\bar 1,\bar2,\bar 3,2,3,1)&\mapsto&(1,2,3,4,5,6)\ ,\ 
\\[2mm]
\Ic_2,\ \ \varphi_2:\;(\bar 1, \bar2,2,\bar 3,3,1)&\mapsto&(1,2,3,4,5,6)\ ,\end{array}
\label{Map1}
\ee
we may cast \req{Sample3} into the following form
\be
F^{(\overline2_\si 3_\si\overline 3_\si)}_{\mathcal{I}_n}=\begin{cases}
\ds{\lf. F^{(\varphi_1(\bar2_{\si}) \varphi_1(3_{\si}),\varphi_1( \overline 3_\si),{4})}\ri|_{\hat s_{ij}\ra \varphi_1^{-1}(\hat s_{ij})}\ ,}&n=1\ ,\\
\ds \lf. F^{(\varphi_2(\bar2_{\si}) \varphi_2(3_{ \si}),\varphi_2( \overline 3_\si),{3})}\ri|_{\hat s_{ij}\ra \varphi_2^{-1}(\hat s_{ij})}\ ,&n=2\ ,
\end{cases}\label{findMAP}
\ee
with:
\bea\label{Sample4}
\ds F^{(2_\si, 3_\si, 5_\si,4)}&=&\ds-\int\limits_{0<z_2<z_3<z_4<1}\mathrm{d}z_{2}\,\mathrm{d}z_3\,\mathrm{d}z_{4}\ \Biggl(\prod_{i<j}|z_{ij}|^{\hat s_{ij}}\Biggl)\ \fc{1}{z_{41}}\ \frac{\hat s_{12_\si}}{z_{12_\si}}\frac{\hat s_{45_\si}}{z_{5_\si 4}}\left(\frac{\hat s_{13_\si}}{z_{13_\si}}+\frac{\hat s_{2_\si 3_\si}}{z_{2_\si 3_\si}}\right)\ ,\\
\ds F^{(2_\si, 4_\si, 5_\si,3)}&=&\ds-\int\limits_{0<z_2<z_3<z_4<1}\mathrm{d}z_{2}\,\mathrm{d}z_3\,\mathrm{d}z_{4}\ \Biggl(\prod_{i<j}|z_{ij}|^{\hat s_{ij}}\Biggl)\ \fc{1}{z_{31}}\ \frac{\hat s_{12_\si}}{z_{12_\si}}\frac{\hat s_{35_\si}}{z_{5_\si 3}}\left(\frac{\hat s_{14_\si}}{z_{14_\si}}+\frac{\hat s_{2_\si4_\si}}{z_{2_\si 4_\si}}\right)\ ,
\eea
respectively. In \req{findMAP} the inverse map $\varphi^{-1}_i$ of \req{Map1} acts on the six--point kinematic invariants $\hat s_{ij}$ to be specified below in \eqref{eq::phi1} and \eqref{eq::phi2}.
The pair of six functions $F^{(abc4)}$ and $F^{(abc3)}$ defined in \req{Sample4} corresponds to a subset of the extended set of $24$ functions  introduced in \cite{npt_2}, respectively relevant for the six point open superstring amplitude for canonical color ordering $(1,2,3,4,5,6)$.  The integrals integrate to triple  hypergeometric functions. In fact, thanks to some (dual) monodromy relations all these $24$ functions can be expressed in terms of a six--dimensional basis $F^{(abc)}\equiv F^{(abc5)}$ as (with $K_i^\ast=(K_i^t)^{-1}$)   \cite{npt_2}
\be
\lf(\begin{matrix}
F^{(2354)}\\
F^{(3254)}\\
F^{(5324)}\\
F^{(3524)}\\
F^{(5234)}\\\
F^{(2534)}\end{matrix}\ri)=K_1^\ast\ 
\lf(\begin{matrix}
F^{(234)}\\
F^{(324)}\\
F^{(432)}\\
F^{(342)}\\
F^{(423)}\\
F^{(243)}\end{matrix}\ri)\ ,\qquad 
\lf(\begin{matrix}
F^{(2453)}\\
F^{(4253)}\\
F^{(5423)}\\
F^{(4523)}\\
F^{(5243)}\\\
F^{(2543)}\end{matrix}\ri)=K_2^\ast\ 
\lf(\begin{matrix}
F^{(234)}\\
F^{(324)}\\
F^{(432)}\\
F^{(342)}\\
F^{(423)}\\
F^{(243)}\end{matrix}\ri)\ ,
\ee
respectively. The $6\times6$ matrices $K_i$ follow from the corresponding (dual) subamplitude
relations and are given by \cite{npt_2}
\be
(K_1)^\si_\pi=\hat s_{46}^{-1}
\begin{pmatrix}
\ss{\s_5-\s_{123}}&\ss{0}&\ss{0}&\ss{0}&\ss{\s_{14}}&\ss{-\hat d_1}\\
\ss{0}&\ss{\s_5-\s_{123}}&\ss{\s_{14}}&\ss{\s_3+\s_{14}}&\ss{0}&\ss{0}\\
\ss{\fc{\s_1\s_4\hat d_0}{\s_{15}\s_{246}}}&\ss{\fc{\s_4\s_{13}(\s_{25}-\s_{46})}{\s_{15}\s_{246}}}&
\ss{\fc{-\s_{13}\s_{14}\s_{25}}{\s_{15}\s_{246}}}&\ss{\fc{-\s_{13}\s_{25}(\s_3+\s_{14})}{\s_{15}\s_{246}}}&
\ss{\fc{\s_{14}(\s_{46}-\s_1)\hat d_0}{\s_{15}\s_{246}}}&
\ss{\fc{\s_1(\s_3+\s_4)\hat d_0}{\s_{15}\s_{246}}}\\
\ss{\fc{-\s_1\s_4}{\s_{246}}}&\ss{\fc{-\s_4(\s_1+\s_2)}{\t_{246}}}&
\ss{\fc{\s_{14}\hat d_4}{\s_{246}}}& 
\ss{\fc{(\s_{14}+\s_3)\hat d_4}{\s_{246}}}  &
\ss{\fc{\s_{14}(\s_1-\s_{46})}{\s_{246}}} &\ss{\fc{-\s_1(\s_3+\s_4)}{\s_{246}}}\\
\ss{\fc{\s_1\s_4(\s_{35}-\s_{46})}{\s_{15}\s_{125}}}&
\ss{\fc{\s_4\s_{13}\hat d_3}{\s_{15}\s_{125}}}&
\ss{\fc{(\s_{46}-\s_{13})\hat d_3\s_{14}}{\s_{15}\s_{125}}}&
\ss{\fc{(\s_4+\s_{24})\s_{13}\hat d_3}{\s_{15}\s_{125}}}&
\ss{\fc{-\s_1\s_{14}\s_{35}}{\s_{15}\s_{125}}}&
\ss{\fc{\s_1\s_{35}\hat d_1}{\s_{15}\s_{125}}}\\
\ss{\fc{\s_4(\s_1-\t_1)}{\s_{125}}}&\ss{\fc{-\s_4\s_{13}}{\s_{125}}}&
\ss{\fc{\s_{14}(\s_{13}-\s_{46})}{\s_{125}}}&\ss{\fc{-\s_{13}(\s_4+\s_{24})}{\s_{125}}}&
\ss{\fc{-\s_{14}\hat d_2}{\s_{125}}}&\ss{\fc{\hat d_1\hat d_2}{
\s_ {125}}}\end{pmatrix}
\ee
and:
\be
(K_2)^\si_\pi=\hat s_{36}^{-1}
\begin{pmatrix}
\ss{\s_{123}-\s_1}& \ss{\s_{13}} & \ss{0}& \ss{0} &\ss{0}&\ss{\hat d_{14}}  \\
\ss{0}&\ss{0}&\ss{\s_3+\s_{13}}&\ss{\s_{13}}&\ss{\hat d_{14}}&\ss{0}\\
\ss{\fc{\s_1 (\s_{345}-\s_4) \hat d_{13}}{\s_{145} \s_{15}}}&\ss{\fc{(\s_{36}-\s_1) \s_{13} \hat d_{13}}{\s_{145} \s_{15}}}&\ss{\fc{-(\s_3+\s_{13}) \s_{14} \s_{25}}{\s_{145} \s_{15}}}&
\ss{\fc{-\s_{13} \s_{14} \s_{25}}{\s_{145} \s_{15}}} &
\ss{\fc{\hat d_8 \s_{14} \s_{35}}{\s_{145} \s_{15} }}&\ss{\fc{\s_1 \s_{35} \hat d_{13}}{\s_{145} \s_{15}}} \\
\ss{\fc{\s_1 (\s_4-\s_{345})}{\s_{145}}}&\ss{\fc{(\s_1-\s_{36}) \s_{13}}{\s_{145}}}&\ss{\fc{(\s_3+\s_{13}) \hat d_5}{\s_{145}}}&\ss{\fc{\s_{13} \hat d_5}{\s_{145}}}&\ss{\fc{-(\s_1+\s_{24}) \s_{35}}{\s_{145}}}&
\ss{\fc{-\s_1 \s_{35}}{\s_{145}}}\\
\ss{\fc{\s_1 \s_4 (\s_1-\s_{123})}{\s_{125} \s_{15}}}&\ss{\fc{-\s_1 \s_4 \s_{13}}{\s_{125} \s_{15}}}&\ss{\fc{ \s_{14} (\s_2+\s_{35})\hat d_3}{\s_{125} \s_{15}}} &\ss{\fc{\s_{13}\hat d_3\hat d_7 }{\s_{125} \s_{15}}} &\ss{\fc{\s_{14} \s_{35}\hat d_3 }{\s_{125} \s_{15}}}&\ss{\fc{\s_1 (\s_4-\s_{36}) \s_{35}}{\s_{125} \s_{15}}}\\
\ss{\fc{(\s_{123}-\s_1)\hat d_6}{\s_{125}}}&\ss{\fc{\s_{13} \hat d_6}{\s_{125}}}&\ss{\fc{-\s_{14} (\s_2+\s_{35})}{\s_{125}}}&\ss{\fc{-\hat d_7 \s_{13}}{\s_{125}}}&\ss{\fc{-\s_{14} \s_{35}}{\s_{125} }}&\ss{\fc{\hat d_1 \s_{35}}{\s_{125}}}\end{pmatrix}
\ee
In the above matrices we have introduced $\hat d_0=\s_{15}+\s_{35},\ 
\hat d_1=\s_3-\s_5+\s_{123}, \hat d_2=\s_1-\s_4-\s_5,\ \hat d_3=\s_3-\s_5-\s_{345},\ \hat d_4=\s_4+\s_5-\s_{13}, \hat d_5=\s_1+\s_{24}-\s_{36},\ \hat d_6=-\s_1+\s_5+\s_{35},\ 
\hat d_7=\s_1-\s_5+\s_{24}-\s_{35},\ \hat d_8=\s_6-\s_4+\s_{13}-\s_{24}, \hat d_{13}=\s_{15}+\s_{45}, \hat d_{14}=\s_{123}-\s_1+\s_3$ and
$\s_{ijk}= \s_{ij}+\s_{ik}+\s_{jk}$.  

Let us now turn to the $\ap$--expansion of  the six integrals \req{Sample1} and likewise the $\ap$--expansion of  twelve integrals \req{findMAP} specified by \req{Sample4}. 
With the above preparations the latter are expressed by the $\ap$--expansion of the six functions $F^{(abc)}$ as:
\be
\lf(\begin{matrix}
F^{(\bar2\bar33)}_{\Ic_i}\\
F^{(\bar3\bar23)}_{\Ic_i}\\
F^{(3\bar3\bar2)}_{\Ic_i}\\
F^{(\bar33\bar2)}_{\Ic_i}\\
F^{(3\bar2\bar3)}_{\Ic_i}\\
F^{(\bar23\bar3)}_{\Ic_i}\end{matrix}\ri)=K_i^\ast\ 
\lf.\lf(\begin{matrix}
F^{(234)}\\
F^{(324)}\\
F^{(432)}\\
F^{(342)}\\
F^{(423)}\\
F^{(243)}\end{matrix}\ri)\ri|_{\hat s_{ij}\ra \varphi_i^{-1}(\hat s_{ij})}\ ,\quad i=1,2\ .\label{alpharel}
\ee
Note, that the inverse maps $\varphi_i^{-1}$ act on the six--point kinematic invariants $\hat s_{ij}$ as
follows
\be
\varphi_1^{-1}:\lf\{\begin{matrix}
\hat s_{12} \mapsto s_{12}& \\
\hat s_{23} \mapsto s_{23},& \hat s_{123}\mapsto s_{12}+s_{23}+s_{13},\\
\hat s_{34} \mapsto s_{2\bar3},&\hat s_{234}\mapsto s_{23}+s_{2\bar3}+s_{2\bar2},\\
\hat s_{45} \mapsto s_{23},&\hat s_{345}\mapsto s_{23}+s_{2\bar3}+s_{3\bar3},\\
\hat s_{56} \mapsto s_{13},&\\
\hat s_{61} \mapsto s_{1\bar1},&\\
\end{matrix}\ri. \label{eq::phi1}
\ee
and
\be
\varphi_2^{-1}:\lf\{\begin{matrix}
\hat s_{12} \mapsto s_{12}& \\
\hat s_{23} \mapsto s_{2\bar2},& \hat s_{123}\mapsto s_{12}+s_{1\bar2}+s_{2\bar2},\\
\hat s_{34} \mapsto s_{2\bar3},&\hat s_{234}\mapsto s_{23}+s_{2\bar3}+s_{2\bar2},\\
\hat s_{45} \mapsto s_{3\bar3},&\hat s_{345}\mapsto s_{23}+s_{2\bar3}+s_{3\bar3},\\
\hat s_{56} \mapsto s_{13},&\\
\hat s_{61} \mapsto s_{1\bar1},&\\
\end{matrix}\ri.  \label{eq::phi2}
\ee
respectively subject to the constraints \req{kin1}--\req{kin3}.
The methods to find the low energy expansion of the latter has been pioneered in \cite{Oprisa:2005wu} and subsequently been applied and systematized  in \cite{Stieberger:2006te,Broedel:2013tta}.
By applying these techniques we find, cf. also \cite{npt_2}:
\begin{align}
 F^{(234)}&=1-\z_2\ (s_{45}s_{56}+s_{12}s_{61}-s_{45}s_{123}-s_{12}s_{345}+s_{123}s_{345})+\z_3(\ldots)+\Oc(\ap^4)\ ,\label{exp1}\\
 F^{(324)}&=-\z_2\ s_{13}(s_{23}-s_{61}+s_{345})+\z_3(\ldots)+\Oc(\ap^4) \\
 F^{(432)}&=-\z_2\ s_{14}s_{25}+\z_3\ s_{14}s_{25}\ (-s_{23}-s_{34}+s_{56}+s_{61}+s_{123}+s_{234}+s_{345})+\Oc(\ap^4)\ ,\\
 F^{(342)}&=\zeta_2\ s_{13}s_{25}+\zeta_3\ s_{13}s_{25}\ (-s_{12} + s_{23} + 2s_{34} - s_{16} -s_{123} - 2s_{234} -s_{345})+\Oc(\ap^4)\ ,\\
 F^{(423)} &=\zeta_2\ s_{14}s_{35}+\zeta_3\ s_{14}s_{35}\ (2s_{23} + s_{34}-s_{45}-s_{56}-s_{123}-2s_{234}-s_{345})+\Oc(\ap^4)\ ,\\
  F^{(243)} &=-\zeta_2\ s_{35}(s_{34}-s_{56}+s_{123})+\zeta_3\ s_{35}\ [-2s_{12}s_{23}-2s_{12}s_{34}+s_{34}^2+s_{34}s_{45}-s_{45}s_{56}\nonumber\\
&-s_{56}^2+s_{123}(2s_{23}+s_{45}+s_{123})+2s_{12}s_{234}+s_{345}(s_{34}-s_{56}+s_{123})]+\Oc(\ap^4)\ .\label{exp6}
\end{align}
Eventually, in \req{eq::partial_1} and \eqref{eq::partial_2} the required combination of \req{alpharel} 
comprises
\begin{align}
&\hskip1.5cm\sin(\pi s_{23})\lf(\begin{matrix}
F^{(\bar2\bar33)}_{\Ic_1}\\[2mm]
F^{(\bar3\bar23)}_{\Ic_1}\\[2mm]
F^{(3\bar3\bar2)}_{\Ic_1}\\[2mm]
F^{(\bar33\bar2)}_{\Ic_1}\\[2mm]
F^{(3\bar2\bar3)}_{\Ic_1}\\[2mm]
F^{(\bar23\bar3)}_{\Ic_1}\end{matrix}\ri)+\sin[\pi( s_{23}+s_{2\bar3})]\lf(\begin{matrix}
F^{(\bar2\bar33)}_{\Ic_2}\\[2mm]
F^{(\bar3\bar23)}_{\Ic_2}\\[2mm]
F^{(3\bar3\bar2)}_{\Ic_2}\\[2mm]
F^{(\bar33\bar2)}_{\Ic_2}\\[2mm]
F^{(3\bar2\bar3)}_{\Ic_2}\\[2mm]
F^{(\bar23\bar3)}_{\Ic_2}\end{matrix}\ri)=\pi\;\lf(\begin{matrix} 
0\\[3mm] 
s_{23}\\[3mm]  
0\\[3mm] 
s_{23}+s_{2\bar3}\\[3mm] 
0\\[3mm] 
0\\[3mm] 
\end{matrix}\ri)\nonumber\\[5mm]
&+\pi\zeta_2\;\tiny\lf(\begin{matrix} 
-s_{12}s_{23}\;(s_{13}+s_{1\bar3}-s_{2\bar3})\\[3mm] 
s_{23}[s_{12}^2 +s_{12} s_{1\bar2}+s_{12} s_{13} +s_{1\bar2} s_{13} +s_{13}^2 +s_{13} s_{1\bar3} +2 s_{12} s_{23}+s_{1\bar2} s_{23}+s_{13} s_{23}+s_{23}^2+s_{12}  s_{2\bar3}+s_{23}s_{2\bar3}] \\[3mm] 
-s_{1\bar3} (s_{12} s_{23}+s_{1\bar2} s_{23}-s_{23} s_{2\bar3}-s_{2\bar3}^2)\\[3mm]  
(s_{23}+s_{2\bar3})[s_{12}^2+s_{12} s_{1\bar2}+s_{13}^2+s_{13} s_{1\bar3}+2 s_{12} s_{23}+s_{1\bar2} s_{23}+s_{13} s_{23}+s_{23}^2+(s_{12}+s_{23}) s_{2\bar3}]+s_{13} s_{23}(s_{12}+s_{1\bar2}) \\[3mm] 
s_{1\bar3}s_{2\bar3}^2\\[3mm] 
-s_{12}s_{2\bar3}^2\\[3mm] 
\end{matrix}\ri)\nonumber\\[5mm]
&\hskip-1cm+\pi\zeta_3\tiny\lf(\begin{matrix} 
s_{12} s_{23} [(s_{13}-s_{1\bar3}) (s_{13}+s_{1\bar3}+s_{23})-(s_{12}+2 s_{1\bar2}) (s_{13}+s_{1\bar3}-s_{2\bar3})-s_{23} s_{2\bar3}+s_{2\bar3}^2]\\[3mm] 
\ldots\\[3mm]  
s_{1\bar3} [s_{12}^2 s_{23}-s_{1\bar2}^2 s_{23}-s_{1\bar2} s_{23} (s_{13}+2 s_{1\bar3}+s_{23})+2 s_{1\bar2} s_{2\bar3}^2+s_{2\bar3} (s_{23}+s_{2\bar3}) (s_{13}+2 s_{1\bar3}+s_{23}+2 s_{2\bar3})+s_{12} (-s_{13} s_{23}-2 s_{1\bar3} s_{23}+s_{23}^2+2 s_{23} s_{2\bar3}+2 s_{2\bar3}^2)]\\[3mm]  
\ldots\\[3mm] 
s_{1\bar3} s_{2\bar3}\; [2 s_{12} (s_{23}+s_{2\bar3})+s_{2\bar3} (2 s_{1\bar2}+s_{13}+2 s_{1\bar3}+s_{23}+2 s_{2\bar3})]\\[3mm] 
-s_{12} s_{2\bar3}\; [s_{1\bar3} (-2 s_{23}+s_{2\bar3})+s_{2\bar3} (s_{12}+2 s_{1\bar2}+s_{13}-s_{23}+s_{2\bar3})]\\[3mm] 
\end{matrix}\ri)\nonumber\\
&\hskip1.5cm+\Oc(\ap^5)\ ,\label{eq::ap-expansion}
\end{align}
after inserting \req{exp1}--\req{exp6}.

\goodbreak

\subsection{Some interpretational remarks}\label{InterpretLowEXP}
\def\ss{{\small}}
\def\si{\sigma}
\def\t{\hat t}

\begin{figure}
\begin{center}
\includegraphics[width=0.5\textwidth]{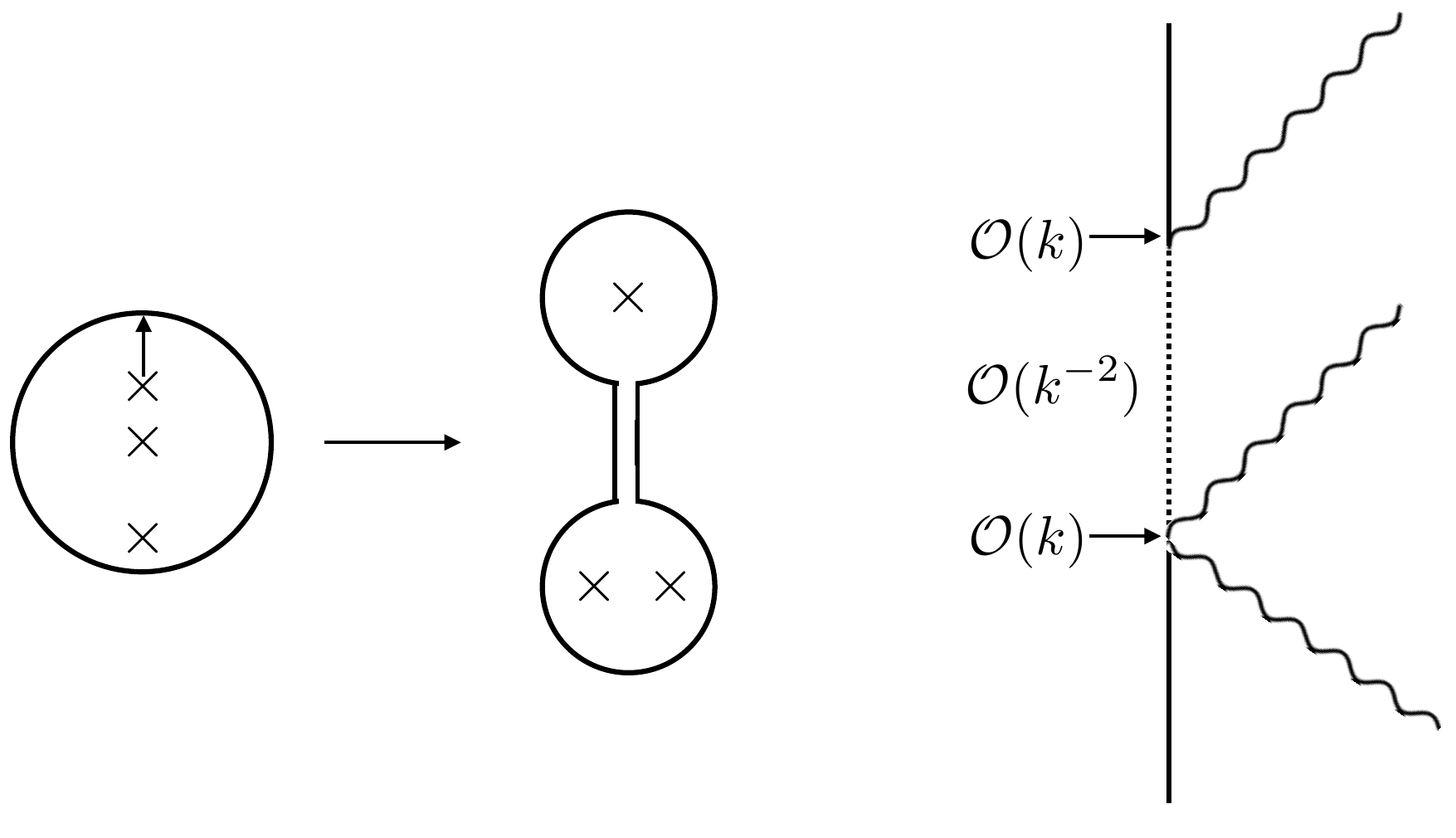} \\[1cm]
\includegraphics[width=0.5\textwidth]{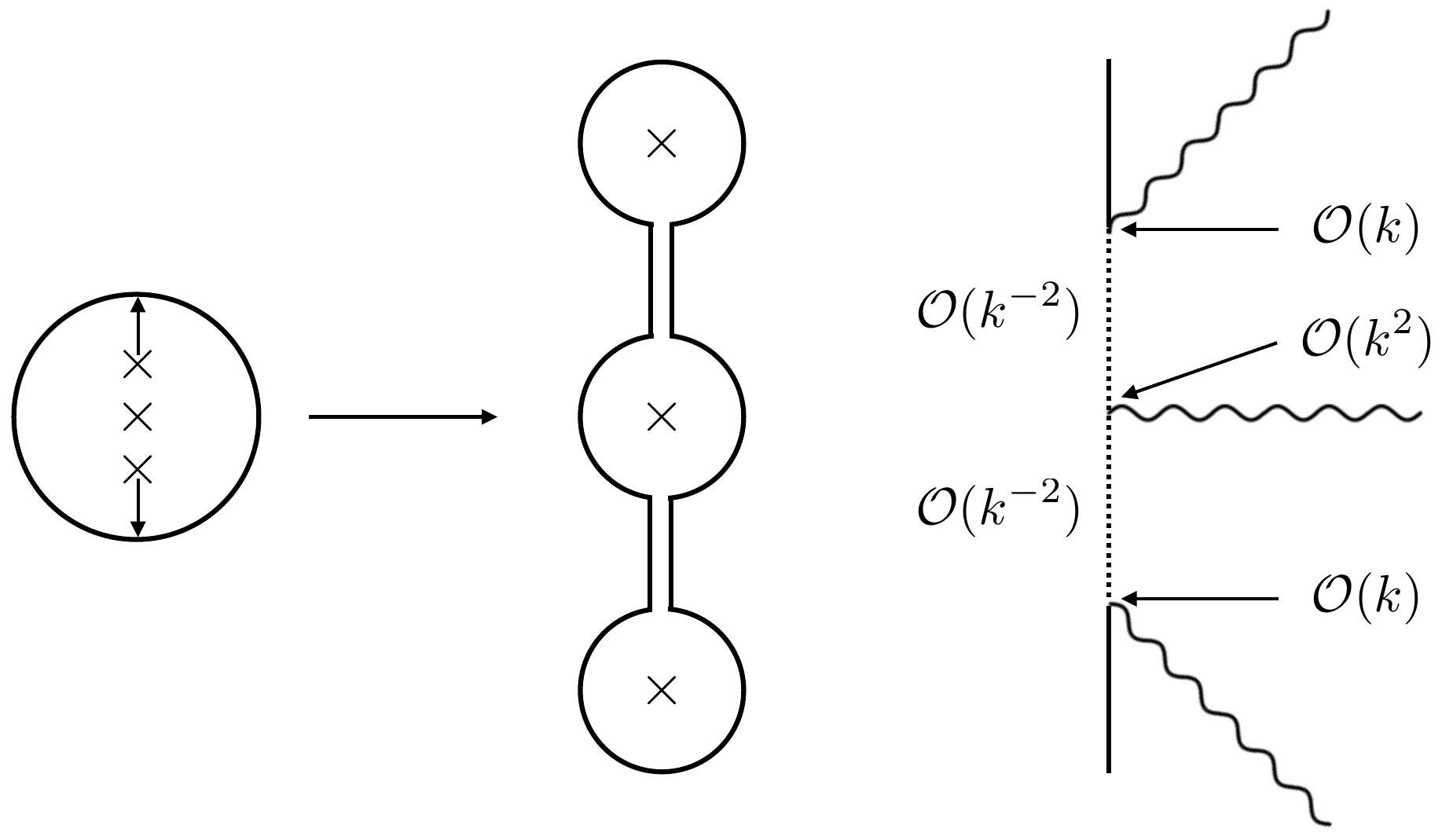} \\[1cm]
\caption{Degeneration limits with no vertex collisions. \label{degenerate_0_collide}}
\end{center}
\end{figure}

\begin{figure}
\begin{center}
\includegraphics[width=0.5\textwidth]{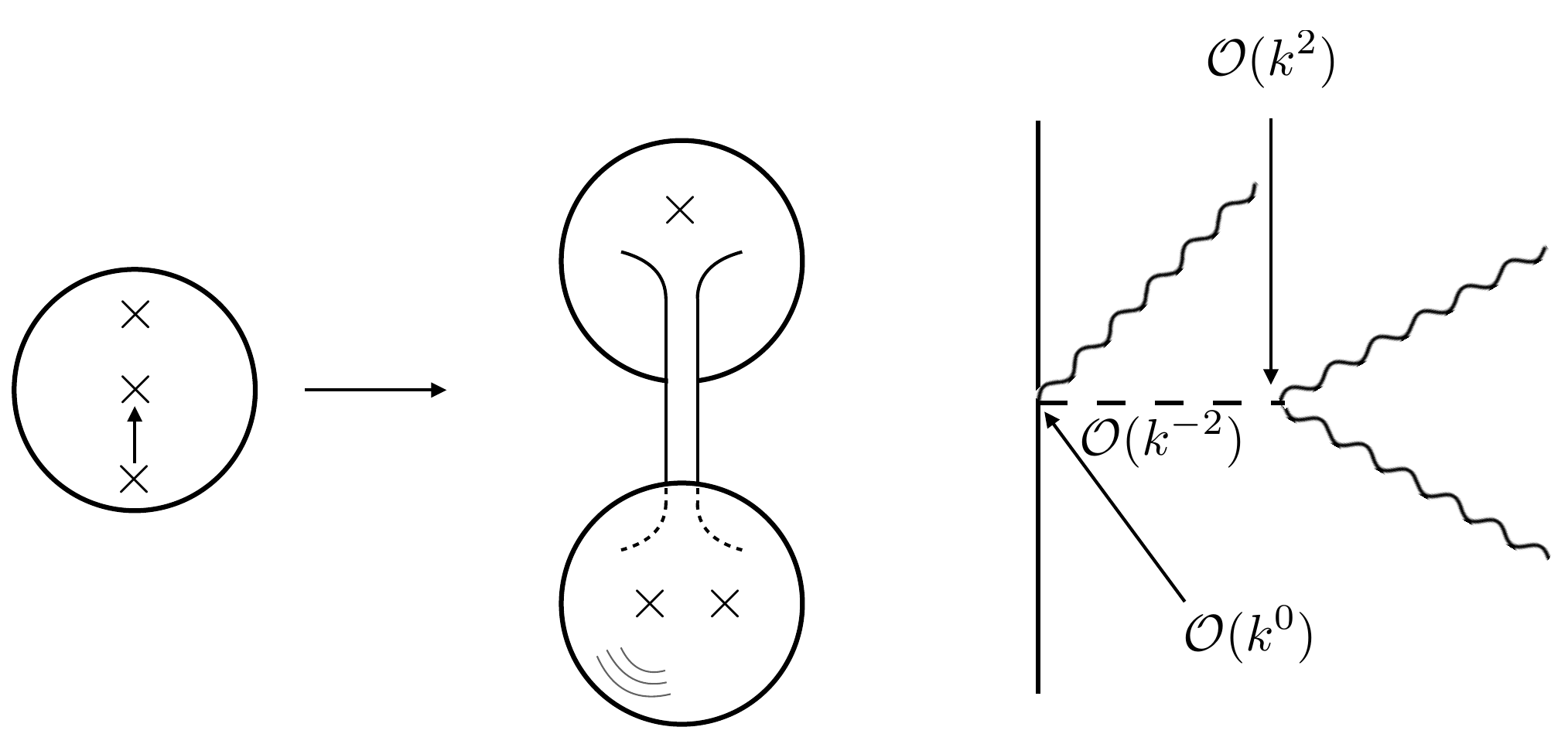} \\[1cm]
\includegraphics[width=0.5\textwidth]{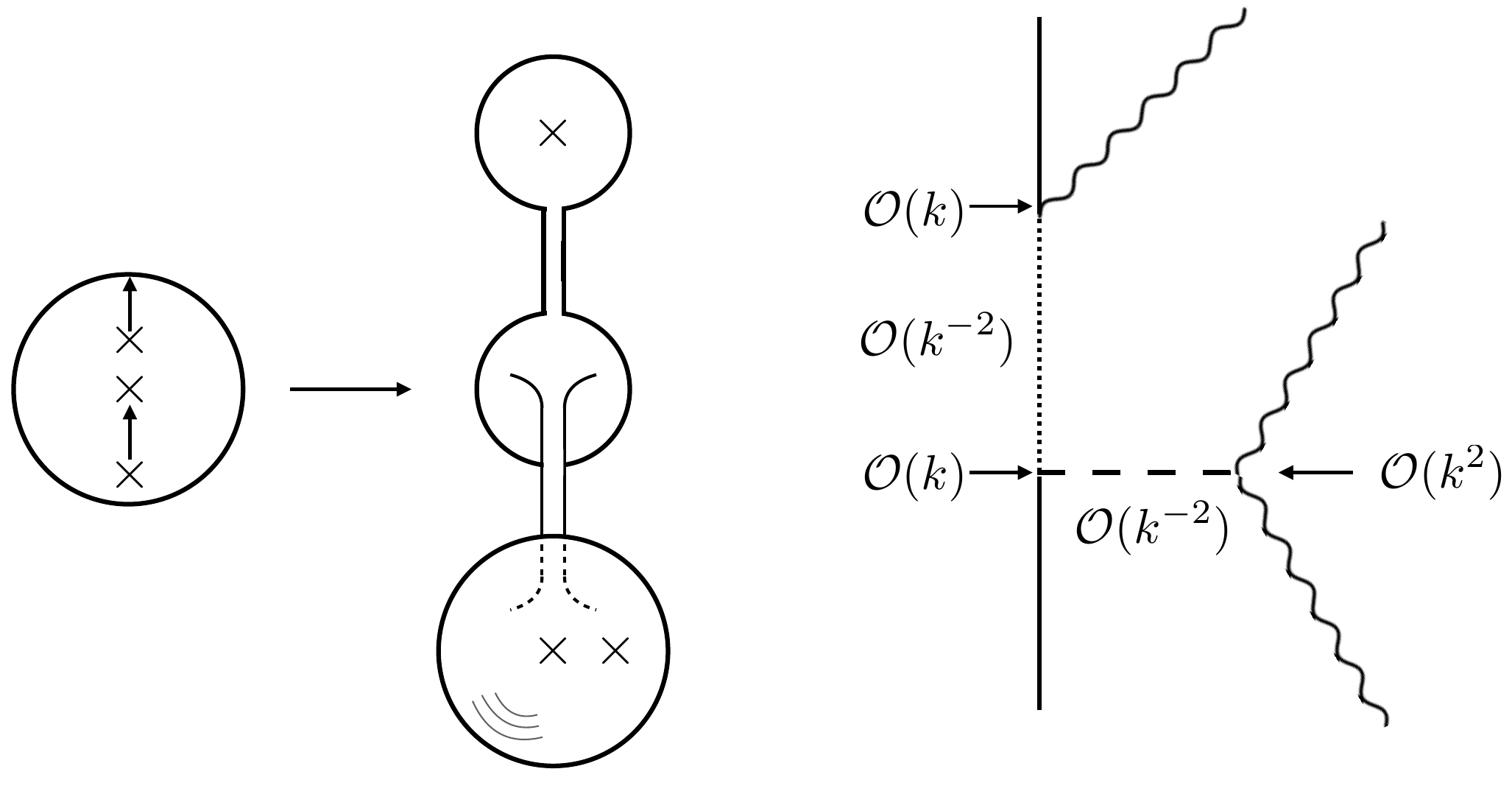}
\caption{Degeneration limits with one vertex collision. \label{degenerate_1_collide}}
\end{center}
\end{figure}

\begin{figure}
\begin{center}
\includegraphics[width=0.5\textwidth]{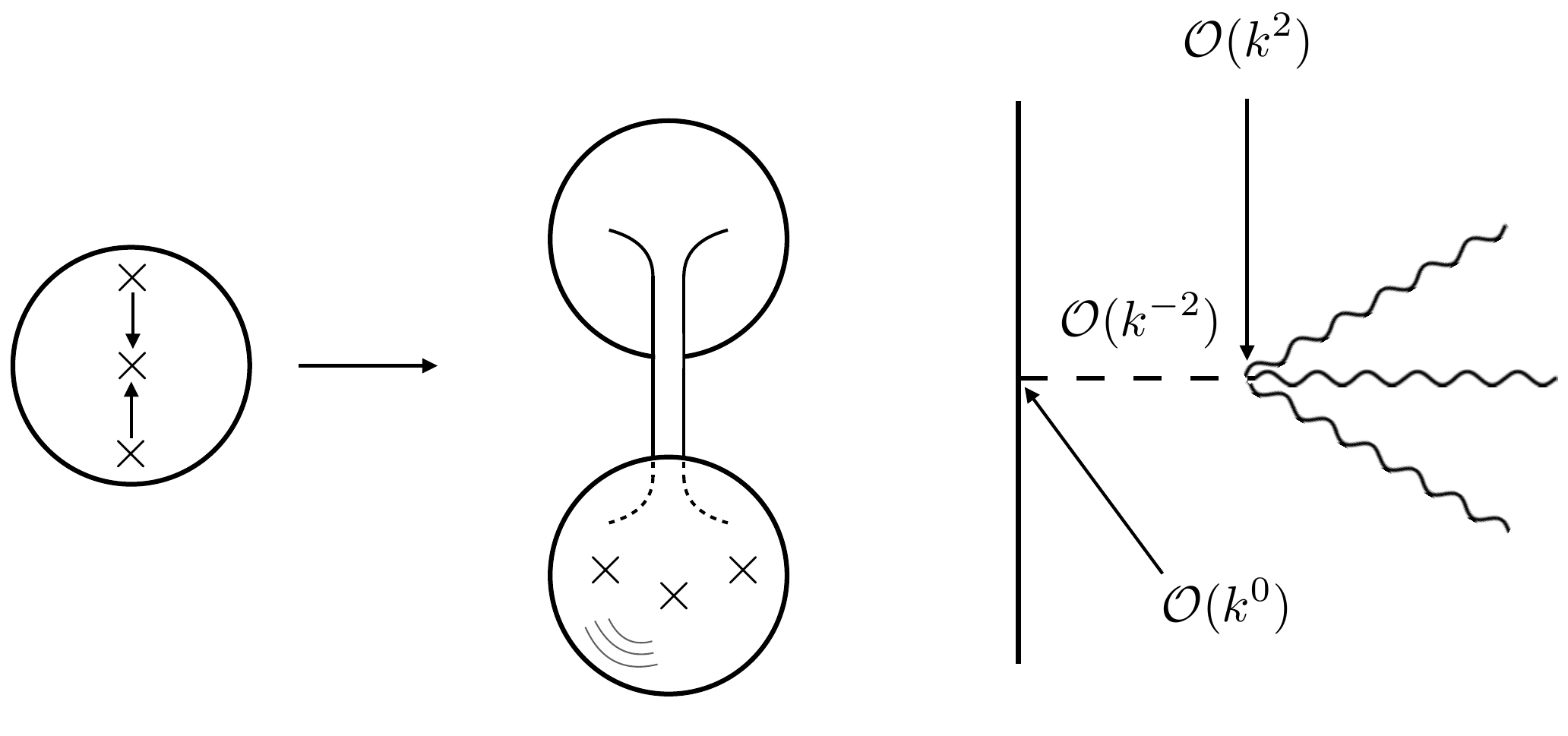} \\[1cm]
\includegraphics[width=0.5\textwidth]{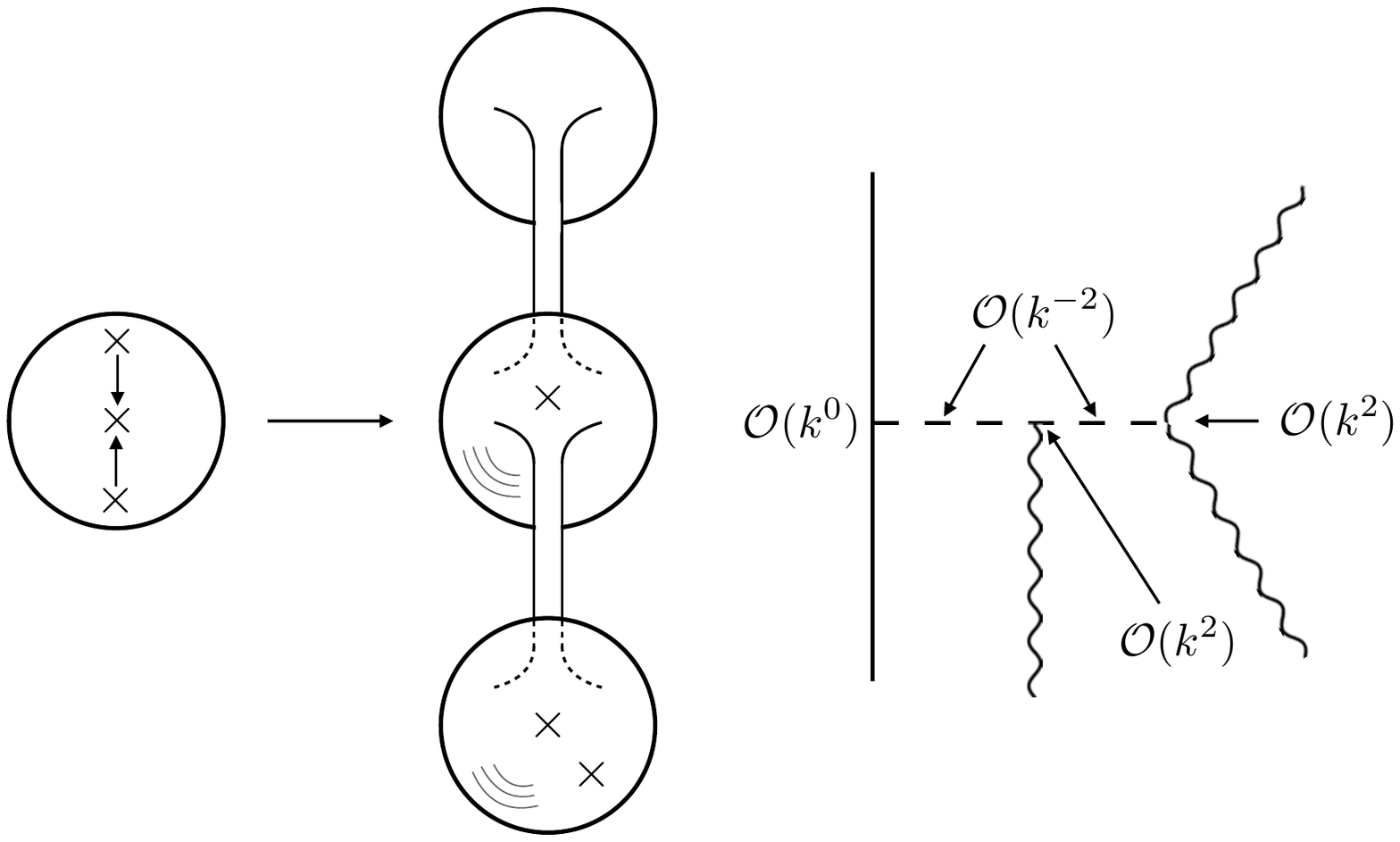}
\caption{Degeneration limits with two vertex collisions. In the upper diagram the three vertex operators approach each other at a uniform pace whereas in the lower diagram the lower two vertex operators collide first and are then approached by the third one. \label{degenerate_2_collide}}
\end{center}
\end{figure}

As can be seen from \eqref{eq::A1} together with \eqref{eq::M1}-\eqref{eq::M3}, our result displays a rich structure of poles. These can be understood as coming from the boundaries of moduli space where vertex operators approach each other or the boundary of the disk, respectively. We do not discuss the pole-structure in detail, but the relevant regions in moduli space can be depicted graphically as in figs.\ \ref{degenerate_0_collide} - \ref{degenerate_2_collide}, which are the analogs of fig.\ 2 in \cite{Hashimoto:1996bf}. All the corresponding contributions scale as ${\cal O}(k^0)$ at the leading order in an expansion for small momenta, where $k$ stands schematically for some combination of the external momenta.

The dotted lines stand for the propagation of massless open string excitations of the D-branes, i.e.\ either of the position scalars $X^i$ or the vector fields $A_a$ living on the D-brane world-volume. The dashed lines on the other hand represent the propagation of massless closed string excitations. Depending on the choice of external states, these could be a dilaton, a graviton, a Kalb-Ramond $B$-field or an $RR$ field. However, if the external states are all dilaton and graviton excitations, the same also holds for the internal (dashed) lines. Given that the propagating fields are all massless, their propagators all scale as ${\cal O}(k^{-2})$ with some combination of external momenta. In order to understand the scaling of the vertices in the figures \ref{degenerate_0_collide} - \ref{degenerate_2_collide}, one has to perform an analysis as in \cite{Garousi:1996ad} and \cite{Garousi:1998fg}. The vertices in the bulk (i.e.\ off the D-brane) can be read off from the bulk action in the Einstein-frame,
\be
S_{\rm NS-NS} = \int d^{10} x \sqrt{-g} \left[ \frac{1}{2 \kappa^2} R - \frac12 (\nabla \phi)^2 - \frac32 e^{-\sqrt{2} \kappa \phi} H^2 \right] + \ldots \ ,
\ee
where the Einstein-frame metric $g_{\mu \nu}$ is related to the string-frame metric via $G_{\mu \nu} = e^{\Phi/2} g_{\mu \nu}$, the dots stand for higher derivative $\alpha'$-corrections and we restricted to the fields from the NSNS-sector (in the RNS formalism). Given that all the terms involve at least two derivatives, the bulk vertices scale as ${\cal O}(k^{2})$. The same holds true when including the RR-fields in the discussion.

The vertices on the brane can be read off from the DBI-action
\be \label{SDBI}
S_{\rm DBI} = -T_p \int d^{p+1} \sigma\, {\rm Tr} \left( e^{\frac{p-3}{4} \Phi} \sqrt{{-\rm det}(\tilde g_{ab} + e^{-\Phi / 2} \tilde B_{ab} + 2 \pi \ell_s^2\, e^{-\Phi / 2} F_{ab}) }\right) 
\ee
with 
\be \label{pullback_metric}
\tilde g_{ab} = g_{ab} + 2 g_{i(a} \partial_{b)}X^i + g_{ij} \partial_{a} X^i \partial_{b} X^j
\ee
the pull-back of the bulk metric in the Einstein-frame and there is an analogous formula to \eqref{pullback_metric} for the pull-back of the Kalb-Ramond $B$-field. At leading order the number of derivatives in the interaction vertices is directly related to the number of open string legs attached to the vertex, i.e.\ ${\cal O}(k^n)$ for a vertex with $n$ open string legs, cf.\ eqs.\ (6)-(11) and (21) in \cite{Garousi:1998fg}. As explained there, this relation arises either directly from the terms in the DBI action \eqref{SDBI}, including the pull-back prescription \eqref{pullback_metric}, or from Taylor-expansion of the closed string fields in the transverse D-brane coordinates. 

\begin{figure}
\begin{center}
\includegraphics[width=0.5\textwidth]{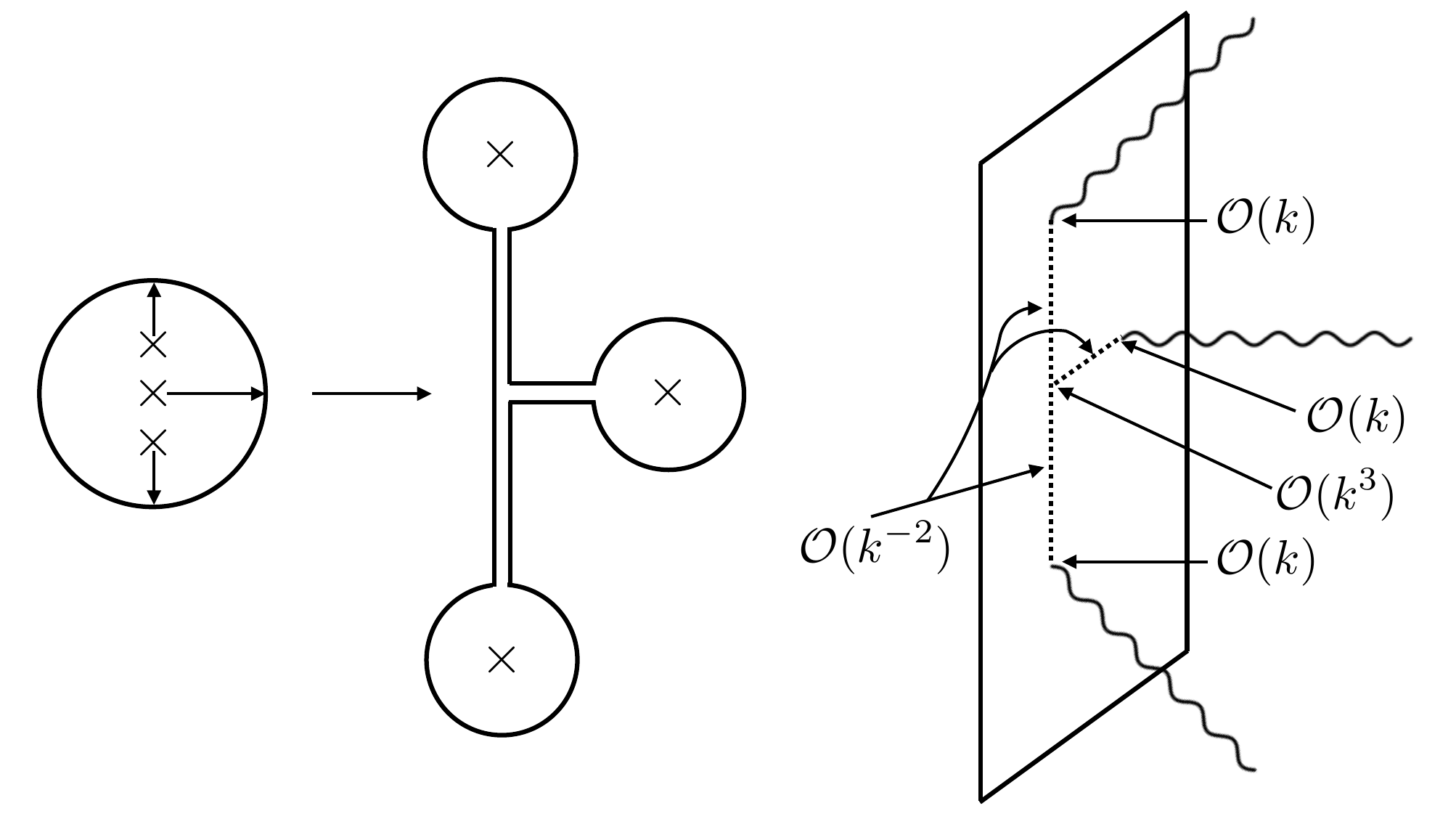} \\[1cm]
\caption{This degeneration limit does not contribute as there is no 3-point vertex for the $U(1)$ or center-of-mass fluctuations of the branes. \label{degenerate_0_collide_2}}
\end{center}
\end{figure}

We should also mention that in the case of a stack of several D-branes describing a non-Abelian gauge group the DBI action \eqref{SDBI} receives additional terms involving commutators of the non-Abelian D-brane scalars, cf.\ \cite{Myers:1999ps}. Moreover, in the non-Abelian case the pullback \eqref{pullback_metric} has to be performed with the covariant derivative of the D-brane scalars, cf.\ \cite{Hull:1997jc}. However, all the open string excitations propagating along the D-branes in figures \ref{degenerate_0_collide} and \ref{degenerate_1_collide} are $U(1)$ or center-of-mass fluctuations, given that they couple linearly to external closed strings \cite{Garousi:1998fg}. This is also the reason why we do not have any contribution from the degeneration shown in figure \ref{degenerate_0_collide_2}. As verified explicitly in sec.\ 3 of \cite{Garousi:1998fg}, there is no non-vanishing 3-point vertex for the $U(1)$ or center-of-mass fluctuations on the D-branes. As a consequence, the closed string disk amplitude should exhibit only terms with at most two poles. For simplicity we just consider the subset of terms containing no contractions of momenta with polarization tensors (for external states from the NSNS sector). At leading order in the $\alpha'$ expansion we indeed find 
\begin{IEEEeqnarray*}{l}
		\lim\limits_{\ap\to0}\mathcal{A}\Big\rvert_{\zeta_i{\cdot}k_j \rightarrow 0}=\\
		g_c^3T_p\biggl[-\left(\frac{s_{12} s_{13}-s_{1\overline3} s_{2\overline3}-s_{1\overline2} (s_{1\overline3}+s_{2\overline3})}{s_{1\overline1} s_{23}}+\frac{s_{12} s_{13}}{s_{23}(s_{12}+s_{13}+s_{23})}\right) \Tr(D{\cdot}\epsilon _1) \Tr(\epsilon _2^T{\cdot}\epsilon _3)\\
		-\frac{s_{23} \Tr(\epsilon_1{\cdot}\epsilon _2^T{\cdot}D{\cdot}\epsilon _3^T)}{s_{12}+s_{13}+s_{23}}-\frac{s_{23} \Tr(\epsilon_1{\cdot}\epsilon_3^T{\cdot}D{\cdot}\epsilon _2^T)}{s_{12}+s_{13}+s_{23}}+\frac{(s_{1\overline2}+s_{1\overline3})  }{s_{1\overline1}}\Tr(D{\cdot}\epsilon _1)\Tr(D{\cdot}\epsilon _2{\cdot}D{\cdot}\epsilon_3)\\
		+\frac13\Tr(\epsilon _1{\cdot}D{\cdot}\epsilon _2{\cdot}D{\cdot}\epsilon _3{\cdot}D)+\frac13\Tr(\epsilon _1^T{\cdot}D{\cdot}\epsilon _2^T{\cdot}D{\cdot}\epsilon _3^T{\cdot}D)\\
		+\left(\frac13+\frac{s_{12}+s_{13}-s_{23}}{2 s_{1\overline1}}+\frac{s_{12} s_{13}}{s_{2\overline2} s_{3\overline3}}+\frac{s_{1\overline1}\left(s_{12}+s_{13}+s_{23}\right)}{4 s_{2\overline2} s_{3\overline3}}\right)\Tr(D{\cdot}\epsilon _1)\Tr(D{\cdot}\epsilon _2) \Tr(D{\cdot}\epsilon _3)\biggl]\\
		+\{1\leftrightarrow2\}+\{1\leftrightarrow3\}\ .\IEEEyesnumber\label{eq::low_energy_result}
\end{IEEEeqnarray*}
This was obtained by plugging in the expansion for $F$ in \eqref{eq::ap-expansion} and the explicit expression of $A_\text{YM}$ (taken from \cite{PSS_components}) into \eqref{eq::A3}.

For the case of three gravitons we would now like to compare \eqref{eq::low_energy_result} with the field theory result based on the DBI action \eqref{SDBI}. As before, we also restrict the field theory calculation to those terms where the polarization tensors are completely contracted among themselves, i.e.\ $\zeta_i{\cdot}k_j \rightarrow 0$. Moreover, again for simplicity, we consider the case in which the gravitons have orthogonal polarizations, i.e.\ ${(\epsilon_i)^\mu}_\nu {(\epsilon_j)^\nu}_\rho=0$. Due to these two restrictions the bulk Einstein-Hilbert term does not contribute and the only diagrams depicted in figures \ref{degenerate_0_collide}--\ref{degenerate_2_collide} that are non-vanishing in the case at hand are the two diagrams in figure \ref{degenerate_0_collide}. 

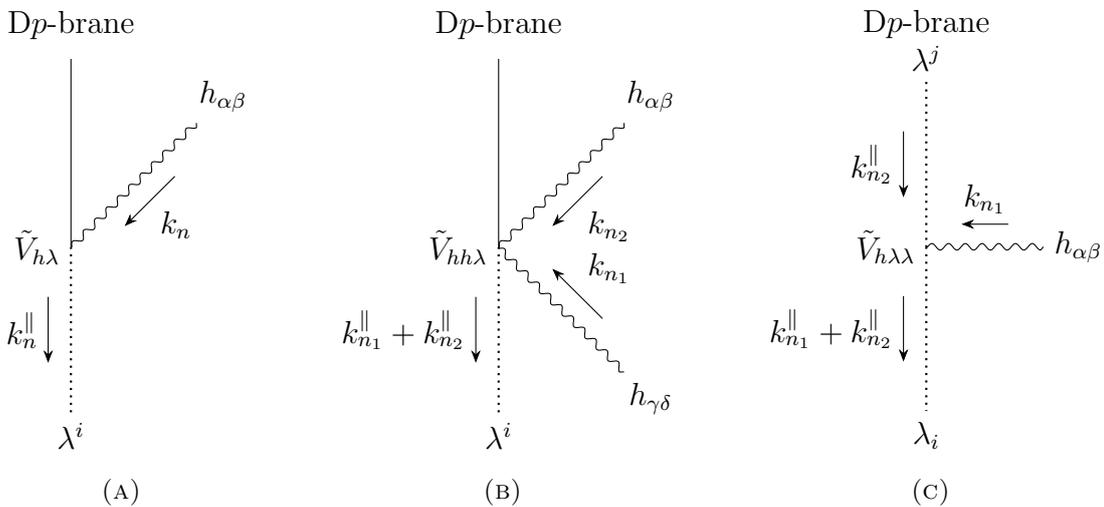
\begin{figure}[h]
	\centering\hfill
	\subcaptionbox{}{
		\begin{tikzpicture}
			\begin{feynman}
				\vertex (c) at (0,0) ;
				\vertex (b) at (0,-2.5);
				\vertex (a) at (0,-5) {$\lambda^i$};
				\vertex (d) at (2,-0.5){$h_{\alpha\beta}$};
				\node at (0,0.45){D$p$-brane};
				\node[left] at (0,-2.5) {$\tilde{V}_{h\lambda}$};
				\diagram* {
					(a) --[ghost,rmomentum={[arrow shorten=0.3]$k^\parallel_n$}] (b) -- (c),
					(b) --[boson,rmomentum'={[arrow shorten=0.3]$k_n$}] (d)
				};
			\end{feynman}
	\end{tikzpicture}}\hfill
	\subcaptionbox{}{
		\begin{tikzpicture}
			\begin{feynman}
				\vertex (c) at (0,0)  ;
				\vertex (b) at (0,-2.5) ;
				\vertex (a) at (0,-5){$\lambda^i$};
				\vertex (d) at (2,-4.5){$h_{\gamma\delta}$};
				\vertex (e) at (2,-0.5){$h_{\alpha\beta}$};
				\node at (0,0.45){D$p$-brane};
				\node[left] at (0,-2.5) {$\tilde{V}_{hh\lambda}$} ;
				\diagram {
					(a) --[ghost, rmomentum={[arrow shorten=0.3]$k^\parallel_{n_1}+k^{\parallel}_{n_2}$}] (b) -- (c),
					(b) --[boson, rmomentum={[arrow shorten=0.3]$k_{n_1}$}] (d),
					(b) --[boson,rmomentum'={[arrow shorten=0.3]$k_{n_2}$}] (e)
				};
			\end{feynman}
	\end{tikzpicture}}\hfill
	\subcaptionbox{}{
		\begin{tikzpicture}
			\begin{feynman}
				\vertex (a) at (0,0) {$\lambda^j$} ;
				\vertex (b) at (0,-2.5) ;
				\vertex (c) at (0,-5) {$\lambda_i$};
				\vertex (d) at (2,-2.5){$h_{\alpha\beta}$};
				\node at (0,0.45){D$p$-brane};
				\node[left] at (0,-2.5) {$\tilde{V}_{h\lambda\lambda}$} ;
				\diagram {
					(a) --[ghost, momentum'={[arrow shorten=0.3]$k^\parallel_{n_2}$}] (b)  --[ghost, momentum'={[arrow shorten=0.3]$k^\parallel_{n_1}+k^\parallel_{n_2}$}] (c),
					(b) --[boson, rmomentum={[arrow shorten=0.3]$k_{n_1}$}] (d)
				};
			\end{feynman}
	\end{tikzpicture}}\hfill\
\caption{Interactions of one and two gravitons with open string excitations on the D$p$-brane. \label{fig::interaction}}
\end{figure}

\begin{figure}[h]
	\centering
	\begin{tikzpicture}
		\begin{feynman}
			\vertex (a) at (0,0) ;
			\vertex (b) at (0,-2.5);
			\vertex (c) at (0,-5);
			\vertex (d) at (2,-4.5){$h_{\mu\nu}$};
			\vertex (e) at (2,-0.5){$h_{\alpha\beta}$};
			\vertex (f) at (2,-2.5){$h_{\gamma\delta}$};
			\node at (0,0.45){D$p$-brane};
			\node[left] at (0,-2.5) {$\tilde S_{hhh}$};
			\diagram* {
				(a) -- (b) -- (c),
				(b) --[boson] (d),
				(b) --[boson] (e),
				(b) --[boson] (f)		
			};
		\end{feynman}
	\end{tikzpicture}
	\caption{Three gravitons sourced from a D$p$-brane.\label{fig::contact}}
\end{figure}
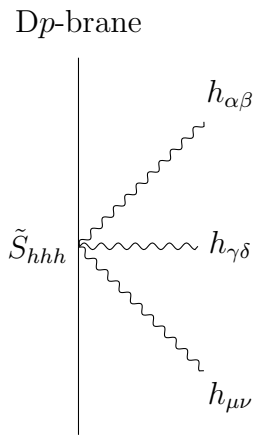

For the scattering of three gravitons, we only need to consider the part of \eqref{SDBI} 
\begin{IEEEeqnarray}{rCl}
	S_{\text{DBI}}^{gravity}&=&-T_p\int\mathrm{d}^{p+1}\sigma\,\text{Tr}\left(\sqrt{-\text{det}(\tilde{g}_{ab})}\right)\ ,\label{eq::SDBIg}
\end{IEEEeqnarray}
that describes the gravitational interaction of D$p$-branes. We can use
\begin{IEEEeqnarray}{rCl}
	\sqrt{-\text{det}({\delta^a}_b+{M^a}_b)}&=&1+\frac12{M^a}_a-\frac14{M^a}_b{M^b}_a+\frac18({M^a}_a)^2+\frac16{M^a}_b{M^b}_c{M^c}_a\nonumber\\
	&&-\frac18{M^a}_a{M^b}_c{M^c}_b+\frac{1}{48}({M^a}_a)^3+\ldots\ .\label{detM}
\end{IEEEeqnarray}
to expand the Lagrangian of the action \eqref{eq::SDBIg} around a flat background $g_{\mu \nu}=\eta_{\mu \nu}+2\kappa h_{\mu \nu}$. This leads to the following terms that are relevant for computing the two diagrams in figure \ref{degenerate_0_collide}:
\begin{IEEEeqnarray}{rCl}
	\mathcal{L}&=&-\kappa\left[T_p {h^a}_a + \sqrt{T_p}\lambda^i\partial_i{h^a}_a+\frac{1}{2}\lambda^i\lambda^j\partial_i\partial_j{h^a}_a+\frac{1}{2}(\partial\lambda)^2{h^a}_a\right]\nonumber\\
	&&-\kappa^2\left[-T_p{h^a}_b{h^b}_a+\frac{1}{2}\sqrt{T_p}\lambda^i\partial_i({h^a}_a)^2-\sqrt{T_p}\lambda^i\partial_i({h^a}_b{h^b}_a)\right]\nonumber\\
	&&-\kappa^3T_p\left[\frac16({h^a}_a)^3-{h^a}_a{h^b}_c{h^c}_b+\frac43{h^a}_b{h^b}_c{h^c}_a\right]+\ldots\ ,\label{eq::LDBI}
\end{IEEEeqnarray}
where we have normalized the open string modes along the D$p$-brane as $X^i=\frac{1}{\sqrt{T_p}}\lambda^i$. This Lagrangian describes the interactions of one or two gravitons with open string excitations on the D$p$-brane depicted in figure \ref{fig::interaction}. The vertices of these interactions obtained from \eqref{eq::LDBI} are given by
\begin{IEEEeqnarray}{rCl}
	\tilde V_{h\lambda}^{\alpha\beta;i}&=&\kappa \sqrt{T_p}V^{\alpha\beta}k^i_n\ ,\nonumber\\
	\tilde V_{hh\lambda}^{\alpha\beta,\gamma\delta;i}&=&\kappa^2 \sqrt{T_p}(k^i_{n_1}+k^i_{n_2})\left(\frac12V^{\alpha\beta}V^{\gamma\delta}-V^{\alpha\delta}V^{\beta\gamma}\right)\ ,\nonumber \\
	\tilde V_{h\lambda\lambda}^{\alpha\beta;i,j}&=&\frac i2\kappa\left(k^i_{n_1} k^j_{n_1}-(k_{n_1}+k_{n_2})\cdot V\cdot k_{n_2} N^{ij}\right) V^{\alpha\beta}\ ,\label{eq::vertex_DBI}
\end{IEEEeqnarray}
where we have introduced the projector into the subspace parallel to the D$p$-brane $V_{\mu\nu}=\frac12(\eta_{\mu\nu}+D_{\mu\nu})$ and we can also define a projector into the subspace transverse to the D$p$-brane $N_{\mu\nu}=\frac12(\eta_{\mu\nu}-D_{\mu\nu})$, see appendix A in \cite{Garousi:1996ad} for more details on these projectors.
\\
Moreover, the contact terms of three gravitons in the Lagrangian \eqref{eq::LDBI} gives rise to the diagram in figure \ref{fig::contact}. 
The source of this contact term can be described as
	\begin{IEEEeqnarray}{rCl}
	\tilde S^{\alpha\beta,\gamma\delta,\mu\nu}_{hhh}&=&-i\kappa^3 T_p\left(\frac16V^{\alpha\beta}V^{\gamma\delta}V^{\mu\nu}-V^{\alpha\beta}V^{\gamma\nu}V^{\mu\delta}+\frac{4}{3}V^{\alpha\nu}V^{\gamma\beta}V^{\mu\delta}\right)\ . \label{eq::contact_term}
\end{IEEEeqnarray}
With the vertices in \eqref{eq::vertex_DBI} we are now ready to calculate the two diagrams in figure \ref{degenerate_0_collide}. The first diagram is given by 
\begin{IEEEeqnarray}{rCl}
	A_{hhh}&=&\sum_{\sigma\in S_3}\epsilon_{\alpha\beta}^{1_\sigma}\epsilon_{\gamma\delta}^{2_\sigma}\epsilon_{\mu\nu}^{3_\sigma}\tilde{V}^{\alpha\beta;i}_{h\lambda}G^\lambda_{ij}\tilde{V}^{\gamma\delta;j,m}G^\lambda_{mn}\tilde{V}^{\mu\nu;n}\nonumber\\
	&=&-i\kappa^3T_p\sum_{\sigma\in S_3}\left(\frac{2k_{1_\sigma}{\cdot}N{\cdot}k_{2_\sigma}k_{2_\sigma}{\cdot}N{\cdot}k_{3_\sigma}+k_{1_\sigma}{\cdot}N{\cdot}k_{3_\sigma}k_{2_\sigma}{\cdot}V{\cdot}k_{2_\sigma}}{4k_{1_\sigma}{\cdot}V{\cdot}k_{1_\sigma}k_{3_\sigma}{\cdot}V{\cdot}k_{3_\sigma}}\right.\nonumber\\
	&&\left.-\frac{k_{1_\sigma}{\cdot}N{\cdot}k_{3_\sigma}}{4k_{1_\sigma}{\cdot}V{\cdot}k_{1_\sigma}}-\frac{k_{1_\sigma}{\cdot}N{\cdot}k_{3_\sigma}}{4k_{3_\sigma}{\cdot}V{\cdot}k_{3_\sigma}}\right)\Tr(\epsilon_{1_\sigma}{\cdot}V)\Tr\left(\epsilon_{2_\sigma}{\cdot}V\right)\Tr(\epsilon_{3_\sigma}{\cdot}V)\ ,
\end{IEEEeqnarray}
where 
\be
G^\lambda_{mn} = -i \frac{N^{mn}}{k \cdot V \cdot k} \label{eq::Glambdaij}
\ee
is the propagator of the open string scalars \cite{Garousi:1998fg} and $k$ in \eqref{eq::Glambdaij} is the momentum of the propagating scalar. The second diagram evaluates to
\begin{IEEEeqnarray}{rCl}
	A_{hh^2}&=&\sum_{\sigma\in S_3}\epsilon_{\alpha\beta}^{1_\sigma}\epsilon_{\gamma\delta}^{2_\sigma}\epsilon_{\mu\nu}^{3_\sigma}\tilde{V}^{\alpha\beta;i}_{h\lambda}G^\lambda_{ij}\tilde{V}^{\gamma\delta,\mu\nu;j}\nonumber\\
	&=&-i\kappa^3T_p\sum_{\sigma\in S_3}\left(\frac{k_{1_\sigma}{\cdot}N{\cdot}k_{2_\sigma}+k_{1_\sigma}{\cdot}N{\cdot}k_{3_\sigma}}{2k_{1_\sigma}{\cdot}V{\cdot}k_{1_\sigma}}\Tr(\epsilon_{1_\sigma}{\cdot}V)\Tr\left(\epsilon_{2_\sigma}{\cdot}V\right)\Tr(\epsilon_{3_\sigma}{\cdot}V)\right.\nonumber\\
	&&\left.-\frac{k_{1_\sigma}{\cdot}N{\cdot}k_{2_\sigma}+k_{1_\sigma}{\cdot}N{\cdot}k_{3_\sigma}}{k_{1_\sigma}{\cdot}V{\cdot}k_{1_\sigma}}\Tr(\epsilon_{1_\sigma}{\cdot}V)\Tr(\epsilon_{2_\sigma}{\cdot}V{\cdot}\epsilon_{3_\sigma}{\cdot}V)\right)\ .
\end{IEEEeqnarray}
Together with the contact term
\begin{IEEEeqnarray}{rCl}
	A_{h^3}&=&\sum_{\sigma\in S_3} \epsilon_{\alpha\beta}^{1_\sigma}\epsilon_{\gamma\delta}^{2_\sigma}\epsilon_{\mu\nu}^{3}\tilde S^{\alpha\beta,\gamma\delta,\mu\nu}_{hhh}\nonumber\\
	&=&-i\kappa^3T_p\Big[\Tr(\epsilon_{1}{\cdot}V)\Tr\left(\epsilon_{2}{\cdot}V\right)\Tr(\epsilon_{3}{\cdot}V)-2\Tr(\epsilon_{1}{\cdot}V)\Tr(\epsilon_{2}{\cdot}V{\cdot}\epsilon_{3}{\cdot}V)\nonumber\\
	&&-2\Tr(\epsilon_{2}{\cdot}V)\Tr(\epsilon_{1}{\cdot}V{\cdot}\epsilon_{3}{\cdot}V)-2\Tr(\epsilon_{3}{\cdot}V)\Tr(\epsilon_{1}{\cdot}V{\cdot}\epsilon_{2}{\cdot}V)\nonumber\\
	&&+8\Tr(\epsilon_{1}{\cdot}V{\cdot}\epsilon_{2}{\cdot}V{\cdot}\epsilon_{3}{\cdot}V)\Big]
\end{IEEEeqnarray}
we obtain for the total interaction of three gravitons with orthogonal polarizations for the field theory calculation the following result
\begin{IEEEeqnarray}{rCl}
 A_{hhh}+A_{hh^2}+A_{h^3}&=&-i\kappa^3T_p\left[-\left(2+\frac{2k_{1}{\cdot}N{\cdot}k_{2}}{k_{1}{\cdot}V{\cdot}k_{1}}+\frac{2k_{1}{\cdot}N{\cdot}k_{3}}{k_{1}{\cdot}V{\cdot}k_{1}}\right)\Tr(\epsilon_{1}{\cdot}V)\Tr(\epsilon_{2}{\cdot}V{\cdot}\epsilon_{3}{\cdot}V)\right.\nonumber\\
 &&+\left(\frac13+\frac{k_{1}{\cdot}N{\cdot}k_{3}k_{2}{\cdot}N{\cdot}k_{3}}{2k_{1}{\cdot}V{\cdot}k_{1}k_{2}{\cdot}V{\cdot}k_{2}}+\frac{k_{1}{\cdot}N{\cdot}k_{2}k_{2}{\cdot}N{\cdot}k_{3}}{2k_{1}{\cdot}V{\cdot}k_{1}k_{3}{\cdot}V{\cdot}k_{3}}+\frac{k_{1}{\cdot}N{\cdot}k_{2}k_{3}{\cdot}V{\cdot}k_{3}}{4k_{1}{\cdot}V{\cdot}k_{1}k_{2}{\cdot}V{\cdot}k_{2}}\right.\nonumber\\
 &&\left.+\frac{k_{1}{\cdot}N{\cdot}k_{3}k_{2}{\cdot}V{\cdot}k_{2}}{4k_{1}{\cdot}V{\cdot}k_{1}k_{3}{\cdot}V{\cdot}k_{3}}+\frac{k_{1}{\cdot}N{\cdot}k_{2}}{2k_{1}{\cdot}V{\cdot}k_{1}}+\frac{k_{1}{\cdot}N{\cdot}k_{3}}{2k_{1}{\cdot}V{\cdot}k_{1}}\right)\Tr(\epsilon_{1}{\cdot}V)\Tr\left(\epsilon_{2}{\cdot}V\right)\Tr(\epsilon_{3}{\cdot}V)\nonumber\\
 &&+\left.\frac{8}{3}\Tr(\epsilon_{1}{\cdot}V{\cdot}\epsilon_{2}{\cdot}V{\cdot}\epsilon_{3}{\cdot}V)\right]+\{1\leftrightarrow2\}+\{1\leftrightarrow3\}\ .\label{eq::ft_res}
\end{IEEEeqnarray}
In the low energy limit the result obtained from the scattering of three closed strings off a D$p$-brane in \eqref{eq::amp_mon_4a} should reproduce the field theory result above. To compare these we can use that $D_{\mu\nu}=2V_{\mu\nu}-\eta_{\mu\nu}$ and
\begin{IEEEeqnarray}{rCl}
	s_{12}&=&\frac{1}{2}k_3{\cdot}V{\cdot}k_3-\frac{1}{2}k_1{\cdot}V{\cdot}k_1-\frac{1}{2}k_2{\cdot}V{\cdot}k_2+k_1{\cdot}N{\cdot}k_2\ ,\nonumber\\
	s_{13}&=&\frac{1}{2}k_2{\cdot}V{\cdot}k_2-\frac{1}{2}k_1{\cdot}V{\cdot}k_1-\frac{1}{2}k_3{\cdot}V{\cdot}k_3+k_1{\cdot}N{\cdot}k_3\ ,\\
	s_{23}&=&\frac{1}{2}k_1{\cdot}V{\cdot}k_1-\frac{1}{2}k_2{\cdot}V{\cdot}k_2-\frac{1}{2}k_3{\cdot}V{\cdot}k_3+k_2{\cdot}N{\cdot}k_3\ .\nonumber
\end{IEEEeqnarray}
Moreover, when all three external states are described by gravitons with orthogonal polarizations, we find for \eqref{eq::low_energy_result}
\begin{IEEEeqnarray}{rCl}
	\lim\limits_{\ap\to0}\mathcal{A}\Big\rvert_{\zeta_i{\cdot}k_j \rightarrow 0}&=&g_c^3T_p\biggl[\frac{(s_{1\overline2}+s_{1\overline3})  }{s_{1\overline1}}\Tr(D{\cdot}\epsilon _1)\Tr(D{\cdot}\epsilon _2{\cdot}D{\cdot}\epsilon
	_3)+\left(\frac13+\frac{s_{12}+s_{13}-s_{23}}{2 s_{1\overline1}}+\frac{s_{12} s_{13}}{s_{2\overline2} s_{3\overline3}}\right.\nonumber\\
	&&\left.+\frac{\left(s_{12}+s_{13}+s_{23}\right)s_{1\overline1}}{4 s_{2\overline2} s_{3\overline3}}\right)\Tr(D{\cdot}\epsilon _1)\Tr(D{\cdot}\epsilon _2) \Tr(D{\cdot}\epsilon _3)\nonumber\\
	&&+\frac23\Tr(D{\cdot}\epsilon _1{\cdot}D{\cdot}\epsilon _2{\cdot}D{\cdot}\epsilon _3)\biggl]+\{1\leftrightarrow2\}+\{1\leftrightarrow3\}\\
	&=&g_c^3T_p\left[-\left(4+\frac{4k_{1}{\cdot}N{\cdot}k_{2}}{k_{1}{\cdot}V{\cdot}k_{1}}+\frac{4k_{1}{\cdot}N{\cdot}k_{3}}{k_{1}{\cdot}V{\cdot}k_{1}}\right)\Tr(\epsilon_{1}{\cdot}V)\Tr(\epsilon_{2}{\cdot}V{\cdot}\epsilon_{3}{\cdot}V)\right.\nonumber\\
	&&+\left(\frac23+\frac{k_{1}{\cdot}N{\cdot}k_{3}k_{2}{\cdot}N{\cdot}k_{3}}{k_{1}{\cdot}V{\cdot}k_{1}k_{2}{\cdot}V{\cdot}k_{2}}+\frac{k_{1}{\cdot}N{\cdot}k_{2}k_{2}{\cdot}N{\cdot}k_{3}}{k_{1}{\cdot}V{\cdot}k_{1}k_{3}{\cdot}V{\cdot}k_{3}}+\frac{k_{1}{\cdot}N{\cdot}k_{2}k_{3}{\cdot}V{\cdot}k_{3}}{2k_{1}{\cdot}V{\cdot}k_{1}k_{2}{\cdot}V{\cdot}k_{2}}\right.\nonumber\\
	&&\left.+\frac{k_{1}{\cdot}N{\cdot}k_{3}k_{2}{\cdot}V{\cdot}k_{2}}{2k_{1}{\cdot}V{\cdot}k_{1}k_{3}{\cdot}V{\cdot}k_{3}}+\frac{k_{1}{\cdot}N{\cdot}k_{2}}{k_{1}{\cdot}V{\cdot}k_{1}}+\frac{k_{1}{\cdot}N{\cdot}k_{3}}{k_{1}{\cdot}V{\cdot}k_{1}}\right)\Tr(\epsilon_{1}{\cdot}V)\Tr\left(\epsilon_{2}{\cdot}V\right)\Tr(\epsilon_{3}{\cdot}V)\nonumber\\
	&&+\left.\frac{16}{3}\Tr(\epsilon_{1}{\cdot}V{\cdot}\epsilon_{2}{\cdot}V{\cdot}\epsilon_{3}{\cdot}V)\right]+\{1\leftrightarrow2\}+\{1\leftrightarrow3\}\ , \label{eq::string_low_energy_res}
\end{IEEEeqnarray}
Hence, comparing the field theory and string result gives
\begin{IEEEeqnarray}l
	\mathcal{A} \sim (A_{hhh}+A_{hh^2}+A_{h^3})\ ,
\end{IEEEeqnarray}
which is a very non-trivial consistency check of our result. 

Finally, we would like to comment on the $\alpha'$-expansion of section \ref{alphaprime_expand}. The fact that the $\alpha'$-corrections start at order ${\cal O}(\alpha'^2)$ relative to the lowest order, i.e.\ relative to the field theory contribution, is consistent with the results of the closed string disk 2-point function \cite{Garousi:1996ad, Hashimoto:1996bf, Garousi:1998fg}. It confirms that the $\alpha'$-corrections to the DBI action start at the 4-derivative level, as analysed in \cite{Bachas:1999um} for the $R^2$-terms. In particular, there is no indication for a disk-contribution to the Einstein-Hilbert term. This requires an explanation due to the discussion of \cite{Green:2016tfs} where indirect arguments (using heterotic - type I duality) were given for the presence of an $\epsilon_{10} \epsilon_{10} R^4$ term in the worldvolume theory of a D9-brane. Upon compactification on a Calabi-Yau manifold with non-vanishing Euler number such a term is expected to lead to a correction of the Einstein-Hilbert term in 4 dimensions, cf.\ \cite{Antoniadis:1997eg}. Given that the disk amplitude of 3 gravitons with four dimensional polarization tensors is agnostic about the form of the additional 6 dimensions, one could have expected to find an Einstein-Hilbert term on the D9-brane worldvolume.\footnote{Note that the on-shell graviton 2-point function vanishes for a D9-brane \cite{Garousi:1996ad, Hashimoto:1996bf, Garousi:1998fg} and, thus, the 2-point amplitude seems to be too degenerate to draw any conclusion about the presence of an Einstein-Hilbert term.} However, our result speaks against it. 

One possible explanation could be the following: In addition to the mentioned $e^{-\Phi} \epsilon_{10} \epsilon_{10} R^4$-term, there are further higher derivative terms at disk level \cite{Green:2016tfs,Tseytlin:1995bi}. The $e^{-\Phi} \epsilon_{10} \epsilon_{10} R^4$-term does not correct the 10-dimensional dilaton equation of motion  \cite{Antoniadis:1997eg} because the epsilon tensors in this term imply that at least one of the Riemann-tensors has to be the one of the flat non-compact spacetime. This is not true for the additional $R^4$-terms at disk level. Hence they can lead to a correction of the 10-dimensional dilaton at disk level. This correction would have to be taken into account when compactifying the 10-dimensional Einstein-Hilbert term $e^{-2 \Phi} R$ to 4 dimensions, similar to the analysis in \cite{Becker:2002nn}. It might be that these two disk level contributions to the 4-dimensional Einstein-Hilbert term cancel each other. It would be interesting to investigate this question further. 

	\section{Concluding remarks}

\subsection[On higher multiplicity of closed strings on the disk]{On higher multiplicity of closed strings on the disk}

For a given color ordering $\rho$ the pure open superstring $n_o$--point amplitude can be expressed as linear combination  of $(n_o-3)!$ kinematical building blocks \cite{npt_1}
\be\label{opsbuild}
A_{YM}(1,2_\si,3_\si,\ldots,(n_o-2)_\si,n_o-1,n_o) \ ,\qquad \si\in S_{n_o-3}
\ee
multiplying  certain string integrands $F^{(2_\si,3_\si,\ldots,(n_o-2)_\si)}_{\Ic_\rho}$ to be integrated over the domain $\Ic_\rho$ subject to the color ordering $\rho$.
Likewise, the pure closed superstring $n_c$--point amplitude on the disk world--sheet can
be expanded in terms of the $(2n_c-3)!$ kinematical building blocks
\be\label{closedbuild}
A_{YM}(\bar 1,\bar 2_\si,3_\si,\bar 3_\si,\ldots,(n_c)_\si,(\overline{n_c})_\si,2,1) \ ,\qquad\si\in S_{2n_c-3}
\ee
inherited from the open superstring \req{opsbuild}.
While in the pure open string case each kinematical building  block \req{opsbuild} is dressed with a single form factor $F^{(2_\si,3_\si,\ldots,(n_o-2)_\si)}_{\Ic_\rho}$ referring to the color ordering $\rho$ under consideration, in the closed string case for each \req{closedbuild} there are $L_{n_c}$ form factors $F^{(\bar 2_\si,3_\si,\bar 3_\si,\ldots,(n_c)_\si,(\overline{n_c})_\si)}_{\Ic_{\rho_l}}$, each one integrated over different color orderings $\rho_l$ ($l=1,\ldots,L_{n_c}$)  and multiplied by a chain of $n_c\!-\!2$ $\sin$--factors as a consequence  of disentangling  left and right--movers on the disk.

Concretely, for $n_c=2$ we have $L_2=1$ \cite{2pt}
\be
\mathcal{A}_2 \sim g_c^2T_pF^{(\bar 2)}_{\Ic_1}\  A_{YM}(\overline1,\overline2,2,1)\ ,
\ee
with the form factor:
\be
F^{(\bar 2)}_{\Ic_1}=\int\mathrm{d}z_{\overline2}\,\prod_{i<j}|z_{ij}|^{s_{ij}}\frac{s_{12}}{z_{\overline{1}\overline{2}}}=\frac{\Gamma(1+s_{12})\Gamma(1+s_{2\overline2})}{\Gamma(1+s_{12}+s_{2\overline2})}\ .
\ee
Furthermore, for   $n_c=3$ we have $L_3=2$  and our result for the
scattering of three closed strings has been presented in
\eqref{eq::A3}, cf. also \req{eq::partial_1} and \req{eq::partial_2}
\be		
\mathcal{A}_3 \sim g_c^3T_p\sum_{\sigma\in S_3}\lf\{\sum_{l=1}^2 \sin(\pi s_{\rho_l})\; F^{(\overline2_\sigma3_\sigma\overline3_\sigma)}_{\mathcal{I}_{\rho_l}}\right\}\ 
A_{YM}(\overline1,\overline2_\sigma,3_\sigma,\overline3_\sigma,2,1)\ ,
\ee	
with the form factors $F^{(\overline2_\sigma3_\sigma\overline3_\sigma)}_{\mathcal{I}_{\rho_l}}$ given in \req{Sample1}, the permutations $\rho_1=(\overline 1, \overline 2, \overline 3, 2, 3, 1),\; \rho_2=(\overline 1, \overline 2, 2, \overline 3, 3, 1)$ and $s_{\rho_1}=s_{23},\; s_{\rho_2}=s_{23}+s_{2\bar3}$. 
For general $n_c\geq 4$ the result may be summarized in the following way:
\begin{align}		
\mathcal{A}_{n_c}& \sim g_c^nT_p \sum_{\sigma\in S_{2n_c-3}}\lf\{\sum_{l=1}^{L_{n_c}} 
\lf[\prod_{k=1}^{n_c-2}\sin(\pi s_{\rho_{l,k}})\ri]\  F^{(\bar 2_\si,3_\si,\bar 3_\si,\ldots,(n_c)_\si,(\overline{n}_c)_\si)}_{\mathcal{I}_{\rho_l}}\right\}\nonumber\\[3mm]
&\times A_{YM}(\bar 1,\bar 2_\si,3_\si,\bar 3_\si,\ldots,(n_c)_\si,(\overline{n}_c)_\si,2,1)\ .\label{Ansatz}
\end{align}		
Above the angles $s_{\rho_{l,k}}$ are linear combinations of kinematic invariants.	
In \req{Ansatz}, the sum over integration domains $\rho_l$ encompasses $L_{n_c}$ terms as a result of applying  $2n_c$--point open string monodromy relations. As a consequence $L_{n_c}\leq (2n_c-3)!$. In fact, in \req{eq::amp_mon_4a} for $n_c=3$ we have found   $L_3=2$. Further results and details for $n_c\geq4$ will be presented elsewhere.
	
\subsection{Summary and further directions}

In this work we have computed the \emph{complete} tree--level disk amplitude
involving any three closed string states in the NSNS, RR, RNS or NSR sectors. Our main result can be found in
\req{eq::amp_mon_4a}. Generalizations to an arbitrary number of closed strings are
written in the ansatz \req{Ansatz}.
Our results are interesting both from the conceptual and physical
point of view. We could express our findings in
terms of a basis of six--point open string subamplitudes and thereby showed that one can connect this closed string amplitude on the disk via KLT-like relations and a $PSL(2,\mathbb R)$ transformation to the scattering of open strings on the disk as expected. Surprisingly, however, our main result \eqref{eq::amp_mon_4a} can be written in terms of only two six--point open string subamplitudes (the basis of these subamplitudes contains six elements and therefore one might think that also the scattering of three closed strings is given in terms of six subamplitudes). We conjecture that this pattern continues for a closed string $N$-point function, see \eqref{Ansatz}. In order to derive our result we introduced monodromy relations for closed strings in section \ref{sec::analytic} that contain six terms instead of only five in open string subamplitude relations \cite{ovsc}.\par 
In addition we have written the three closed string amplitude on the disk for NSNS-states in terms of Yang-Mills amplitudes using the identification in section \ref{sec::analytic} such that we could use the relations in \eqref{eq::partial_1} and \eqref{eq::partial_2}, which are in a similar form already known for open strings. To verify that these relations hold also for closed strings on the disk, we have explicitly computed the closed string correlator in appendix \ref{sec::correlator}. Usually in the PSF correlators are computed by using OPE contractions \cite{6pt,npt_1,5pt}. We chose a different path and for the first time (to our knowledge) computed a correlator in the PSF using Wick's theorem. \par
In the limit $\ap\to0$ we checked that for a subset of terms our result \eqref{eq::string_low_energy_res} from the string theory calculation agrees with the field theory results \eqref{eq::ft_res} that one can obtain from the DBI action. This is an important consistency check. Moreover, in the scattering of three closed strings on the double cover of the disk, i.e.\ the sphere, there are no poles in $s_{i\overline{\imath}}$ present in the low energy expansion. Nevertheless, these poles are expected to be found in the physical amplitude and indeed we could show that they are present, cf.\  \eqref{eq::low_energy_result}. It is not surprising that they are absent on the double cover and a simple argument was presented in appendix \ref{sec:symmetrization} for the closed string $2$-point function. This argument should also hold for an $N$-point function.\par
Generalizing our computations to one--loop would be of considerable
interest. This amounts to closed  string scattering on a cylinder
world--sheet and tackling the world--sheet cylinder integrations along
the prescription developed in \cite{Stieberger:2021daa,Stieberger:2022lss} will prove to be useful.
Furthermore, taking into account massive strings in the spirit of \cite{Pouria1, Pouria2} would also be very interesting. Finally, it would be worthwhile to better understand why our calculation does not see any hints of an Einstein-Hilbert term on the disk which is expected to arise in 4 dimensions from compactifying an $\epsilon_{10} \epsilon_{10} R^4$ term in the worldvolume of a D9-brane, cf.\ our discussion at the end of section \ref{InterpretLowEXP}.

	\ \\
{\bf Acknowledgements:}
We thank Alice Aldi, Ralph Blumenhagen, 	Ulf Danielsson, Carlos Mafra, Gianfranco Pradisi, and  Oliver Schlotterer for useful discussions. The work of M.H.\ is supported by the Origins Excellence Cluster in Munich.
	
	\appendix
	\section{BRST building blocks}\label{sec::block}
In string theory physical states are identified as elements of the cohomology of the BRST operator $Q$ and only those states contribute to the scattering amplitude \eqref{eq::3pt}. In the pure spinor formalism the BRST operator takes a simple form $Q=\lambda^\alpha D_\alpha$, which enables an efficient method to organize the computation of scattering amplitudes. Hence, in this section we want to discuss the BRST properties of the objects naturally appearing in a scattering process. We will define the composite superfields $\tilde L_{jikipi},\tilde L_{jiki}$ and $\tilde L_{ji}$ arising from the OPEs of the vertex operators in \eqref{eq::3pt} and derive their BRST properties. Due to their recursive definition these superfields generically contain terms that originate from a BRST exact term, which don't contribute to the end result of a scattering amplitude, because BRST exact terms/states are not in the cohomology of $Q$. Note, that terms originating from a BRST exact expression aren't necessarily BRST exact themselves. Nevertheless, these terms should drop out of any physical scattering amplitude, which was explicitly shown for the scattering of up to six open strings on the disk \cite{npt_1} and also for two closed strings on the disk in \cite{2pt}. It was conjectured that this pattern should persist also for higher-point amplitudes \cite{Schlotterer:2011psa,Mafra:2022wml}, but this was not shown explicitly yet. Nevertheless, these terms are crucial for the conformal invariance of the CFT correlator, even if they do not contribute to the scattering amplitude, cf.\ appendix \ref{sec::correlator}. After the BRST exact parts are eliminated from the correlator by integration by parts relations, which is a tedious but straightforward computation, we obtain the composite superfields $L_{jikipi},L_{jiki}$ and $L_{ji}$ that transform covariantly under the BRST charge \cite{Mafra:2010ir}. Together with the corrections of the double pole integrals to the superfields $L$ we will be able to define the BRST building blocks $T_{ijkp},T_{ijk}$ and $T_{ij}$ \cite{npt_1}. Let us also mention that in the literature the superfields $\tilde L$ and $L$ are often both denoted by $L$, but in our discussion in the following it is useful to clearly distinguish between the two. 

\subsection{Contractions of vertex operators}\label{sec::contraction}
Because the conformal primaries with dimension $h=1$ have no zero modes on the disk, we can utilize Wick's theorem and contract primaries by replacing them by the sum over their singularities with the other fields in the correlator. Wick's theorem requires the sum over all possible contractions of the three integrated and three unintegrated vertex operators and hence, we encounter the same contractions over and over again but with a different labelling of the contracted vertex operators, which makes it convenient to define the composite superfields for the correlator in \eqref{eq::3pt}. We start by applying Wick's theorem to two vertex operators
\vspace{5pt}\begin{IEEEeqnarray}{rCl}
	K_{ji}\equiv z_{ji} \tikzmark{Vj}{V}^{(a)}_{j}(z_{j}) \tikzmark{Vi}{V}^{(b)}_{i}(z_{i})\ .
	\begin{tikzpicture}[overlay,remember picture,>=latex,shorten >=1pt,shorten <=1pt,very thin]
		\draw[->-] (Vj) --++(0,10pt) -| (Vi);
	\end{tikzpicture}
\end{IEEEeqnarray}
This defines the contraction of the $j^{\text{th}}$ vertex operator with the $i^{\text{th}}$ vertex operator: The arrow indicates that we contract the $h=1$ primaries of $V^{(a)}_{j}$ with $V^{(b)}_{i}$, but don't contract the $h=1$ primaries of $V^{(b)}_{i}$ with $V^{(a)}_{j}$. Moreover, this implies that $a=1$, because there are no $h=1$ fields in an unintegrated vertex operator $V_i^{(0)}$. As an example we want to consider the two possibilities $(a,b)\in\{(1,0),(1,1)\}$, which correspond, for instance, to $K_{\overline21}$ and $K_{\overline23}$ for \eqref{eq::3pt}, respectively:
\vspace{5pt}\begin{IEEEeqnarray}{rCl}
	K_{\overline21}&=&z_{\overline21}\tikzmark{Uo2_1}{U}_{\overline2}(z_{\overline2}) \tikzmark{V1_1}{V}_{1}(z_{1})\nonumber\\
	&=&-A^1_m(\lambda\gamma^mW_{\overline{2}})-V^1(i k_1\cdot A_{\overline{2}})+Q(A_1W_{\overline{2}})\ ,\label{eq::K_0}\vspace{5pt}\\
	K_{\overline23}&=&z_{\overline23}\tikzmark{Uo2_2}{U}_{\overline2}(z_{\overline2}) \tikzmark{U3_1}{U}_{3}(z_{3})\nonumber\\
	&=&-(i k_{3} \cdot A_{\overline2})U_{3}+\partial\theta^\alpha D_\alpha A^3_\beta W_{\overline3}^\beta+\Pi^mi k_m^3(A_3W_{\overline2})+(\partial\theta\gamma^mW_3)A^{\overline2}_m\nonumber\\
	&&+\frac14(d\gamma^{mn}W_{\overline3})\mathcal F^3_{mn}+N^{mn}\left(k^3_m(W_3\gamma_nW_{\overline2}+\eta^{ab}\mathcal F^{\overline2}_{ma}\mathcal F^3_{nb})\right)\ .
	\begin{tikzpicture}[overlay,remember picture,>=latex,shorten >=1pt,shorten <=1pt,very thin]
		\draw[->-] (Uo2_1) --++(0,10pt) -| (V1_1);
		\draw[->-] (Uo2_2) --++(0,10pt) -| (U3_1);
	\end{tikzpicture}
\end{IEEEeqnarray}
Note, that each superfield above still depends on the according vertex operator position, i.e$.$ each superfield $\mathcal V_i$ and conformal primary still depends on $z_i$ even after the contraction, which is different from the literature. That is important when defining $K$s involving more than one contraction (for \eqref{eq::3pt} we have three integrated vertex operators and therefore three contractions per term). Let's give some examples with two and three integrated vertex operators
\vspace{5pt}\begin{IEEEeqnarray}{rCl}
	K_{\overline213\overline2}&=&z_{\overline21}z_{3\overline2}\tikzmark{Uo2_5}{U}_{\overline2}(z_{\overline2})\tikzmark{U3_4}{U}_{3}(z_{3})\tikzmark{V1_3}{V}_1(z_1)\nonumber
	\begin{tikzpicture}[overlay,remember picture,>=latex,shorten >=1pt,shorten <=1pt,very thin]
		\draw[->-] (Uo2_5) --++(0,10pt) -| (V1_3);
		\draw[->-] (U3_4) --++(0,-10pt) -| (Uo2_5);
	\end{tikzpicture}\vspace{5pt}\\
	&=&\left[(i k_1\cdot A_{\overline2})(i k_{\overline2}\cdot A_3)-ik^1_m(W_{\overline2}\gamma^m W_3)+s_{1\overline2}(A_2W_3)\right]V_1+(\lambda\gamma^mW_{\overline2})(i k_{\overline2}\cdot A_3)A^1_m\nonumber\\
	&&-\frac14(\lambda\gamma^m\gamma^{pq}W_3)A^1_m\mathcal F^{\overline2}_{pq}-s_{\overline23}(A_1W_{\overline2})V_3-Q\left[(i k_{\overline2}\cdot A_3)(A_1W_{\overline2})-\frac14(A_1\gamma^{mn}W_3)\mathcal F^{\overline2}_{mn}\right]\ ,\nonumber\\\label{eq::K_1}\\
	K_{\overline213\overline2\overline31}&=&z_{\overline21}z_{3\overline2}z_{\overline31}\tikzmark{Uo2_3}{U}_{\overline2}(z_{\overline2})\tikzmark{U3_2}{U}_{3}(z_{3})\tikzmark{Uo3_1}{U}_{\overline3}(z_{\overline3})\tikzmark{V1_2}{V}_1(z_1)
	\begin{tikzpicture}[overlay,remember picture,>=latex,shorten >=1pt,shorten <=1pt,very thin]
		\draw[->-] (Uo2_3) --++(0,14pt) -| (V1_2);
		\draw[->-] (U3_2) --++(0,-10pt) -| (Uo2_3);
		\draw[->-] (Uo3_1) --++(0,10pt) -| (V1_2);
	\end{tikzpicture}\nonumber\vspace{5pt}\\
	&=&\left[(i k_1\cdot A_{\overline2})(i k_{\overline2}\cdot A_3)-ik^1_m(W_{\overline2}\gamma^mW_3)+s_{1\overline2}(A_{\overline2}W_3)\right]K_{\overline31}+\biggl[(\lambda\gamma^mW_{\overline2})(i k_{\overline2}\cdot A_3)\nonumber\\
	&&-\frac14(\lambda\gamma^m\gamma^{pq}W_3)\mathcal F^{\overline2}_{pq}\biggl]\left(-(i k_1\cdot A_{\overline 3})A^1_m+(W_1\gamma^mW_{\overline3})+k^1_m(A_1W_{\overline3})\right)\nonumber\\
	&&+\frac18\biggl[(\lambda\gamma^{rs}\gamma^mW_{\overline2})(i k_{\overline2}\cdot A_3)-\frac14(\lambda\gamma^{rs}\gamma^m\gamma^{pq}W_3)\biggl]A^1_m\mathcal F^{\overline2}_{pq}\mathcal F^{\overline3}_{rs}+s_{\overline23}\biggl[(i k_1\cdot A_{\overline3})(A_1W_{\overline2})V_3\nonumber\\
	&&+D_\alpha A^1_\beta W^\beta_{\overline2}W^\alpha_{\overline3}V_3\biggl]-s_{1\overline3}\left[(i k_{\overline2}\cdot A_{3})(A_1W_{\overline2})+\frac14(A_1\gamma^{mn}W_3)\mathcal F^{\overline2}_{mn}\right]V_{\overline3}\ ,\label{eq::K_2}
	\vspace{9pt}\\
	K_{\overline233\overline2\overline3\overline2}&=&z_{\overline23}z_{3\overline2}z_{\overline3\overline2}\tikzmark{Uo2_4}{U}_{\overline2}(z_{\overline2})\tikzmark{U3_3}{U}_{3}(z_{3})\tikzmark{Uo3_2}{U}_{\overline3}(z_{\overline3})\nonumber
	\begin{tikzpicture}[overlay,remember picture,>=latex,shorten >=1pt,shorten <=1pt,very thin]
		\draw[-<-] (Uo2_4) --++(0,10pt) -| (U3_3);
		\draw[->-] (Uo2_4) --++(0,-10pt) -| (U3_3);
		\draw[-<-] (Uo2_4) --++(0,14pt) -| (Uo3_2);
	\end{tikzpicture}\vspace{5pt}\\
	&=&(1-s_{\overline23})\left\{(i k_{\overline2}\cdot A_{\overline3})\left[(A_{\overline2}W_3)+(A_3W_{\overline2})-(A_{\overline2}\cdot A_3)\right]-D_\alpha A^{\overline2}_\beta W^\beta_3W^\alpha_{\overline3}\vphantom{\frac14}\right.\nonumber\\
	&&\left.-\frac14(A_3\gamma^{mn}W_{\overline3})\mathcal F^{\overline2}_{mn}+(W_{\overline2}\gamma^mW_{\overline3})A^3_m+(i k_{\overline2}\cdot A_3)(A_{\overline2}W_{\overline3})\right\}\ .\label{eq::K_3}
\end{IEEEeqnarray}
\vspace{5pt}Concretely the arrow notation for Wick contractions means the following: In \eqref{eq::K_1} for example the contraction  $\tikzmark{P1_1}{U}_{\overline2}(z_{\overline2}) \tikzmark{P1_2}{V}_{1}(z_{1})
\begin{tikzpicture}[overlay,remember picture,>=latex,shorten >=1pt,shorten <=1pt,very thin]
	\draw[->-] (P1_1) --++(0,10pt) -| (P1_2);
\end{tikzpicture}$ is given by \eqref{eq::K_0}. As we stressed above, the fields with index $\overline2$ still depend on $z_{\overline2}$ (and only those fields). The second contraction with $U_{3}(z_3)$, i.e.\ \vspace{5pt}$\tikzmark{P1_3}{U}_{\overline2}(z_{\overline2}) \tikzmark{P1_4}{U}_{3}(z_{3})
\begin{tikzpicture}[overlay,remember picture,>=latex,shorten >=1pt,shorten <=1pt,very thin]
\draw[->-] (P1_4) --++(0,-10pt) -| (P1_3);
\end{tikzpicture}$ is meant to contract the $h=1$ primaries of $U_3(z_3)$ only with those terms in ${U}_{\overline2}(z_{\overline2}){V}_{1}(z_{1})$ which depend on $z_{\overline2}$. Similar remarks hold for \eqref{eq::K_2} and \eqref{eq::K_3}.\par  
Here we are looking at Wick contractions and have to sum over all possible contractions: We can contract each integrated vertex operator with the five other vertex operators in the correlator such that the overall number of contractions is given by $5\times 5\times 5=125$. Schematically, we can write all possible 125 contractions for the correlator in \eqref{eq::3pt} as $K_{\overline2i3j\overline3k}$ with $i,j,k\in\{1,\overline1,2,\overline2,3,\overline3\}$, but we don't contract a vertex operator with itself. Compared to six open strings on the disk \cite{6pt} at first we find a larger number of terms, because in \cite{6pt} the vertex operator contractions are computed via an OPE method. With this method they already combine some of the Wick contractions such that they end up with a total of 34 terms, see also footnote \ref{fn::1} below for more details. Moreover, we explicitly relate the building blocks of \cite{6pt} with Wick contractions $K_{\overline2i3j\overline3k}$ in \eqref{eq::K} below.

\subsection{Composite superfields $\tilde L_{jikipi},\tilde L_{jiki}$ and $\tilde L_{ji}$}\label{sec::blockL}
To employ the methods of previous scattering amplitude calculations we have to relate the composite superfields $K$ and $\tilde L$: Using partial fractioning the composite superfields $K$ can be combined to form the superfields $\tilde L$, cf.\ equation \eqref{eq::K} and appendix \ref{sec::correlator} in general. The  $\tilde L$ can be recursively defined as\footnote{We will only need the composite superfields $\tilde L_{ji},\tilde L_{jiki}$ and $\tilde L_{jikipi}$, because the composite superfields containing contractions between integrated vertex operators are either corrections to \eqref{eq::comp_super}, such that those can be written in terms of the BRST building blocks $T$ of appendix \ref{sec::building_block_T} or can be express through the superfields in \eqref{eq::comp_super} by using equation \eqref{eq::ident_order}, see for example \cite{6pt, npt_1,Mafra:2022wml} for more details.} \cite{5pt,6pt,npt_1}
\begin{IEEEeqnarray}{C}
	\begin{IEEEeqnarraybox}C
	\IEEEstrut
	\lim_{z_{\overline{2}}\to z_{1}}V_{1}(z_{1})U_{\overline{2}}(z_{\overline{2}})\to\frac{\tilde L_{\overline{2}1}(z_1)}{z_{\overline{2}1}}\ ,
	\qquad \lim_{z_{3}\to z_{1}}\tilde L_{\overline{2}1}(z_{1})U_{3}(z_{3})\to\frac{\tilde L_{\overline{2}131}(z_1)}{z_{31}}\ ,
	\\\lim_{z_{\overline 3}\to z_{1}}\tilde L_{\overline{2}131}(z_{1})U_{\overline{3}}(z_{\overline{3}})\to\frac{\tilde L_{\overline{2}131\overline{3}1}(z_1)}{z_{\overline{3}1}}\ .
	\IEEEstrut
	\end{IEEEeqnarraybox}
	\label{eq::comp_super}
\end{IEEEeqnarray}
By definition all $\tilde L$s above only depend on $z_1$ and not on the other world sheet coordinates any more. At first glance this might suggest that we can't express the superfields $K$ through $\tilde L$, but after all $h=1$ fields are integrated out, only the zero modes, which don't depend on world sheet coordinates, contribute. Hence, it is possible to find a relation between $K$ and $\tilde L$ inside a correlator. Integrating out the dimension 1 fields via \eqref{eq::comp_super} is a priori different from using Wick contractions. In order to make the distinction we sometimes refer to \eqref{eq::comp_super} as the OPE method or as OPE contractions. \par
The composite superfields can be expressed in terms of the usual SYM superfields of the pure spinor formalism by using the OPEs in \eqref{psfOPE} of section \ref{sec::psf}:
\begin{IEEEeqnarray}{rCl}
	\tilde L_{\overline{2}1}&=&\lim_{z_{\overline{2}}\to z_1}z_{\overline{2}1}V_{1}(z_1)U_{\overline{2}}(z_{\overline{2}})\nonumber\\
	&=&-A^1_m(\lambda\gamma^mW_{\overline{2}})-V^1(i k_1\cdot A_{\overline{2}})+Q(A_1W_{\overline{2}})\nonumber\\
	&=&L_{\overline{2}1}+Q(A_1W_{\overline{2}})\ .\label{eq::L21}\
\end{IEEEeqnarray}
For the scattering of two closed strings on the disk (and with the same reasoning also for the tree-level scattering of four open strings) the composite superfields $\tilde L_{ji}$ contain the BRST exact part $Q(A_iW_j)$ as shown in \cite{2pt,Mafra:2008ar}. However, $\langle Q(A_i W_j)V_k V_l\rangle=0$, because $QV_i=0$. For higher point amplitudes this term becomes multiplied by further unintegrated vertex operators, i.e.\  $Q(A_iW_j)U_{m_1}\ldots U_{m_n}V_kV_l$, which are not BRST exact at the level of the correlator, but rather satisfy $QU=\partial V$ such that
\begin{IEEEeqnarray}{l}
	\langle Q(A_iW_j)U_{m_1}\ldots U_{m_n}V_kV_l\rangle=-\sum_{p=1}^n\langle(A_iW_j)U_{m_1}\ldots\partial V_{m_p}\ldots U_{m_n}V_kV_l\rangle\neq0\ .
\end{IEEEeqnarray}
Nevertheless, we will call $Q(A_iW_j)(\ldots)$ and similar terms BRST exact in line with the literature \cite{5pt,6pt,npt_1}, because once we integrate over them in the amplitude they become BRST exact
\begin{IEEEeqnarray}{l}
	\int\mathrm{d}z_{m_1}\ldots\int\mathrm{d}z_{m_n}\,\langle Q(A_iW_j)U_{m_1}\ldots U_{m_n}V_kV_l\rangle=0\ .
\end{IEEEeqnarray}
With similar calculations as in \eqref{eq::L21} and following \cite{6pt, npt_1} we find expressions for the other superfields $\tilde L$ in \eqref{eq::comp_super}
\begin{IEEEeqnarray}{rCl}
	\tilde L_{\overline{2}131}&=&L_{\overline{2}131}-s_{1\overline{2}}[(A_{1} W_3)V_{\overline{2}}-(A_{\overline{2}} W_3)V_{1}]-(s_{13}+s_{\overline{2}3})(A_{1} W_{\overline{2}})V_3\nonumber\\
	&&-Q[(i k_{1}\cdot A_{\overline{2}})(A_{1} W_3)]-Q[A^{1}_m(W_{\overline{2}}\gamma^mW_3)]-Q[U_3\{A_{1} W_{\overline{2}}\}]\ ,\label{eq::tL2}\\
	\tilde L_{\overline{2}131\overline{3}1}&=&L_{\overline{2}131\overline{3}1}+(A_{1}W_{\overline{3}})[s_{1\overline{2}}V_{\overline{2}}(i k_{1}\cdot A_3)-s_{1\overline{2}}L_{3\overline{2}}+(s_{13}+s_{\overline{2}3})(i k_{1}\cdot A_{\overline{2}})V_3]\nonumber\\
	&&+(A_{\overline{2}}W_{\overline{3}})[-s_{1\overline{2}}(i k_{\overline{2}}\cdot A_3)V_{1}+s_{1\overline{2}}L_{31}]+(s_{13}+s_{\overline{2}3})(A_3W_{\overline{3}})L_{\overline{2}1}\nonumber\\
	&&-(s_{13}+s_{\overline{2}3})(W_{\overline{2}}\gamma^{m}W_{\overline{3}})A_m^{1}V_3+(W_3\gamma^mW_{\overline{3}})[-s_{1\overline{2}}A^{\overline{2}}_mV_{1}+s_{1\overline{2}}A_m^{1}V_{\overline{2}}]\nonumber\\
	&&-s_{1\overline{2}}[U_{\overline{3}}(A_{1}W_3)V_{\overline{2}}-U_{\overline{3}}\{A_{\overline{2}}W_3\}V_{1}+(A_{1}W_3)\tilde L_{\overline{3}\overline{2}}-(A_{\overline{2}}W_3)\tilde L_{\overline{3}1}]\nonumber\\
	&&-(s_{13}+s_{\overline{2}3})[U_{\overline{3}}(A_{1}W_{\overline{2}})V_3+(A_{1}W_{\overline{2}})\tilde L_{\overline{3}3}]\nonumber\\
	&&+(s_{1\overline{3}}+s_{\overline{2}\overline{3}}+s_{3\overline{3}})[(i k_{1}\cdot A_{\overline{2}})(A_{1}W_3)+(W_{\overline{2}}\gamma^mW_3)A_m^{1}-U_3\{A_{1}W_{\overline{2}}\}]V_{\overline3}\ ,\nonumber\\\label{eq::tL3}
\end{IEEEeqnarray}
where $L_{jiki}$ and $L_{jikili}$ are defined below and the expression $U_i\{A_jW_k\}$ is an abbreviation of the contraction of $U_i$ with $(A_jW_k)$ which is given by
\begin{IEEEeqnarray}l
	U_i\{A_jW_k\}=-(i k_{jk}\cdot A_i)(A_jW_k)+D_\alpha A^j_\beta W_k^\beta W_i^\alpha+\frac14(A_j\gamma^{mn}W_i)\mathcal F^{k}_{mn}\ ,
\end{IEEEeqnarray}
where $k_{jk}^\mu=k_j^\mu+k_k^\mu$. To obtain the expression in \eqref{eq::tL2} and \eqref{eq::tL3} we had to integrate the BRST charge by parts. Hence, we already used that $\langle Q(\ldots)\rangle=0$ for a generic superfield expression $(\ldots)$ and dropped terms of the form $Q(\ldots)$ in  \eqref{eq::tL3}, where the BRST operator acts on the complete superfield expression. 
For $\tilde L_{jikipi}$ we omitted this part already, because terms containing $\tilde L_{jikipi}$ can only have two more unintegrated vertex operators, which are BRST closed.
After discarding BRST exact pieces from $\tilde L$ (which are all terms in \eqref{eq::tL2} and \eqref{eq::tL3} that are not contained in $L$) we obtain the superfields $L$. The open string equivalents\footnote{The composite superfields for open and closed strings have the same structure, i.e$.$ their explicit form in terms of the superfields is the same, but the labelling of these superfields is different. For closed strings we can also have overlined labels of the right-movers -- as described in section \ref{sec_Vops} -- which means that some of the superfields have their polarisation vector and momentum multiplied by the matrix $D$, which accounts for either Neumann or Dirichlet boundary conditions.} 
 of the composite superfields $L$ can also be found for example in \cite{6pt,npt_1}. The OPE contractions result in the following form for those superfields\footnote{Note, that we deviate from the usual pure spinor convention where the momenta are defined as $k_\text{PSF}=ik$. Hence, we have different signs and additional factors of $i$ compared to the literature \cite{npt_1}.}
\begin{IEEEeqnarray}{rCl}
	L_{\overline2131}&=&-L_{\overline{2}1}(i k_{1\overline{2}}\cdot A_3)+(\lambda\gamma^mW_3)\left[\mathcal F^{\overline{2}}_{mn}A^n_1+(i k_1\cdot A_{\overline{2}})A^1_m-(W_1\gamma_mW_{\overline{2}})\right],\\
	L_{\overline{2}131\overline{3}1}&=&-L_{\overline{2}131}(i k_{1\overline{2}3}\cdot A_{\overline{3}})+(\lambda\gamma^mW_{\overline{3}})\Big[\mathcal F^{\overline{2}}_{nm}A^n_1(i k_{1\overline{2}}\cdot A_3)-(A_1\cdot i k_{\overline{2}})(W_{\overline{2}}\gamma_mW_3)\nonumber\\
	&&+\mathcal F^{\overline{2}}_{pq}A^p_1A^q_3k^3_m-\mathcal F^{\overline{2}}_{pq}A^q_{1}k^p_3A^3_m+\mathcal{F}^3_{mn}(W_{1}\gamma^nW_{\overline{2}})-\mathcal F^{3}_{mn}A^{n}_{1}(i k_{1}\cdot A_{\overline{2}})\nonumber\\
	&&+(W_{1}\gamma_mW_3)(i k_{1}\cdot A_{\overline2})+\big[(W_{1}\gamma_mW_{\overline2})-A^{1}_m(i k_1\cdot A_{\overline2})\big](i k_{1\overline{2}}\cdot A_3)\nonumber\\
	&&+\frac14(W_{\overline2}\gamma_{pq}\gamma_mW_3)\mathcal{F}^{pq}_{1}-\frac14(W_{1}\gamma_{pq}\gamma_mW_3)\mathcal{F}^{pq}_{\overline2}\Big]\ .
\end{IEEEeqnarray}
As mentioned before the superfields transform covariantly under the action of the BRST charge, i.e$.$ the BRST variation of higher order composite superfields can be written in terms of lower order composite superfields \cite{6pt}. To see this, we will make use of the recursive definition of the composite superfields instead of acting with $Q$ on their explicit expressions. We can utilize $QV=0$ and $QU=\partial V$ to obtain
\begin{IEEEeqnarray}{rCl}
	QL_{\overline21}&=&-s_{1\overline2}V_1V_{\overline2},\label{eq::QL1}\\
	QL_{\overline 2131}&=&\lim_{z_{3}\to z_1}z_{31}\Big[(QL_{\overline21})(z_1)U^{3}(z_{3})-L_{\overline21}\partial V^{3}(z_{3})\Big]\nonumber\\
	&=&-(s_{13}+s_{\overline23})L_{\overline21}V_3-s_{1\overline2}(L_{31}V_{\overline2}+V_1L_{3\overline2})\ ,\label{eq::QL2}\\
	QL_{\overline 2131\overline31}&=&\lim_{z_{\overline3}\to z_1}z_{\overline31}\Big[(QL_{\overline2131})(z_1)U^{\overline3}(z_{\overline3})-L_{\overline2131}\partial V^{\overline3}(z_{\overline3})\Big]\nonumber\\
	&=&-(s_{1\overline3}+s_{\overline2\overline3}+s_{3\overline3})L_{\overline2131}V_{\overline3}-(s_{13}+s_{\overline23})(L_{\overline21}L_{\overline33}+L_{\overline21\overline31}V_3)\nonumber\\
	&&-s_{1\overline2}(L_{31\overline31}V_{\overline2}+L_{31}L_{\overline3\overline2}+L_{\overline31}L_{3\overline2}+V_1L_{3\overline2\overline3\overline2})\ .\label{eq::QL3}
\end{IEEEeqnarray}
We evaluate the first term of the first line in \eqref{eq::QL2} and \eqref{eq::QL3} by using the recursive definition of the composite superfields \eqref{eq::comp_super} (after the BRST exact terms were discarded). For the second term of \eqref{eq::QL2} and \eqref{eq::QL3} we can just contract $\partial V=(\partial\lambda^\alpha)A_\alpha+\Pi^mi k_mV+\partial\theta^\alpha D_\alpha V$ with the composite superfield by applying the OPEs in equation \eqref{psfOPE}.\par
The action of the BRST charge on the composite superfields $L$ suggests that after discarding $Q(A_iW_j)$ in $L_{ji}$ we must also drop all the BRST exact terms in \eqref{eq::tL2}, because  \eqref{eq::QL2} only holds for the $L_{jiki}$ and not for $\tilde L_{jiki}$ and similar for the higher order composite superfields.

\subsection{BRST building blocks $T_{ijkp},T_{ijk}$ and $T_{ij}$}\label{sec::building_block_T}
As described in \cite{npt_1} for open strings the composite superfields can be substituted by the BRST building blocks $T_{1\overline23\ldots q}$ (with $q\in\{\overline 2, 3, \overline 3 \ldots\}$ if one chooses the three
unintegrated vertex operators as $V_1, V_{\overline1}$ and $V_2$ as we did) in two steps
\begin{IEEEeqnarray}{r}
	L_{\overline2131\dots q1}\overset{(i)}{\longrightarrow}\tilde T_{1\overline23\ldots q}\overset{(ii)}{\longrightarrow}T_{1\overline23\ldots q}\ .
\end{IEEEeqnarray}
The purpose of this substitution is to remove BRST exact terms\footnote{This procedure will make the amplitude manifestly invariant under BRST transformations.} that still appear in the superfields $L_{jikipi},L_{jiki}$ and $L_{ji}$ and after carrying out the first step $(i)$ also in $\tilde T_{ijkp},\tilde T_{ijk}$ and $\tilde T_{ij}$ while simultaneously preserving the BRST variation identities \eqref{eq::QL1}--\eqref{eq::QL3} but for $T_{ijkp},T_{ijk}$ and $T_{ij}$, i.e.\ we want the equations \eqref{eq::QL1}--\eqref{eq::QL3} to hold when substituting the composite superfields $L$ by their corresponding building blocks $T$.\par
To define the BRST building blocks for closed strings we will closely follow \cite{npt_1} for the open string discussion. As a first step $(i)$ we redefine the superfields $L_{\overline2131\ldots q1}\to\tilde T_{1\overline 23\ldots q}$ such that $Q\tilde T_{1\overline 23\ldots q}$ is written in terms of the building blocks $T_{1\overline23\ldots p}$ with lower rank $p<q$. Explicitly, we find for the relevant building blocks
\begin{IEEEeqnarray}{rCl}
	Q\tilde{T}_{1\overline23}&=&-s_{1\overline2}(T_{13}V_2+V_1T_{\overline23})-(s_{13}+s_{\overline23})T_{1\overline2}V_3\ ,\label{eq::QtT2}\\
	Q\tilde{T}_{1\overline23\overline3}&=&-(s_{1\overline3}+s_{\overline{23}}+s_{3\overline3})T_{1\overline23}V_{\overline3}-(s_{13}+s_{\overline23})(T_{1\overline2}T_{3\overline3}+T_{1\overline{23}}V_3)\nonumber\\
	&&-s_{1\overline2}(T_{13\overline3}V_{\overline2}+T_{13}T_{\overline{23}}+T_{1\overline3}T_{\overline23}+V_1T_{\overline23\overline3})\ .\label{eq::QtT3}
\end{IEEEeqnarray}
With \eqref{eq::QtT2} and \eqref{eq::QtT3} we can check that specific combinations of $\tilde T$s are BRST closed. For example we find $Q(\tilde T_{1\overline23}+\tilde T_{\overline231}+\tilde T_{31\overline2})=0$.\par
We are now finally ready to remove the remaining BRST exact parts: So for the second step $(ii)$ we take the sums below of the BRST building blocks $\tilde T_{ijk}$ and $\tilde T_{ijkp}$ 
\begin{IEEEeqnarray}{rCl}
	\sum_{\text{perm.}}\tilde T_{ijk}&=&QR^{(I)}_{ijk}\ ,\qquad I=1,2\ ,\label{eq::T_perm_1}\\
	\sum_{\text{perm.}}\tilde T_{ijkp}&=&QR^{(I)}_{ijkp}\ ,\qquad I=1,2,3\label{eq::T_perm_2}
\end{IEEEeqnarray}
which are BRST closed such that we have to subtract the corresponding BRST exact part from $\tilde T_{ijk}$ and $\tilde T_{ijkp}$, see \eqref{eq::T123} and \eqref{eq::T3} for the explicit expression of $T_{ijk}$ and $T_{ijkp}$. Thereby, the BRST closed sums of $\tilde T$s will become BRST symmetries for the BRST building blocks $T$
\begin{IEEEeqnarray}l
	\sum_{\text{perm.}}T_{ij}=\sum_{\text{perm.}}T_{ijk}=\sum_{\text{perm.}}T_{ijkp}=0\ ,\label{eq::T_perm}
\end{IEEEeqnarray}
which will be the condition that we can use to construct $T_{ijkp},T_{ijk}$ and $T_{ij}$ from $\tilde T_{ijkp},\tilde T_{ijk}$ and $\tilde T_{ij}$. The building blocks defined in this way transform under the action of the BRST charge as follows
\begin{IEEEeqnarray}{rCl}
	QT_{1\overline2}&=&-s_{12}V_1V_{\overline2}\ ,\\
	QT_{1\overline23}&=&-s_{1\overline2}(T_{13}V_2+V_1T_{\overline23})-(s_{13}+s_{\overline23})T_{1\overline2}V_3\ ,\\
	QT_{1\overline23\overline3}&=&-(s_{1\overline3}+s_{\overline{23}}+s_{3\overline3})T_{1\overline23}V_{\overline3}-(s_{13}+s_{\overline23})(T_{1\overline2}T_{3\overline3}+T_{1\overline{23}}V_3)\nonumber\\
	&&-s_{1\overline2}(T_{13\overline3}V_{\overline2}+T_{13}T_{\overline{23}}+T_{1\overline3}T_{\overline23}+V_1T_{\overline23\overline3})\ .
\end{IEEEeqnarray}
During the first step $(i)$ we have not discussed the redefinition of $L_{ji}$ to $\tilde T_{ij}$, due to the fact that $\tilde T_{ij}\equiv L_{ji}$. This building block only requires the second step: The action of the BRST charge in \eqref{eq::QL1} together with the equations of motion \eqref{eq::eoms_superfields} suggest that $Q(\tilde T_{\overline21}+\tilde T_{1\overline2})=-s_{1\overline2}(V_1V_{\overline2}+V_{\overline2}V_1)=0$, i.e.\ $\tilde T_{\overline21}+\tilde T_{1\overline2}$  is BRST closed and moreover also BRST exact \cite{Mafra:2010ir}
\begin{IEEEeqnarray}l
	\tilde T_{\overline21}+\tilde T_{1\overline2}=-Q(A_1\cdot A_{\overline2})\equiv-QD_{1\overline 2}\ .
\end{IEEEeqnarray}
We can construct $T_{1\overline2}$ by using \eqref{eq::T_perm}, i.e.\ $T_{1\overline2}+T_{\overline21}=0$, which is achieved by
\begin{IEEEeqnarray}l
	T_{1\overline2}=\tilde T_{[\overline21]}=\tilde T_{\overline21}+\frac12QD_{1\overline2}\ .
\end{IEEEeqnarray}
Next we want to construct $T_{1\overline 23}$ by carrying out step $(ii)$. First, we substitute $L_{ji}=\tilde T_{ij}= T_{ij}-\frac12QD_{ij}$ in $QL_{\overline2131}$ in equation \eqref{eq::QL2}. Demanding that this results in \eqref{eq::QtT2} we find that $\tilde T_{1\overline23}$ takes the following form
\begin{IEEEeqnarray}l
	\tilde T_{1\overline23}=L_{\overline2131}-\frac12s_{1\overline2}[D_{13}V_{\overline2}-D_{\overline23}V_1]-\frac12(s_{13}+s_{\overline23})D_{1\overline2}V_3\ .
\end{IEEEeqnarray}
Further, we consider the two BRST closed combinations of $\tilde T_{ijk}$ to determine the remaining BRST exact terms
\begin{IEEEeqnarray}l
	Q(\tilde T_{1\overline23}+\tilde T_{\overline213})=0,\qquad Q(\tilde T_{1\overline23}+\tilde T_{31\overline2}+\tilde T_{\overline231})=0\label{eq::9}\ .
\end{IEEEeqnarray}
Note, that the higher rank building blocks inherit all symmetries from the lower order symmetries in their first indices, which follows from the recursive definition of the composite superfields $\tilde L$ in \eqref{eq::comp_super}. This is the origin of the first sum in \eqref{eq::9}, which goes back to antisymmetry of $T_{ij}$. Given that the BRST cohomology for the composite superfields is empty, the BRST closed combinations are also BRST exact \cite{npt_1}, i.e.
\begin{IEEEeqnarray}l
	\tilde T_{1\overline23}+\tilde T_{\overline213}=QR^{(1)}_{1\overline23},\qquad \tilde T_{1\overline23}+\tilde T_{31\overline2}+\tilde T_{\overline231}=QR^{(2)}_{1\overline23}\ ,\label{eq::BRST_exact_part}
\end{IEEEeqnarray}
where the ghost number zero superfields are given by $R^{(1)}_{1\overline23}=D_{1\overline2}(i k_{1\overline2}\cdot A_3)$ and $R^{(2)}_{1\overline23}=D_{1\overline2}(i k_{\overline2}\cdot A_3)+\text{cyclic}(1\overline23)$, which are motivated by the residues of the double pole contractions of two and three integrated vertex operators, respectively. Subtracting the  BRST exact parts in equation \eqref{eq::BRST_exact_part} we obtain the new BRST building block
\begin{IEEEeqnarray}{rCl}
	T_{1\overline23}&=&\tilde T_{1\overline23}-QS^{(1)}_{1\overline23}\nonumber\\
	&=&\frac13\left(\tilde T_{1\overline23}-\tilde T_{\overline213}\right)+\frac16\left(\tilde T_{3\overline21}-\tilde T_{31\overline2}+\tilde T_{13\overline2}-\tilde T_{\overline231}\right)\label{eq::T123}
\end{IEEEeqnarray}
with $S^{(1)}_{1\overline23}=\frac12R^{(1)}_{1\overline23}+\frac16\left(R^{(2)}_{1\overline23}-R^{(2)}_{\overline213}\right)$. The building block $T_{1\overline23}$ satisfies the following BRST symmetries:
\begin{IEEEeqnarray}{rCl}
	T_{1\overline23}+T_{\overline213}=T_{1\overline23}+T_{31\overline2}+T_{\overline231}=0\ ,
\end{IEEEeqnarray}
which correspond to \eqref{eq::T_perm}. The remaining building block $T_{ijkp}$ is obtained similarly and requires the redefinition of $L_{ji}$ and $L_{jiki}$: Executing step $(i)$ first demands to substitute $L_{ji}\to T_{ij}$ and $L_{jiki}\to T_{ijk}$ in the right hand side of \eqref{eq::QL3} to find $\tilde T_{1\overline23\overline3}$
\begin{IEEEeqnarray*}{rCl}
	\tilde T_{1\overline23\overline3}&=&L_{\overline2131\overline31}+\frac14\left[(s_{13}+s_{\overline23})D_{1\overline2}QD_{3\overline3}+s_{1\overline2}(D_{13}QD_{\overline{23}}+D_{1\overline3}QD_{\overline23})\right]\\
	&&-\frac12\left[(s_{13}+s_{\overline23})(D_{1\overline2}T_{3\overline3}-D_{3\overline3}T_{1\overline2})+s_{1\overline2}(D_{13}T_{\overline{23}}+D_{1\overline3}T_{\overline23}-D_{\overline23}T_{1\overline3}-D_{\overline{23}}T_{13})\right]\\
	&&+(s_{1\overline3}+s_{\overline{23}}+s_{3\overline3})S^{(1)}_{1\overline23}V_{\overline3}+(s_{13}+s_{\overline23})S^{(1)}_{1\overline{23}}V_3-s_{1\overline2}(S^{(1)}_{\overline23\overline4}V_1-S^{(1)}_{13\overline3}V_{\overline2})\ ,\IEEEyesnumber
\end{IEEEeqnarray*}
whose BRST variation is as required given by \eqref{eq::QtT3}. Moreover, the first three labels of $\tilde T_{1\overline23\overline3}$ inherit the two lower order identities of $\tilde T_{1\overline23}$ in \eqref{eq::9} and there is also a further BRST identity, which involves also the fourth index
\begin{IEEEeqnarray}{C}
	Q(\tilde T_{1\overline23\overline{3}}+\tilde T_{\overline 213\overline{3}})=0\ ,\qquad Q(\tilde T_{1\overline23\overline{3}}+\tilde T_{31\overline2\overline{3}}+\tilde T_{\overline231\overline{3}})=0\ ,\nonumber\\ Q(\tilde T_{1\overline23\overline3}-\tilde T_{1\overline2\overline33}+\tilde T_{3\overline31\overline2}-\tilde T_{3\overline3\overline21})=0\ .
\end{IEEEeqnarray}
The BRST exact form of these equations can be obtained by using the equations of motion of the superfields
\begin{IEEEeqnarray}l
	\tilde T_{1\overline23\overline{3}}+\tilde T_{\overline 213\overline{3}}=QR^{(1)}_{1\overline23\overline3}\ ,\label{eq::tT3_brst_1}\\
	\tilde T_{1\overline23\overline{3}}+\tilde T_{31\overline2\overline{3}}+\tilde T_{\overline231\overline{3}}=QR^{(2)}_{1\overline23\overline3}\ ,\\
	\tilde T_{1\overline23\overline3}-\tilde T_{1\overline2\overline33}+\tilde T_{3\overline31\overline2}-\tilde T_{3\overline3\overline21}=QR^{(3)}_{1\overline23\overline3}\ ,\label{eq::tT3_brst_3}
\end{IEEEeqnarray}
where we have defined $R^{(i)}_{1\overline23\overline3}$ as
\begin{IEEEeqnarray}{rCl}
	R^{(1)}_{1\overline23\overline3}&=&-R^{(1)}_{1\overline23}(i k_{1\overline23}\cdot A_{\overline3})+\frac14s_{1\overline2}[D_{13}D_{\overline{23}}+D_{1\overline3}D_{\overline23}]\ ,\\
	R^{(2)}_{1\overline23\overline3}&=&-R^{(2)}_{1\overline23}(i k_{1\overline23}\cdot A_{\overline3})+\frac14[s_{1\overline2}D_{\overline23}D_{1\overline{3}}+s_{\overline23}D_{\overline{23}}D_{13}+s_{13}D_{3\overline3}D_{1\overline2}]\ ,\\
	R^{(3)}_{1\overline23\overline3}&=&(i k_1\cdot A_{\overline2})[D_{1\overline3}(i k_{\overline3}\cdot A_3)-D_{13}(i k_{3}\cdot A_{\overline3})]-(i k_{\overline2}\cdot A_{1})[D_{\overline2\overline3}(i k_{\overline3}\cdot A_3)-D_{\overline23}(i k_{3}\cdot A_{\overline3})]\nonumber\\
	&&-\frac14D_{1\overline2}D_{3\overline3}(s_{1\overline3}+s_{\overline23}-s_{13}-s_{\overline{23}})+D_{1\overline2}[(i k_{\overline3}\cdot A_3)(i k_{\overline2}\cdot A_{\overline3})-(i k_{3}\cdot A_{\overline3})(i k_{\overline2}\cdot A_{3})]\nonumber\\
	&&+D_{3\overline3}[(i k_{\overline2}\cdot A_1)(i k_{\overline3}\cdot A_{\overline2})-(i k_{1}\cdot A_{\overline2})(i k_{\overline3}\cdot A_{1})]+(W_1\gamma^mW_{\overline2})(W_3\gamma_mW_{\overline3})\ ,
\end{IEEEeqnarray}
where $k_{mnp}=k_m+k_n+k_p$. In the second step $\tilde T_{1\overline23\overline3}\overset{(ii)}{\longrightarrow}T_{1\overline23\overline3}$ we want to remove these BRST exact terms, which can be realized by defining
\begin{IEEEeqnarray}l
	T_{1\overline23\overline3}=\tilde T_{1\overline23\overline3}-QS^{(2)}_{1\overline23\overline3}\ .\label{eq::T3}
\end{IEEEeqnarray}
In $T_{1\overline23\overline3}$ we introduced the recursively defined field $S^{(2)}_{1\overline23\overline3}$ as
\begin{IEEEeqnarray}{rCl}
	S^{(2)}_{1\overline23\overline3}&=&\frac34S^{(1)}_{1\overline23\overline3}+\frac14\left(S^{(1)}_{1\overline2\overline33}-S^{(1)}_{3\overline31\overline2}+S^{(1)}_{3\overline3\overline21}\right)+\frac14R^{(3)}_{1\overline23\overline3}\ ,\\
	S^{(1)}_{1\overline23\overline3}&=&\frac12R^{(1)}_{1\overline23\overline3}+\frac16\left(R^{(2)}_{1\overline23\overline3}-R^{(2)}_{\overline213\overline3}\right)\ ,
\end{IEEEeqnarray}
which satisfies the following three identities
\begin{IEEEeqnarray}{l}
	S^{(2)}_{1\overline23\overline3}+S^{(2)}_{\overline213\overline3}=R^{(1)}_{1\overline23\overline3}\ ,\\
	S^{(2)}_{1\overline23\overline3}+S^{(2)}_{31\overline2\overline3}+S^{(2)}_{\overline231\overline3}=R^{(2)}_{1\overline23\overline3}\ ,\\
	S^{(2)}_{1\overline23\overline3}-S^{(2)}_{1\overline2\overline33}+S^{(2)}_{3\overline31\overline2}-S^{(2)}_{3\overline3\overline21}=R^{(3)}_{1\overline23\overline3}\ .
\end{IEEEeqnarray}
With these identities the building block in \eqref{eq::T3} obeys the BRST symmetries following from \eqref{eq::tT3_brst_1}--\eqref{eq::tT3_brst_3}
\begin{IEEEeqnarray}{rCl}
	 T_{1\overline23\overline{3}}+ T_{\overline 213\overline{3}}=
	 T_{1\overline23\overline{3}}+ T_{31\overline2\overline{3}}+ T_{\overline231\overline{3}}=
	 T_{1\overline23\overline3}- T_{1\overline2\overline33}+ T_{3\overline31\overline2}- T_{3\overline3\overline21}=0\ .
\end{IEEEeqnarray}
To conclude this section we want to remark that we don't need to perform step $(ii)$ for $L_{jikipi}$, because after we arrive at $L_{jikipi}\overset{(i)}{\longrightarrow}\tilde T_{ijkp}$ in the correlator this building block is already BRST exact: The building block $\tilde T_{ijkp}$ always appears with two unintegrated vertex operators in the CFT correlator of the amplitude in \eqref{eq::3pt}. Hence, in the correlator we can drop the BRST exact terms in  $\tilde T_{ijkp}$ without step $(ii)$ because after integrating the BRST charge by part, we can use $QV=0$ such that the BRST exact parts of $\tilde T_{ijkp}$ vanish. Moreover also due to the same reasoning the identities \eqref{eq::tT3_brst_1}--\eqref{eq::tT3_brst_3} lead to vanishing results within correlators, i.e.
\begin{IEEEeqnarray}l
	\left\langle\left(\tilde T_{1\overline23\overline{3}}+\tilde T_{\overline 213\overline{3}}\right)V_{\overline1} V_2\right\rangle=\left\langle QR^{\left(1\right)}_{1\overline23\overline3}V_{\overline1} V_2\right\rangle=0\ ,\\
	\left\langle\left(\tilde T_{1\overline23\overline{3}}+\tilde T_{31\overline2\overline{3}}+\tilde T_{\overline231\overline{3}}\right)V_{\overline1} V_2\right\rangle=\left\langle QR^{\left(2\right)}_{1\overline23\overline3}V_{\overline1} V_2\right\rangle=0\ ,\\
	\left\langle\left(\tilde T_{1\overline23\overline3}-\tilde T_{1\overline2\overline33}+\tilde T_{3\overline31\overline2}-\tilde T_{3\overline3\overline21}\right)V_{\overline1} V_2\right\rangle=\left\langle QR^{\left(3\right)}_{1\overline23\overline3}V_{\overline1} V_2\right\rangle=0\ .
\end{IEEEeqnarray}
Performing the second step $(ii)$ for the scattering of three closed strings on the disk is not strictly necessary and in some sense obsolete, because $\langle \tilde T_{ijkp}V_mV_n\rangle=\langle T_{ijkp}V_mV_n\rangle$. 
	\section{The CFT correlator of three closed strings on the disk}\label{sec::correlator}
	We can take \eqref{eq::amp_mon_4a} and use $(2.5)$ in \cite{npt_2} to obtain \eqref{eq::partial_1} and \eqref{eq::partial_2} as we did in section \ref{sec::integration}. However, here we would like to show how to obtain \eqref{eq::A3} by performing the Wick contractions in \eqref{eq::q3} and $PSL(2,\mathbb R)$ transforming the result as a consistency check.\par
	As we can write the scattering of three closed strings on the disk in terms of six open strings on the disk, we would at first expect a similar evaluation of the correlator as in \cite{6pt}. For the six-point function of open strings on the disk contractions with the vertex operator whose position is fixed to infinity do not contribute to the correlator because they go as $\lim_{z\to\infty}\frac1z=0$. But for three closed strings on the disk the natural gauge fixing procedure, cf$.$ equation \eqref{eq::3pt}, does not result in a correlator with one vertex operator position at infinity. Hence, we expect to find contractions of all vertex operators in the correlator, which is different compared to \cite{6pt}.\par
	In the following we will address this issue: For that, we make use of the fact that we can write the correlator of three closed strings as six open strings. Moreover, we will performed a $PSL(2,\mathbb R)$ transformation that maps the fixed vertex operator positions $(-1,y,1)$ to $(0,1,\infty)$. For this purpose we will utilize that a correlator of holomorphic and antiholomorphic fields on the unit disk is invariant under $PSL(2,\mathbb R)$ transformations.\par
	Afterwards, following \cite{6pt}, we use some of the features of the pure spinor formalism which allow us to perform a rather simple evaluation of this correlator. For the computation we consider the interplay between the kinematic building blocks and integration by parts of their associated integrals. In the end we will obtain a compact and simple result written in pure spinor superspace, which is organized using the BRST building blocks described in appendix \ref{sec::block}.
	
	\subsection{The correlator of three closed strings as the correlator of six open strings}\label{sec::correlator1}
	
	The correlator is given in terms of all possible contractions of the integrated vertex operators with each other and with the unintegrated vertex operators. According to appendix \ref{sec::contraction} the sum over all possible contractions can be organised in terms of the composite superfields $K_{\overline2i3j\overline3k}$. Using the OPEs in \eqref{psfOPE} to perform these contractions the three-point amplitude of closed strings is given by\footnote{For simplicity, we have not given the non-contracted unintegrated vertex operators, but they are still there.}
	\begin{IEEEeqnarray}{rCl}
		\left\langle V_1(1)V_{\overline1}(-1)V_2(y)U_{\overline2}(-y)U_3(\xi)U_{\overline3}(\eta)\right\rangle=\sum_{i,j, k}\KN(y,\xi,\eta)\frac{\langle K_{\overline2i3j\overline3k}\rangle}{z_{\overline2i}z_{3j}z_{\overline3k}}\ ,
	\end{IEEEeqnarray}
	where the sum runs over $i,j,k\in\{1,\overline1,2,\overline2,3,\overline3\}$ with $i\neq\overline2,j\neq3$ and $k\neq\overline3$. Using partial fractioning and the relations between the superfields $K$ and $\tilde L$, which are given by\footnote{These relations can be found by comparing the OPE method for vertex operator contractions of \cite{npt_1} with vertex operator contractions one obtains using Wick's theorem.}
	\begin{IEEEeqnarray}{l}
		\begin{IEEEeqnarraybox}[][c]{rCl}
		\IEEEstrut
		\tilde L_{\overline2i3i\overline3i}&=& K_{\overline2i3i\overline3i}+K_{\overline2i3i\overline3\overline2}+K_{\overline2i3i\overline33}+K_{\overline2i3\overline2\overline3i}+K_{\overline2i3\overline2\overline3\overline2}+K_{\overline2i3\overline2\overline33}\ ,\\
		\tilde L_{\overline233i\overline3i}&=& K_{\overline233i\overline3i}+K_{\overline233i\overline3\overline2}+K_{\overline233i\overline33}-K_{\overline2i3\overline2\overline3i}-K_{\overline2i3\overline2\overline3\overline2}-K_{\overline2i3\overline2\overline33}\ ,\\
		\tilde L_{\overline233\overline3\overline3i}&=& K_{\overline233\overline3\overline3i}-K_{\overline233i\overline33}-K_{\overline233i\overline3\overline2}+K_{\overline2i3\overline2\overline3\overline2}+K_{\overline2i3\overline2\overline33}-K_{\overline2\overline33\overline2\overline3i}\ ,\\
		\tilde L_{\overline23\overline233\overline3} &=&K_{\overline233\overline2\overline3\overline2}-K_{\overline233\overline3\overline3\overline2}-K_{\overline2\overline33\overline2\overline33}-K_{\overline2\overline33\overline2\overline3\overline2}+K_{\overline2\overline33\overline3\overline33}\ ,\\
		\tilde L_{\overline2i3\overline3\overline3i}&=& K_{\overline2i3\overline3\overline3i}-K_{\overline2i3i\overline33}-K_{\overline2i3\overline2\overline33}+K_{\overline2i3\overline3\overline3\overline2}\ ,\\
		\tilde L_{\overline2i3\overline3\overline3j} &=& K_{\overline2i3\overline3\overline3j}-K_{\overline2i3j\overline33}\ ,\qquad\tilde L_{\overline2i3i\overline3j}= K_{\overline2i3i\overline3j}+K_{\overline2i3\overline2\overline3j}\ ,\\
		\tilde L_{\overline2\overline33\overline33\overline3}&=& 
		 K_{\overline233\overline2\overline33}+K_{\overline233\overline2\overline3\overline2}\ \ ,\qquad \tilde L_{\overline2i3\overline33\overline3} = K_{\overline2i3\overline3\overline33}\ ,\\
		 \tilde L_{\overline2i3j\overline3k}&=&K_{\overline2i3j\overline3k}
		 \IEEEstrut
		\end{IEEEeqnarraybox}\label{eq::K}
	\end{IEEEeqnarray}
	and by similar relations but with permutations of the integrated vertex operators, we can reorganise the kinematic terms according to the recursive definitions of the building blocks of appendix \ref{sec::blockL} and write the correlator as a sum of single- and double-pole integrands\footnote{Comparing this to the literature, in \cite{6pt} one vertex operator is fixed to $\infty$ and therefore there are 34 different terms in (3.2) in \cite{6pt}. Because we have not fixed any vertex operator to $\infty$ the expression corresponding to (3.2) in \cite{6pt} contains 76 terms in this different vertex operator position fixing in \eqref{eq8.0}.\label{fn::1}}  
	\begin{IEEEeqnarray}{rCl}
		\IEEEeqnarraymulticol{3}{l}{\left\langle V_1(1)V_{\overline1}(-1)V_2(y)U_{\overline2}(-y)U_3(\xi)U_{\overline3}(\eta)\right\rangle=}\nonumber\\
		&=&\KN(y,\xi,\eta)\biggl\langle\smashoperator{\sum_{\substack{i,j,k\\\in\{1,\overline1,2\}}}}\epsilon_{ijk}\biggl(\frac{  \tilde{L}_{\overline2i3j\overline3k} }{z_{\overline2i}z_{3j}z_{\overline3k}}+\frac{  \tilde{L}_{\overline2i3i\overline3j}V_k }{z_{\overline2i}z_{3i}z_{\overline3j}}+\frac{  \tilde{L}_{\overline2i3j\overline3i}V_k }{z_{\overline2i}z_{3j}z_{\overline3i}}+\frac{  \tilde{L}_{\overline2j3i\overline3i}V_k}{z_{\overline2j}z_{3i}z_{\overline3i}}+\frac{  \tilde{L}_{\overline2i3\overline3\overline3j}V_k}{z_{\overline2i}z_{3\overline3}z_{\overline3j}}\nonumber\\
		&&+\frac{  \tilde{L}_{\overline2\overline33i\overline3j}V_k}{z_{\overline2\overline3}z_{3i}z_{\overline3j}}+\frac{ \tilde{L}_{\overline233i\overline3j}V_k}{z_{\overline23}z_{3i}z_{\overline3j}}+\frac{ \tilde{L}_{\overline2i3i\overline3i}V_jV_k}{z_{\overline2i}z_{3i}z_{\overline3i}}+\frac{  \tilde{L}_{\overline2i3\overline3\overline3i}V_j V_k}{z_{\overline2i}z_{3\overline3}z_{\overline3i}}+\frac{  \tilde{L}_{\overline2\overline33i\overline3i}V_j V_k}{z_{\overline2\overline3}z_{3i}z_{\overline3i}}+\frac{  \tilde{L}_{\overline233i\overline3i}V_j V_k}{z_{\overline23}z_{3i}z_{\overline3i}}\nonumber\\
		&&+\frac{ \tilde{L}_{\overline2\overline33\overline33i}V_j V_k}{z_{\overline2\overline3}z_{3 \overline3}z_{3i}}+\frac{ \tilde{L}_{\overline233\overline3\overline3i}V_j V_k}{z_{\overline23}z_{3 \overline3}z_{\overline3i}}+\frac{ \tilde{L}_{\overline2 i3\overline33\overline3}V_j V_k}{z_{\overline2 i}z_{3 \overline3}^2}+\frac{ \tilde{L}_{\overline2 \overline3\overline2 \overline33i}V_j V_k}{z_{\overline2 \overline3}^2 z_{3i}}+\frac{ \tilde{L}_{\overline2 3\overline2 3\overline3i}V_j V_k}{z_{\overline2 3}^2z_{\overline3i}}\biggl)\nonumber\\
		&&+\frac{ \tilde{L}_{\overline2 33\overline33\overline3}V_{\overline1}V_1V_2}{z_{\overline2 3}z_{3\overline3}^2}+\frac{ \tilde{L}_{\overline2 \overline33\overline33\overline3}V_{\overline1}V_1V_2}{z_{\overline2 \overline3}z_{3\overline3}^2}+\frac{ \tilde{L}_{\overline2 \overline3\overline2 \overline33\overline3}V_{\overline1}V_1V_2}{z_{\overline2 \overline3}^2z_{3\overline3}}+\frac{ \tilde{L}_{\overline2 3\overline2 33\overline3}V_{\overline1}V_1V_2}{z_{\overline2 3}^2z_{3\overline3}}\biggl\rangle\ ,\label{eq8.0}
	\end{IEEEeqnarray}
	where we have chosen the convention $\epsilon_{1\overline12}=1$ and $KN(y,\xi,\eta)$ is the Koba-Nielsen factor that takes the following form after contracting the plane wave factors
	\begin{IEEEeqnarray}{rCl}
		\KN(y,\xi,\eta)&=&\langle e^{ik_1{\cdot}X(1)}  e^{i D \cdot k_1{\cdot} X(-1)} e^{ik_2{\cdot}X(y)} e^{i D \cdot k_2{ \cdot}X(-y)} e^{ik_3{\cdot}X(\xi)} e^{i D \cdot k_3{\cdot}X(\eta)} \rangle\nonumber\\
		&=&2^{s_{1\overline1}} |1-y|^{s_{12}} |1+y|^{s_{1\overline2}} |1-\xi|^{s_{13}} |1-\eta|^{ s_{1\overline3}}|1+y|^{s_{1\overline2}}\nonumber                \\
		&   & \times|1-y|^{s_{12}}|1+\xi|^{ s_{1\overline3}}|1+\eta|^{s_{13}}|2y|^{ s_{2\overline2}}|y-\xi|^{s_{23}}\nonumber \\
		&   &\times|y-\eta|^{ s_{2\overline3}} |y+\xi|^{ s_{2\overline3}}|y+\eta|^{s_{23}} |\xi-\eta|^{ s_{3\overline3}}\ ,\label{eq::KN}
	\end{IEEEeqnarray}
	where $s_{ij}=k_i{\cdot}k_j$ and $s_{i\overline\jmath}=k_i{\cdot}D{\cdot}k_j$.\par
	In string theory we can benefit from the independence of the CFT correlator of the order in which we integrate out the conformal dimension one fields using the OPE method described at the beginning of section \ref{sec::blockL}. We have chosen an explicit order of contracting the vertex operators meaning that we start with $\overline2$, then $3$ and in the end we contract the vertex operator $\overline 3$. The result of the OPE contractions should not depend on this particular order.\footnote{The end result is independent of the order of contraction. Nevertheless, during the computation one has to choose an order and stick to it.} Instead we could also start with $3$ or $\overline 3$. Hence, we find identities between the kinematic factors by comparing different orders in which the conformal weight one fields are integrated out by demanding that in the end they should give the same result \cite{5pt}. For a different order of contraction than the one in \eqref{eq8.0} the kinematic factors $\tilde L$ and their worldsheet dependent numerators $\frac{1}{z_{ij}z_{mn}z_{rs}}$ can be obtained just by relabelling them according to the new order in which the vertex operators are contracted.\par
	For the building blocks the relabelling is straightforward, but if we look at the $z_{ij}$ dependence we recognize that the relabelling introduces different poles in the integrand, which are not present in the original expression \eqref{eq8.0}. This makes comparing different orders of contraction to find relations between kinematic building blocks non-straightforward. However, we can use partial fractioning
	\begin{IEEEeqnarray}l
		\frac{1}{z_{ji}z_{ki}}+\frac{1}{z_{jk}z_{ji}}=\frac{1}{z_{jk}z_{ki}}
	\end{IEEEeqnarray}
	so that subtracting amplitudes with different ordering of contraction gives rise to relations of the form \cite{5pt,6pt}
	\begin{IEEEeqnarray}{rCl}
		\tilde{L}_{\overline233i\overline3i}&=&\tilde{L}_{3i\overline2i\overline3i}-\tilde{L}_{\overline2i3i\overline3i}\ ,\nonumber\\
		\tilde{L}_{\overline2i3\overline3\overline3i}&=&\tilde{L}_{\overline2i\overline3i3i}-\tilde{L}_{\overline2i3i\overline3i}\ ,\nonumber\\
		\tilde{L}_{\overline233\overline3\overline3i}&=&\tilde{L}_{\overline3i3i\overline2i}-\tilde{L}_{\overline3i\overline2i3i}+\tilde{L}_{\overline2i3i\overline3i}-\tilde{L}_{3i\overline2i\overline3i}\ ,\label{eq::ident_order}\\
		\tilde{L}_{\overline233i\overline3j}&=&\tilde{L}_{3i\overline2i\overline3j}-\tilde{L}_{\overline2i3i\overline3j}\ ,\nonumber\\
		\tilde{L}_{\overline2j3\overline3\overline3i}&=&\tilde{L}_{\overline2j\overline3i3i}-\tilde{L}_{\overline2j3i\overline3i}\ .\nonumber
	\end{IEEEeqnarray}
	Already applying this method here will reduce the amount of superfield manipulations later, because we can reduce the number of different building blocks we have to consider. For example, the OPEs leading to $\tilde{L}_{\overline233\overline3\overline3i}$ are rather tedious to compute, because we have to contract unintegrated vertex operators $U_k(z_k)$ with each other, whereas the building blocks $\tilde{L}_{\overline3i3i\overline2i}, \tilde{L}_{\overline3i\overline2i3i}, \tilde{L}_{\overline2i3i\overline3i}$ and $\tilde{L}_{3i\overline2i\overline3i}$ are simpler to calculate, due to the fact that we are con\-si\-de\-ring contractions between $U_k(z_k)$ and $V_i(z_i)$ only.\par
	We can then simplify the expression in \eqref{eq8.0} by using the relations \eqref{eq::ident_order} between the building blocks and relabelling of those relations such that we can write the correlator as
	\begin{IEEEeqnarray}{rCl}
		\IEEEeqnarraymulticol{3}{l}{\left\langle V_1(1)V_{\overline1}(-1)V_2(y)U_{\overline2}(-y)U_3(\xi)U_{\overline3}(\eta)\right\rangle=}\nonumber\\
		&=&\KN(z_2,z_3,z_{\overline{3}})\biggl\langle\smashoperator{\sum_{\substack{i,j,k\\\in\{1,\overline1,2\}}}}\epsilon_{ijk}\biggl(\frac{  \tilde{L}_{\overline2i3j\overline3k} }{z_{\overline2i}z_{3j}z_{\overline3k}}+\frac{ \tilde{L}_{3i\overline2i\overline3j}V_k}{z_{\overline23}z_{3i}z_{\overline3j}}+\frac{ \tilde{L}_{\overline2j\overline3i3i}V_k}{z_{\overline2j}z_{3\overline3}z_{\overline3i}}+\frac{ \tilde{L}_{\overline3i3j\overline2i}V_k}{z_{\overline23}z_{3\overline3}z_{\overline3i}}-\frac{ \tilde{L}_{\overline2i3i\overline3j}V_k}{z_{\overline23}z_{\overline2i}z_{\overline3j}}\nonumber\\
		&&-\frac{ \tilde{L}_{\overline2i3j\overline3i}V_k}{z_{\overline2\overline3}z_{\overline2i}z_{3j}}-\frac{ \tilde{L}_{\overline2j3i\overline3i}V_k}{z_{3\overline3}z_{3i}z_{\overline2j}}+\frac{ \tilde{L}_{\overline2i3i\overline3i}V_jV_k}{z_{\overline23}z_{3\overline3}z_{\overline2i}}+\frac{ \tilde{L}_{3i\overline3i\overline2i}V_jV_k}{z_{\overline2\overline3}z_{\overline33}z_{3i}}+\frac{ \tilde{L}_{\overline3i3i\overline2i}V_jV_k}{z_{\overline23}z_{3\overline3}z_{\overline3i}}-\frac{ \tilde{L}_{3i\overline2i\overline3i}V_jV_k}{z_{\overline23}z_{\overline2\overline3}z_{3i}}\nonumber\\
		&&-\frac{ \tilde{L}_{\overline2i\overline3i3i}V_jV_k}{z_{\overline2\overline3}z_{3\overline3}z_{\overline2i}}-\frac{ \tilde{L}_{\overline3i\overline2i3i}V_jV_k}{z_{\overline23}z_{\overline2\overline3}z_{\overline3i}}+\frac{ \tilde{L}_{\overline2 i3\overline33\overline3}V_jV_k}{z_{\overline2 i}z_{3 \overline3}^2}+\frac{ \tilde{L}_{\overline2 \overline3\overline2 \overline33i}V_jV_k}{z_{\overline2 \overline3}^2 z_{3i}}+\frac{ \tilde{L}_{\overline2 3\overline2 3\overline3i}V_jV_k}{z_{\overline2 3}^2z_{\overline3i}}\biggl)\nonumber\\
		&&+\frac{ \tilde{L}_{\overline2 33\overline33\overline3}V_{\overline1}V_1V_2}{z_{\overline2 3}z_{3\overline3}^2}+\frac{ \tilde{L}_{\overline2 \overline33\overline33\overline3}V_{\overline1}V_1V_2}{z_{\overline2 \overline3}z_{3\overline3}^2}+\frac{ \tilde{L}_{\overline2 \overline3\overline2 \overline33\overline3}V_{\overline1}V_1V_2}{z_{\overline2 \overline3}^2z_{3\overline3}}+\frac{ \tilde{L}_{\overline2 3\overline2 33\overline3}V_{\overline1}V_1V_2}{z_{\overline2 3}^2z_{3\overline3}}\biggl\rangle\ .\label{eq::correlator}
	\end{IEEEeqnarray}
	In section \ref{sec::integration} we have already $PSL(2,\mathbb{R})$ transformed the integration regions of the seven subamplitudes. Next, we want to discuss the transformation of the correlator in more detail.

	\subsubsection*{\boldmath{$PSL(2,\mathbb R)$} transformation of the correlator}
	In order to make contact to the literature on open string correlators we map the fixed vertex operator positions from $(-1,y,1)$ to $(0,1,\infty)$ and check that the kinematic factors containing the vertex operator, whose position is mapped to infinity, is not present in a composite superfield any more after the transformation, which is not obvious: A priori the composite superfields for three closed strings contain contractions with all vertex operators. By gauge fixing one can put one vertex operator position to infinity and therefore this vertex operator has only vanishing contractions as already stated above. For our choice of gauge fixing the vertex operator $V_1(1)$ will be mapped to infinity by the $PSL(2,\mathbb R)$ transformation in \eqref{eq::q5} so that we would expect all building blocks containing a contraction with $V_1(1)$ to drop out.\par
	Moreover, by just using the transformation in \eqref{eq::q5} we would not get the correct $z_{ij}$ dependencies; we would obtain terms containing for example square roots of vertex operator positions, which are not present in an open string six-point amplitude. By conformal invariance only differences between vertex operator positions can appear after contracting conformal primaries \cite{BLT}. Hence, we have to ensure  that the correlator \eqref{eq::correlator} is invariant under global conformal transformations. As discussed in appendix \ref{sec::psl2r} this correlator has to satisfy the conditions \eqref{eq::04}--\eqref{eq::05}, which in our case become:\footnote{The Koba-Nielsen factor KN satisfies these equations trivially using momentum conservation, see the example at the end of appendix \ref{sec::psl2r}, now with $p_i^2=0$. Moreover, these relations are linear in the derivatives so we only need to consider the vertex operators without a plane wave factor in \eqref{eq::con_vo_1}--\eqref{eq::con_vo_2}.}
	\begin{IEEEeqnarray}{rCl}
		0&=&\sum_{i=1}^3\left(\partial_i+\overline\partial_i\right)\langle V_1(z_1)V_{\overline1}(\overline z_1)V_2(z_2)U_{\overline2}(\overline z_2)U_3(z_3)U_{\overline3}(\overline z_3)\rangle\ ,\label{eq::con_vo_1}\\
		0&=&\sum_{i=1}^3\left(h_i+z_i\partial_i+\overline h_i+\overline z_i\overline \partial_i\right)\langle V_1(z_1)V_{\overline1}(\overline z_1)V_2(z_2)U_{\overline2}(\overline z_2)U_3(z_3)U_{\overline3}(\overline z_3)\rangle\ ,\\
		0&=&\sum_{i=1}^3\left(2h_iz_i+z_i^2\partial_i+2\overline h_i\overline z_i+\overline z_i^2\overline \partial_i\right)\langle V_1(z_1)V_{\overline1}(\overline z_1)V_2(z_2)U_{\overline2}(\overline z_2)U_3(z_3)U_{\overline3}(\overline z_3)\rangle\ .\label{eq::con_vo_2}
	\end{IEEEeqnarray}
	The integrated vertex operators $U_i$ have conformal weight $h_i=1$ and the unintegrated vertex operators $V_i$ have conformal weight $h_i=0$, which also holds for the antiholomorphic part. We can then evaluate the conditions \eqref{eq::con_vo_1}--\eqref{eq::con_vo_2}: Using momentum conservation one can show that the correlator of $\mathcal{A}$ satisfies the conditions in the first two equations, but the third condition \eqref{eq::con_vo_2} is a priori non vanishing
	\begin{IEEEeqnarray}{rCl}
		0&=&\sum_{i=1}^3\left(2h_iz_i+z_i^2\partial_i+2\overline h_i\overline z_i+\overline z_i^2\overline \partial_i\right)\langle V_1(z_1)V_{\overline1}(\overline z_1)V_2(z_2)U_{\overline2}(\overline z_2)U_3(z_4)U_{\overline3}(\overline z_3)\rangle=\nonumber\\
		&=&\frac{1}{z_{\overline1\overline2}z_{\overline23}}\langle \tilde L_{\overline233\overline1\overline3\overline1}V_{2}V_{1}+ V_{\overline1}\tilde L_{\overline233\overline1}\tilde L_{\overline32}V_{1}+ V_{\overline1}V_2\tilde L_{\overline233\overline1}\tilde L_{\overline31}\rangle+\frac{1}{z_{\overline1\overline2}z_{\overline13}}\langle \tilde L_{\overline2\overline13\overline1\overline3\overline1}V_{2}V_{1}+\tilde L_{\overline2\overline13\overline1}\tilde L_{\overline3\overline2}V_{1}\nonumber\\
		&&+\tilde L_{\overline2\overline13\overline1}V_{2}\tilde L_{\overline31}+\tilde L_{\overline233\overline1\overline3\overline1}V_{2}V_{1}+ V_{\overline1}\tilde L_{\overline233\overline1}\tilde L_{\overline32}V_{1}+ V_{\overline1}V_2\tilde L_{\overline233\overline1}\tilde L_{\overline31}\rangle+\ldots\ .\label{eq::third}
	\end{IEEEeqnarray} 
	In general, the fractions $\frac{1}{z_{ij}z_{mn}}$ don't vanish for arbitrary $z_i$ (and $z_{\overline\imath}$). Moreover, the individual fractions $\frac{1}{z_{ij}z_{mn}}$ appearing in \eqref{eq::third} are independent and hence equation \eqref{eq::third} gives rise to relations between the building blocks $\tilde{L}$. By equating coefficients these are given by 
	\begin{IEEEeqnarray}{l}
		\begin{IEEEeqnarraybox}[][c]{rCl}
			\IEEEstrut
			0&=&\langle \tilde L_{\overline2\overline13\overline1\overline3\overline1} V_{2} V_{1} + \tilde L_{\overline2\overline13\overline3\overline3\overline1} V_{2} V_{1} + \tilde L_{\overline233\overline1\overline3\overline1} V_{2} V_{1} + \tilde L_{\overline233\overline3\overline3\overline1} V_{2} V_{1} + \tilde L_{\overline2\overline33\overline1\overline3\overline1} V_{2} V_{1}\\&& + \tilde L_{3\overline1\overline3\overline1} \tilde L_{\overline22} V_{1} + \tilde L_{3\overline3\overline3\overline1} \tilde L_{\overline22} V_{1} + \tilde L_{3\overline1\overline3\overline1} V_{2} \tilde L_{\overline21} + \tilde L_{3\overline3\overline3\overline1} V_{2} \tilde L_{\overline21} \rangle\ ,\\
			0&=&\langle \tilde L_{\overline2\overline13\overline1\overline3\overline1} V_{2} V_{1} + \tilde L_{\overline2\overline13\overline3\overline3\overline1} V_{2} V_{1} + \tilde L_{\overline2\overline1\overline3\overline1} \tilde L_{32} V_{1} + \tilde L_{\overline2\overline1\overline3\overline1} V_{2} \tilde L_{31} + \tilde L_{\overline2\overline33\overline1\overline3\overline1} V_{2} V_{1} \\&&+ \tilde L_{\overline2\overline3\overline3\overline1} \tilde L_{32} V_{1} + \tilde L_{\overline2\overline3\overline3\overline1} V_{2} \tilde L_{31} \rangle\ ,\\
			0&=&\langle \tilde L_{\overline233\overline3\overline3\overline1} V_{2} V_{1} + V_{\overline1} \tilde L_{\overline233\overline3\overline32} V_{1} + V_{\overline1} V_{2} \tilde L_{\overline233\overline3\overline31} - \tilde L_{\overline2\overline3\overline333\overline1} V_{2} V_{1} - V_{\overline1} \tilde L_{\overline2\overline3\overline3332} V_{1} - V_{\overline1} V_{2} \tilde L_{\overline2\overline3\overline3331} \rangle\ ,\\
			0&=&\langle \tilde L_{\overline2\overline13\overline1\overline3\overline1} V_{2} V_{1} + \tilde L_{\overline2\overline13\overline1} \tilde L_{\overline32} V_{1} + \tilde L_{\overline2\overline13\overline1} V_{2} \tilde L_{\overline31} + \tilde L_{\overline233\overline1\overline3\overline1} V_{2} V_{1} + \tilde L_{\overline233\overline1} \tilde L_{\overline32} V_{1} + \tilde L_{\overline233\overline1} V_{2} \tilde L_{\overline31} \rangle\ ,\\
			0&=&\langle \tilde L_{\overline2\overline13\overline1} \tilde L_{\overline32} V_{1} + \tilde L_{\overline233\overline1} \tilde L_{\overline32} V_{1} + \tilde L_{\overline2\overline33\overline1} \tilde L_{\overline32} V_{1} + \tilde L_{3\overline1} \tilde L_{\overline22\overline32} V_{1} + \tilde L_{3\overline1} \tilde L_{\overline32} \tilde L_{\overline21} \rangle\ ,\\
			0&=&\langle \tilde L_{\overline2\overline1} \tilde L_{3\overline33\overline3} V_{2} V_{1} + \tilde L_{\overline233\overline33\overline3} V_{\overline1} V_{2} V_{1} + \tilde L_{\overline2\overline33\overline33\overline3} V_{\overline1} V_{2} V_{1} + V_{\overline1} \tilde L_{\overline22} \tilde L_{3\overline33\overline3} V_{1} + V_{\overline1} V_{2} \tilde L_{\overline21} \tilde L_{3\overline33\overline3} \rangle\ ,\\
			0&=&\langle \tilde L_{\overline2\overline13\overline3\overline3\overline1} V_{2} V_{1} + \tilde L_{\overline233\overline3\overline3\overline1} V_{2} V_{1} - \tilde L_{\overline2\overline3\overline333\overline1} V_{2} V_{1} + \tilde L_{3\overline3\overline3\overline1} \tilde L_{\overline22} V_{1} + \tilde L_{3\overline3\overline3\overline1} V_{2} \tilde L_{\overline21} \rangle\ ,\\
			0&=&\langle \tilde L_{\overline2\overline1\overline3\overline1} \tilde L_{32} V_{1} + \tilde L_{\overline23\overline3\overline1} \tilde L_{32} V_{1} + \tilde L_{\overline2\overline3\overline3\overline1} \tilde L_{32} V_{1} + \tilde L_{\overline3\overline1} \tilde L_{\overline2232} V_{1} + \tilde L_{\overline3\overline1} \tilde L_{32} \tilde L_{\overline21} \rangle\ ,\\
			0&=&\langle \tilde L_{\overline2\overline13\overline1} \tilde L_{\overline32} V_{1} + \tilde L_{\overline2\overline1} \tilde L_{3\overline3\overline32} V_{1} + \tilde L_{\overline2\overline1} \tilde L_{32\overline32} V_{1} + \tilde L_{\overline2\overline1} \tilde L_{\overline32} \tilde L_{31} \rangle\ ,\\
			0&=&\langle \tilde L_{\overline23\overline233\overline3} V_{\overline1} V_{2} V_{1} - \tilde L_{\overline23\overline23} \tilde L_{\overline3\overline1} V_{2} V_{1} - \tilde L_{\overline23\overline23} V_{\overline1} \tilde L_{\overline32} V_{1} - \tilde L_{\overline23\overline23} V_{\overline1} V_{2} \tilde L_{\overline31} \rangle\ ,\\
			0&=&\langle \tilde L_{\overline233\overline1\overline3\overline1} V_{2} V_{1} + \tilde L_{\overline233\overline3\overline3\overline1} V_{2} V_{1} + \tilde L_{\overline3\overline1} \tilde L_{\overline2332} V_{1} + \tilde L_{\overline3\overline1} V_{2} \tilde L_{\overline2331} \rangle\ ,\\
			0&=&\langle \tilde L_{\overline2\overline3\overline2\overline3} \tilde L_{3\overline1} V_{2} V_{1} + \tilde L_{\overline2\overline3\overline2\overline33\overline3} V_{\overline1} V_{2} V_{1} + \tilde L_{\overline2\overline3\overline2\overline3} V_{\overline1} \tilde L_{32} V_{1} + \tilde L_{\overline2\overline3\overline2\overline3} V_{\overline1} V_{2} \tilde L_{31} \rangle\ ,\\
			0&=&\langle \tilde L_{\overline2\overline33\overline1\overline3\overline1} V_{2} V_{1} + \tilde L_{\overline2\overline3\overline32} \tilde L_{\overline32} V_{1} + \tilde L_{3\overline1} V_{2} \tilde L_{\overline2\overline3\overline31} + \tilde L_{\overline2\overline3\overline333\overline1} V_{2} V_{1} \rangle\\
			\IEEEstrut
		\end{IEEEeqnarraybox}\label{eq::c}
	\end{IEEEeqnarray}
	and permutations thereof. Eventually, we have all we need to perform the transformation from one vertex operator position fixing to another: Using the relations \eqref{eq::c} and their permutations together with the $PSL(2,\mathbb R)$ transformation \eqref{eq::q5} we can map the correlator in \eqref{eq::q3} to $\langle V_{\overline1}(0)V_{2}(1)V_1(\infty)U_{\overline2}(x)U_3(\tilde\xi)U_{\overline3}(\tilde\eta)\rangle$. Explicitly, after the transformation the correlator can be written as
	\begin{IEEEeqnarray}{rCl}
		\IEEEeqnarraymulticol{3}{l}{\left\langle V_{\overline1}(0)V_{2}(1)V_1(\infty)U_{\overline2}(x)U_3(\tilde\xi)U_{\overline3}(\tilde\eta)\right\rangle=}\nonumber\\
		&=&\det(\mathcal J)^{-1}\KN(x,\tilde\xi,\tilde\eta)\biggl\langle\sum_{i,j\in\{\overline1,2\}}\biggl(\frac{  \tilde{L}_{\overline2i3i\overline3j} V_{1}}{z_{\overline2i}z_{3i}z_{\overline3j}}+\frac{  \tilde{L}_{\overline2i3j\overline3i} V_{1}}{z_{\overline2i}z_{3j}z_{\overline3i}}+\frac{  \tilde{L}_{\overline2j3i\overline3i}V_{1}}{z_{\overline2j}z_{3i}z_{\overline3i}}+\frac{  \tilde{L}_{\overline2i3\overline3\overline3j}V_{1}}{z_{\overline2i}z_{3\overline3}z_{\overline3j}}\nonumber\\
		&&+\frac{  \tilde{L}_{\overline2\overline33i\overline3j}V_{1}}{z_{\overline2\overline3}z_{3i}z_{\overline3j}}+\frac{ \tilde{L}_{\overline233i\overline3j}V_{1}}{z_{\overline23}z_{3i}z_{\overline3j}}+\frac{ \tilde{L}_{\overline2i3i\overline3i}V_jV_{1}}{z_{\overline2i}z_{3i}z_{\overline3i}}+\frac{  \tilde{L}_{\overline2i3\overline3\overline3i}V_jV_{1}}{z_{\overline2i}z_{3\overline3}z_{\overline3i}}+\frac{  \tilde{L}_{\overline2\overline33i\overline3i}V_jV_{1}}{z_{\overline2\overline3}z_{3i}z_{\overline3i}}+\frac{  \tilde{L}_{\overline233i\overline3i}V_jV_{1}}{z_{\overline23}z_{3i}z_{\overline3i}}\nonumber\\
		&&+\frac{ \tilde{L}_{\overline2\overline33\overline33i}V_jV_{1}}{z_{\overline2\overline3}z_{3 \overline3}z_{3i}}+\frac{ \tilde{L}_{\overline233\overline3\overline3i}V_jV_{1}}{z_{\overline23}z_{3 \overline3}z_{\overline3i}}+\frac{ \tilde{L}_{\overline2 i3\overline33\overline3}V_jV_{1}}{z_{\overline2 i}z_{3 \overline3}^2}+\frac{ \tilde{L}_{\overline2 \overline3\overline2 \overline33i}V_jV_{1}}{z_{\overline2 \overline3}^2 z_{3i}}+\frac{ \tilde{L}_{\overline2 3\overline2 3\overline3i}V_jV_{1}}{z_{\overline2 3}^2z_{\overline3i}}\biggl)\nonumber\\
		&&+\frac{ \tilde{L}_{\overline2 33\overline33\overline3}V_{\overline1}V_{2}V_{1}}{z_{\overline2 3}z_{3\overline3}^2}+\frac{ \tilde{L}_{\overline2 \overline33\overline33\overline3}V_{\overline1}V_{2}V_{1}}{z_{\overline2 \overline3}z_{3\overline3}^2}+\frac{ \tilde{L}_{\overline2 \overline3\overline2 \overline33\overline3}V_{\overline1}V_{2}V_{1}}{z_{\overline2 \overline3}^2z_{3\overline3}}+\frac{ \tilde{L}_{\overline2 3\overline2 33\overline3}V_{\overline1}V_{2}V_{1}}{z_{\overline2 3}^2z_{3\overline3}}\biggl\rangle\ ,\label{eq::101}
	\end{IEEEeqnarray}
	where the vertex operator positions are given by $z_{\overline1}=0, z_1=\infty,z_{\overline2}=x,z_{2}=1,z_{\overline3}=\tilde\eta,z_3=\tilde\xi$ and the determinant of the Jacobi matrix $\mathcal J$ is given by
	\begin{IEEEeqnarray}l
		\det(\mathcal J)=-\frac{4\sqrt{x}}{(1+\sqrt{x})^2(\sqrt{x}+\xi)^2(\sqrt{x}+\eta)^2}\ .
	\end{IEEEeqnarray}
	Comparing this result with the correlator of six open strings on the disk in equation $(3.2)$ in \cite{6pt}, both of them are in agreement with each other after the identification \eqref{eq::identification}
	and up to an overall factor, that will cancel against the Jacobian of the $PSL(2,\mathbb R)$ transformation, when we take also the measure of the worldsheet integrals into account -- i.e$.$ when we consider the complete three point amplitude in the next section.
	
	\subsection{The three–point amplitude with BRST building blocks}
	
	After we have explicitly shown that the scattering of three closed strings on the disk behaves like the scattering of six open strings on the disk, we can exploit the methods presented in \cite{6pt, npt_1, npt_2}, which provide further simplification for this amplitude. Namely, we want to replace the superfield expressions $\tilde L_{ji}, \tilde L_{jiki}$ and $\tilde L_{jikili}$ by their corresponding BRST building blocks $T_{ij},T_{ijk}$ and $T_{ijkl}$.\par
	The correlator in \eqref{eq::101} still contains BRST exact terms, which were dropped in \cite{6pt} from the beginning, because when integrating over the correlator they cancel out. We have not dropped them so far, because otherwise the set of relations between the kinematic terms in \eqref{eq::c} would not hold, even though the BRST exact terms do not contribute to the end result of the amplitude.\par
	Since we do not need those terms any more we can finally get rid of them. After an easy but tedious calculation it can be seen that all the BRST exact terms cancel by using partial fractioning and integration by parts. This then results in the same correlator\footnote{We have also cancelled the prefactor in \eqref{eq::101} against the Jacobian, which we obtained by the transformation of the measure.} as in \eqref{eq::101} but with the substitution $\tilde L\rightarrow L$
	\begin{IEEEeqnarray}{rCl}
		\mathcal{A}&=&i T_p g^3_c\sum_{n=1}^2\Pi_n\int_{\mathcal I_n}\mathrm dx\,\mathrm d\tilde\xi\,\mathrm d\tilde\eta\,\left\langle V_{\overline1}(0)V_{2}(1)V_1(\infty)U_{\overline2}(x)U_3(\tilde\xi)U_{\overline3}(\tilde\eta)\right\rangle\nonumber\\
		&=&i T_p g^3_c\sum_{n=1}^2\Pi_n\int_{\mathcal I_n}\mathrm dx\,\mathrm d\tilde\xi\,\mathrm d\tilde\eta\,\KN(x,\tilde\xi,\tilde\eta)\biggl\langle\sum_{\substack{i,j\in\{\overline1,2\}\\i\neq j}}\epsilon_{ij}\biggl(\frac{  L_{\overline2i3i\overline3j} V_{1}}{z_{\overline2i}z_{3i}z_{\overline3j}}+\frac{  L_{\overline2i3j\overline3i} V_{1}}{z_{\overline2i}z_{3j}z_{\overline3i}}+\frac{  L_{\overline2j3i\overline3i}V_{1}}{z_{\overline2j}z_{3i}z_{\overline3i}}\nonumber\\
		&&+\frac{  L_{\overline2i3\overline3\overline3j}V_{1}}{z_{\overline2i}z_{3\overline3}z_{\overline3j}}+\frac{  L_{\overline2\overline33i\overline3j}V_{1}}{z_{\overline2\overline3}z_{3i}z_{\overline3j}}+\frac{ L_{\overline233i\overline3j}V_{1}}{z_{\overline23}z_{3i}z_{\overline3j}}+\frac{ L_{\overline2i3i\overline3i}V_jV_{1}}{z_{\overline2i}z_{3i}z_{\overline3i}}+\frac{  L_{\overline2i3\overline3\overline3i}V_jV_{1}}{z_{\overline2i}z_{3\overline3}z_{\overline3i}}+\frac{  L_{\overline2\overline33i\overline3i}V_jV_{1}}{z_{\overline2\overline3}z_{3i}z_{\overline3i}}\nonumber\\
		&&+\frac{  L_{\overline233i\overline3i}V_jV_{1}}{z_{\overline23}z_{3i}z_{\overline3i}}+\frac{ L_{\overline2\overline33\overline33i}V_jV_{1}}{z_{\overline2\overline3}z_{3 \overline3}z_{3i}}+\frac{ L_{\overline233\overline3\overline3i}V_jV_{1}}{z_{\overline23}z_{3 \overline3}z_{\overline3i}}+\frac{ L_{\overline2 i3\overline33\overline3}V_jV_{1}}{z_{\overline2 i}z_{3 \overline3}^2}+\frac{ L_{\overline2 \overline3\overline2 \overline33i}V_jV_{1}}{z_{\overline2 \overline3}^2 z_{3i}}+\frac{ L_{\overline2 3\overline2 3\overline3i}V_jV_{1}}{z_{\overline2 3}^2z_{\overline3i}}\biggl)\nonumber\\
		&&+\frac{ L_{\overline2 33\overline33\overline3}V_{\overline1}V_{2}V_{1}}{z_{\overline2 3}z_{3\overline3}^2}+\frac{ L_{\overline2 \overline33\overline33\overline3}V_{\overline1}V_{2}V_{1}}{z_{\overline2 \overline3}z_{3\overline3}^2}+\frac{ L_{\overline2 \overline3\overline2 \overline33\overline3}V_{\overline1}V_{2}V_{1}}{z_{\overline2 \overline3}^2z_{3\overline3}}+\frac{ L_{\overline2 3\overline2 33\overline3}V_{\overline1}V_{2}V_{1}}{z_{\overline2 3}^2z_{3\overline3}}\biggl\rangle\ ,\label{eq::L}
	\end{IEEEeqnarray}
	which means that all BRST exact terms are now gone. Above we have introduced the phase 
\begin{IEEEeqnarray}l
\Pi_1=2i\sin(s_{23}),\qquad\Pi_2=2i\sin(s_{23}+s_{2\overline3}) \label{eq::phasesPi}
\end{IEEEeqnarray}	
	of the subamplitudes in \eqref{eq::amp_mon_4a}. From here on we will look at the complete amplitude \eqref{eq::amp_mon_4a} and not only the correlator. This means that we include the integration over the worldsheet coordinates of the vertex operators and in addition all the prefactors. As we stated already at the end of appendix \ref{sec::correlator1} the amplitude in \eqref{eq::L} is identical to (3.2) in \cite{6pt} when using \eqref{eq::identification}. Hence, after performing the same computation as in \cite{6pt} and \cite{npt_1} we obtain\footnote{The analogous expression for the six-point function in \cite{npt_1} is given in (5.19) therein.} 
	\begin{IEEEeqnarray}{rCl}
		\mathcal{A}&=&-ig_c^3T_p\sum_{n=1}^2\Pi_n\int_{\mathcal I_n}\mathrm dx\,\mathrm d\tilde\xi\,\mathrm d\tilde\eta\,\KN(x,\tilde\xi,\tilde\eta)\biggl\{\frac{s_{\overline{1}\overline2}}{z_{\overline{1}\overline2}}\left(\frac{s_{\overline{1}3}}{z_{\overline{1}3}}+\frac{s_{\overline23}}{z_{\overline23}}\right)\frac{s_{\overline{3}2}}{z_{\overline{3}2}}\langle M_{\overline{1}\overline23\overline{3}}V_{2}V_1\nonumber\\
		&&+M_{\overline{1}\overline23}M_{\overline{3}2}V_1+M_{\overline{1}\overline2}M_{3\overline{3}2}V_1+V_{\overline{1}}M_{\overline23\overline{3}2}V_1\rangle+\mathcal{P}(\overline2,3,\overline{3})\biggl\}\ ,\label{eq::A1}
	\end{IEEEeqnarray}
	where $\mathcal{P}(\overline{2},3,\overline{3})$ are all permutations of $S_3$ and we have introduced the Berends-Giele currents
	\begin{IEEEeqnarray}{rCl}
		M_{\overline{2}\overline{1}}&=&-\frac{T_{\overline{2}\overline{1}}}{s_{\overline{2}\overline{1}}}\label{eq::M1}\\
		M_{\overline{1}\overline{2}3}&=&\frac1{s_{\overline{1}\overline{2}3}}\left(\frac{T_{\overline{1}\overline{2}3}}{s_{\overline{1}\overline{2}}}+\frac{T_{\overline{1}\overline{2}3}-T_{\overline{1}3\overline{2}}}{s_{\overline{2}3}}\right), \label{eq::M2}\\
		M_{\overline{1}\overline{2}3\overline{3}}&=&-\frac{1}{s_{\overline{1}\overline{2}3\overline{3}}}\left(\frac{T_{\overline{1}\overline{2}3\overline{3}}}{s_{\overline{1}\overline{2}}s_{\overline{1}\overline{2}3}}+\frac{T_{\overline{1}\overline{2}3\overline{3}}-T_{\overline{1}3\overline{2}\overline{3}}}{s_{\overline{2}3}s_{\overline{1}\overline{2}3}}+\frac{T_{\overline{1}\overline{2}3\overline{3}}-T_{\overline{1}\overline{2}\overline{3}3}+T_{\overline{1}\overline{3}3\overline{2}}-T_{\overline{1}3\overline{3}\overline{2}}}{s_{3\overline{3}}s_{\overline{2}3\overline{3}}}\right.\nonumber\\
		&&\left.+\frac{T_{\overline{1}\overline{3}3\overline{2}}-T_{\overline{1}\overline{3}\overline{2}3}+T_{\overline{1}\overline{2}3\overline{3}}-T_{\overline{1}3\overline{2}\overline{3}}}{s_{\overline{2}3}s_{\overline{2}3\overline{3}}}+\frac{T_{\overline{1}\overline{2}3\overline{3}}-T_{\overline{1}\overline{2}\overline{3}3}}{s_{\overline{1}\overline{2}} s_{3\overline{3}}}\right)\ . \label{eq::M3}
	\end{IEEEeqnarray}
	Under the identification of closed and open strings in \eqref{eq::identification}, i.e.\ the identification of their momenta 
\eqref{eq::ident_mom} and polarizations \eqref{eq::ident_pol}, the above combination of Berends-Giele currents is equivalent to a Yang-Mills amplitude.\footnote{For the open string case see for example \cite{Mafra:2022wml} for more details.} We can use that $z_{\overline2}=x,z_3=\tilde\xi$ and $z_{\overline3}=\tilde\eta$ to write \eqref{eq::A1} as
	\begin{IEEEeqnarray}{rCl}
		\mathcal{A}&=&-ig_c^3T_p\sum_{n=1}^2\sum_{\sigma\in S_3}\Pi_n\int_{\mathcal I_n}\mathrm dz_{\overline2}\,\mathrm dz_3\,\mathrm dz_{\overline3}\,\KN(z_{\overline2},z_3,z_{\overline3})\nonumber\\
		&&\times\frac{s_{\overline1\overline2_\sigma}}{z_{\overline1\overline2_\sigma}}\frac{s_{\overline3_\sigma2}}{z_{\overline3_\sigma2}}\left(\frac{s_{\overline13_\sigma}}{z_{\overline13_\sigma}}+\frac{s_{\overline2_\sigma3_\sigma}}{z_{\overline2_\sigma3_\sigma}}\right) A_{\rm YM}(\overline1,\overline2_\sigma,3_\sigma,\overline3_\sigma,2,1)\ ,\label{eq::A2}
	\end{IEEEeqnarray}
	where $\sigma\in S_3$ describes the permutations of the labels $(\overline2,3,\overline3)$. We can also find the six hypergeometric basis integrals in \eqref{eq::A2}, which in this case are given by
	\begin{IEEEeqnarray}{rCl}
		F^{(\overline2_\sigma3_\sigma\overline3_\sigma)}_{\mathcal I_n}&=&-\int_{\mathcal{I}_n}\mathrm{d}z_{\overline2}\,\mathrm{d}z_3\,\mathrm{d}z_{\overline3}\,\Biggl(\prod_{i<j}|z_{ij}|^{s_{ij}}\Biggl)\frac{s_{\overline1\overline2_\sigma}}{z_{\overline1\overline2_\sigma}}\frac{s_{\overline3_\sigma2}}{z_{\overline3_\sigma2}}\left(\frac{s_{\overline13_\sigma}}{z_{\overline13_\sigma}}+\frac{s_{\overline2_\sigma3_\sigma}}{z_{\overline2_\sigma3_\sigma}}\right)\ .\label{F1}
	\end{IEEEeqnarray}
	In section \ref{alphaprime_expand} we have also given the relation of the integrals in \eqref{eq::A2} to the corresponding open string integrals in $(2.29)$ of \cite{npt_2}. These are in agreement with the previously found basis functions in $(2.29)$ of \cite{npt_2} again under the identification \eqref{eq::identification} and up to the region of integration. Here we don't integrate over $0<z_{2}<z_{3}<z_{4}<1$, but instead we have different integration regions $\mathcal{I}_n$, which are given by
	\begin{IEEEeqnarray}l
		\begin{IEEEeqnarraybox}[][c]{rlrClCrClCrCl}
			\IEEEstrut
		\mathcal{I}_1&:\quad&0&<z_{\overline2}<&1\ ,&\quad&1&<z_{3}<&\infty\ ,&\quad&z_{\overline2}&<z_{\overline3}<&1\ ,\\
		\mathcal{I}_2&:&0&<z_{\overline2}<&1\ ,&&1&<z_{3}<&\infty\ ,&&1&<z_{\overline3}<&z_{3}\ .
		\IEEEstrut
		\end{IEEEeqnarraybox}\label{domains}
	\end{IEEEeqnarray}
To conclude this section we can write the scattering amplitude of three closed strings on the disk as
	\begin{IEEEeqnarray}{rCl}
		\mathcal{A}&=& -2g_c^3T_p\sum_{\sigma\in S_3}\left\{\sin(\pi s_{23})F^{(\overline2_\sigma3_\sigma\overline3_\sigma)}_{\mathcal{I}_1}+\sin\left[\pi (s_{23}+s_{2\overline3})\right]F^{(\overline2_\sigma3_\sigma\overline3_\sigma)}_{\mathcal{I}_2}\right\}\ A_{\rm YM}(\overline1,\overline2_\sigma,3_\sigma,\overline3_\sigma,2,1)\ ,\nonumber\\ \label{eq::A3_B}
	\end{IEEEeqnarray}
which agrees with \eqref{eq::A3}. We have shown that one can write this scattering process as a scattering of six open strings as proposed in \cite{ovsc}. In fact, the six permutations $\sigma$ of the YM-amplitudes in equation \eqref{eq::A3_B} are the same as for the six-point function of open strings and the expressions are the same up to the phases \eqref{eq::phasesPi} and the regions of integration. Moreover, in section \ref{sec::integration} we have argued that $\mathcal I_1$ and $\mathcal I_2$ correspond to two different open string subamplitudes.

	\section{Complex integration and analytic continuation}\label{sec::phase}
In this appendix we want to present a derivation of the phase factor
in \eqref{eq::phase}. Originally, in \cite{KLT} to find a relation
between open and closed strings they introduced a phase depending only
on the kinematic invariants to account for the correct branch of
integration. Besides the original approach of \cite{KLT} the notes
\cite{KLT2} may be  useful and we will gather the steps presented therein to the scattering of three closed strings on the disk.\par
We start by taking the amplitude and split it into a part containing branch cuts, which is essentially the Koba-Nielsen-factor originating from the plane wave contractions, and a branch cut independent piece $F(y,z,\overline z)$, which comes from the remaining contractions of the vertex operators:
\begin{IEEEeqnarray}{rCl}
	\mathcal{A}&\sim&\int_0^1\mathrm{d}y\int_{\mathbb{H}_+}\mathrm{d}^2z\,\langle V_1(i)V_{\overline1}(-i)V_2(iy)U_{\overline2}(-iy)U_3(z)U_{\overline3}(\overline z)\rangle\nonumber\\
	&=&\int_0^1\mathrm{d}y\int_{\mathbb{H}_+}\mathrm{d}^2z\,F(y,z,\overline z)2^{s_{1\overline1}}|2y|^{s_{2\overline2}}|1-y|^{2s_{12}}|1+y|^{2s_{1\overline2}}|i-z|^{2s_{13}}|i+z|^{2s_{1\overline3}}\nonumber\\
	&&\times|iy-z|^{2s_{23}}|iy+z|^{2s_{2\overline3}}|z-\overline z|^{s_{3\overline3}}\ .\label{eq::phase_1}
\end{IEEEeqnarray}
We can immediately recognize that no branch cuts can arise from $2y, 1-y, 1+y$ and $z-\overline z=2\Im(z)$, because they are always $\geq0$ due to the fact that we integrate $0 \leq y \leq 1$ and $(z,\overline z)$ over the upper half plane ($\Im(z)\geq0$).\par
We can now perform the analytic continuation described in section \ref{sec::analytic} and introduce the variables 
\begin{IEEEeqnarray}l
	z\to iz_1+iz_2=i\xi\ ,\qquad \overline z\to iz_1-iz_2=i\eta\ ,
\end{IEEEeqnarray} 
where $\xi\in\mathbb{R}$ and $\eta\in\mathbb{R}$ which have to satisfy $\xi-\eta\geq0$. We do so by deforming the contour integral of $\Re(z)=z_1$ avoiding the branch cuts as depicted in figure \ref{fig::z2}, which is achieved by 
\begin{IEEEeqnarray}l
	z_1\to ie^{-2i\varepsilon}z_1\approx i(1-2i\varepsilon)z_1=iz_1+2\varepsilon z_1\ ,
\end{IEEEeqnarray}
where $\varepsilon$ is small and $\varepsilon>0$. Thus, after the contour deformation we have for $\lambda\in\mathbb{R}$
\begin{IEEEeqnarray}{rCl}
	|i\lambda-z|^{2s}&=&\left[z_1^2+(\lambda-z_2)^2\right]^s\nonumber\\
	&\to&\left[(iz_1+2\varepsilon z_1)^2+(\lambda-z_2)^2\right]^s\nonumber\\
	&=&\left[(\xi-\lambda-i\varepsilon\delta)(-\eta-\lambda+i\varepsilon\delta)\right]^s\ ,
\end{IEEEeqnarray}
where we have introduced $\delta=2z_1=\xi+\eta$. With the above we can write \eqref{eq::phase_1} as
\begin{IEEEeqnarray}{rCl}
	\mathcal{A}&\sim&\int_0^1\mathrm{d}y\int^\infty_{-\infty}\mathrm{d}\xi\int^\xi_{-\infty}\mathrm{d}\eta\,F(y,\xi,\eta)2^{s_{1\overline{1}}}|2y|^{s_{2\overline2}}|1-y|^{2s_{12}}|1+y|^{2s_{1\overline2}}|\xi-\eta|^{s_{3\overline3}}\nonumber\\
	&&\times\left[(\xi-1-i\varepsilon\delta)(-\eta-1+i\varepsilon\delta)\right]^{s_{13}}\left[(\xi+1-i\varepsilon\delta)(-\eta+1+i\varepsilon\delta)\right]^{s_{1\overline3}}\nonumber\\
	&&\times\left[(\xi-y-i\varepsilon\delta)(-\eta-y-i\varepsilon\delta)\right]^{s_{23}}\left[(\xi+y-i\varepsilon\delta)(-\eta+y+i\varepsilon\delta)\right]^{s_{2\overline3}}\label{eq::phase2_1}\\
	&=&\int_0^1\mathrm{d}y\int^\infty_{-\infty}\mathrm{d}\xi\int^\xi_{-\infty}\mathrm{d}\eta\,F(y,\xi,\eta)2^{s_{1\overline{1}}}|2y|^{s_{2\overline2}}|1-y|^{2s_{12}}|1+y|^{2s_{1\overline2}}|\xi-\eta|^{s_{3\overline3}}\nonumber\\
	&&\times(-\xi+1+i\varepsilon\delta)^{s_{13}}(\eta+1-i\varepsilon\delta)^{s_{13}}(-\xi-1+i\varepsilon\delta)^{s_{1\overline3}}(\eta-1-i\varepsilon\delta)^{s_{1\overline3}}\nonumber\\
	&&\times(-\xi+y+i\varepsilon\delta)^{s_{23}}(\eta+y-i\varepsilon\delta)^{s_{23}}(-\xi-y+i\varepsilon\delta)^{s_{2\overline3}}(\eta-y-i\varepsilon\delta)^{s_{2\overline3}}\ .\label{eq::phase2}
\end{IEEEeqnarray}
In order to get from \eqref{eq::phase2_1} to \eqref{eq::phase2} we use that 
\begin{IEEEeqnarray}l
	(z_1 z_2)^c= (-z_1)^c (-z_2)^c \quad \text{for }\sign(\Im(z_1))=-\sign(\Im(z_2))\ .\label{eq::sign_zz}
\end{IEEEeqnarray}
We choose the branch cut of the power function to lie on the negative real axis, i.e.\ we restrict the power function to $z^c = |z|^c e^{c i \theta}$ with $- \pi < \theta \leq \pi$. Analogously to appendix A of \cite{KLT2} one can show that this implies (for both signs of $\Re(z)$)
\begin{IEEEeqnarray}l
	z^c=\left\{\begin{array}{ll}
		e^{i\pi c}(-z)^c&\text{for }\Im(z)\geq0\ ,\\
		e^{-i\pi c}(-z)^c&\text{for }\Im(z)<0\ ,\\
	\end{array}\right.\label{eq::sign_z}
\end{IEEEeqnarray}
and 
\begin{IEEEeqnarray}l
	(z_1 z_2)^c= z_1^c z_2^c \quad \text{for }\sign(\Im(z_1))=-\sign(\Im(z_2))\ . \label{eq::sign_zz2}
\end{IEEEeqnarray}
Taken together, these imply \eqref{eq::sign_zz}.\par
\begin{figure}[b]
	\begin{center}
		\begin{tikzpicture}
			\draw[dashed,thick,red] (-3.5,1)--(-3.5,-2)node[below]{$-\xi$};
			
			\draw[thick,dashed] (-7.5,0,0) -- (-6.5,0);
			\draw[thick,->] (-6.5,0) -- (6.5,0);
			\draw[thick] (-5,0.1)--(-5,-0.1) node[below]{$-1$};
			\draw[thick] (-1.5,-0.1)--(-1.5,0.1) node[above]{$-y$};
			\draw[thick] (0,-0.1)--(0,0.1) node[above]{$0$};
			\draw[thick] (1.5,-0.1)--(1.5,0.1 )node[above]{$y$};
			\draw[thick] (5,-0.1)--(5,0.1) node[above]{$1$};
			\draw[thick] (3.5,-0.1)--(3.5,0.1) node[above]{$\xi$};
			
			\draw[thick,dashed, blue] (-7.5,0.25) -- (-6.5,0.25);
			\draw[thick, blue] (-6.5,0.25) -- (-5.25,0.25);
			\draw[thick, blue] (-4.75,0.25) -- (-3.5,0.25);
			\draw[thick, blue] (-3.5,0.25) -- (-3.5,-0.25);
			\draw[thick, blue] (-3.5,-0.25) -- (-1.75,-0.25);
			\draw[thick, blue] (-1.25,-0.25) -- (1.25,-0.25);
			\draw[thick, blue] (1.75,-0.25) -- (3.5,-0.25);
			\draw[thick, blue] (3.5,-0.15) -- (3.5,-0.35);
			\draw[thick,red] (3.5,-0.25) -- (4.75,-0.25);
			\draw[thick,red,->] (5.25,-0.25) -- (6.5,-0.25);
			
			\draw[thick,blue] (-5.25,0.25) arc (180:0:0.25);
			\draw[thick,blue] (-1.75,-0.25) arc (-180:0:0.25);
			\draw[thick,blue] (1.25,-0.25) arc (-180:0:0.25);
			\draw[thick,red] (4.75,-0.25) arc (-180:0:0.25);
			
			\draw[thick,red,<->] (-7.5,-1.25) -- (-5.5,-1.25)node[below]{$\delta<0$} -- (-3.5,-1.25);
			\draw[thick,red,<->] (-3.5,-1.25) -- (1.25,-1.25)node[below]{$\delta>0$} -- (6.5,-1.25);
		\end{tikzpicture}
	\end{center}
	\caption{$\eta$-integration contour ($\eta<\xi$ in blue and $\eta>\xi$ in red) for $y<\xi<1$.}\label{fig::phase_3}
\end{figure}
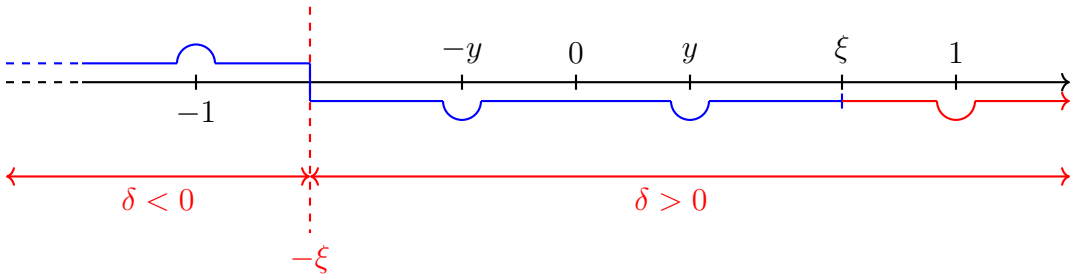

If $\xi<-1$ the behaviour of the imaginary parts in the $\eta$-terms at the branch points is
\begin{IEEEeqnarray*}{ll}
	\eta\approx-1:\quad&\delta=\xi+\eta\approx\xi-1<0\ ,\\
	\eta\approx-y:&\delta\approx\xi-y<0\ ,\\
	\eta\approx y:&\delta\approx\xi+y<0\ ,\\
	\eta\approx1:&\delta\approx\xi+1<0\ .
\end{IEEEeqnarray*}
Similarly, if $\xi>1$ at all branch points $\delta>0$. Thus we integrate along the contours depicted in figure \ref{fig::phase_2}.
\begin{figure}[h]
$\xi<-1$:
\begin{center}
	\begin{tikzpicture}
	\draw[thick,dashed] (-7,0,0) -- (-6,0);
	\draw[thick,->] (-6,0) -- (6,0);
	\draw[thick] (-3.5,0.1)--(-3.5,-0.1) node[below]{$-1$};
	\draw[thick] (-1.5,0.1)--(-1.5,-0.1) node[below]{$-y$};
	\draw[thick] (0,0.1)--(0,-0.1) node[below]{$0$};
	\draw[thick] (1.5,0.1)--(1.5,-0.1 )node[below]{$y$};
	\draw[thick] (3.5,0.1)--(3.5,-0.1) node[below]{$1$};
	\draw[thick] (-5,0.1)--(-5,-0.1) node[below]{$\xi$};
	
	\draw[thick,dashed, blue] (-7,0.25) -- (-6,0.25);
	\draw[thick, red] (-6,0.25) -- (-3.75,0.25);
	\draw[thick, red] (-3.25,0.25) -- (-1.75,0.25);
	\draw[thick, red] (-1.25,0.25) -- (1.25,0.25);
	\draw[thick, red] (1.75,0.25) -- (3.25,0.25);
	\draw[thick, red,->] (3.75,0.25) --(6,0.25);
	\draw[thick,red] (-3.75,0.25) arc (180:0:0.25);
	\draw[thick,red] (-1.75,0.25) arc (180:0:0.25);
	\draw[thick,red] (1.25,0.25) arc (180:0:0.25);
	\draw[thick,red] (3.25,0.25) arc (180:0:0.25);
	
	\draw[thick,dashed, blue] (-7,0.25) -- (-6,0.25);
	\draw[thick, blue] (-6,0.25) -- (-5,0.25);
	\draw[thick, blue] (-5,0.15) -- (-5,0.35);
	\end{tikzpicture}
\end{center}
	\vspace{0.5cm}$\hphantom-\xi>1$:
\begin{center}
	\begin{tikzpicture}
	\draw[thick,dashed] (-7,0,0) -- (-6,0);
	\draw[thick,->] (-6,0) -- (6,0);
	\draw[thick] (-3.5,-0.1)--(-3.5,0.1) node[above]{$-1$};
	\draw[thick] (-1.5,-0.1)--(-1.5,0.1) node[above]{$-y$};
	\draw[thick] (0,-0.1)--(0,0.1) node[above]{$0$};
	\draw[thick] (1.5,-0.1)--(1.5,0.1 )node[above]{$y$};
	\draw[thick] (3.5,-0.1)--(3.5,0.1) node[above]{$1$};
	\draw[thick] (5,-0.1)--(5,0.1) node[above]{$\xi$};
		
	\draw[thick,dashed, blue] (-7,-0.25) -- (-6,-0.25);
	\draw[thick, blue] (-6,-0.25) -- (-3.75,-0.25);
	\draw[thick, blue] (-3.25,-0.25) -- (-1.75,-0.25);
	\draw[thick, blue] (-1.25,-0.25) -- (1.25,-0.25);
	\draw[thick, blue] (1.75,-0.25) -- (3.25,-0.25);
	\draw[thick, blue] (3.75,-0.25) --(5,-0.25);
	\draw[thick, blue] (5,-0.15) -- (5,-0.35);
	\draw[thick,red,->] (5,-0.25) -- (6,-0.25);
	
	\draw[thick,blue] (-3.75,-0.25) arc (-180:0:0.25);
	\draw[thick,blue] (-1.75,-0.25) arc (-180:0:0.25);
	\draw[thick,blue] (1.25,-0.25) arc (-180:0:0.25);
	\draw[thick,blue] (3.25,-0.25) arc (-180:0:0.25);
	\end{tikzpicture}
	\end{center}
	\caption{$\eta$-integration contour ($\eta<\xi$ in blue and $\eta>\xi$ in red) for $\xi<-1$ and $1<\xi$.}\label{fig::phase_2}
\end{figure} 
\noindent Note that the $\eta$-integration always ends at $\xi$, i.e.\ it ranges only over the blue contour in figure \ref{fig::phase_2}. For $\xi\in]-1,1[$ we find that $\delta<0$ for $\eta<-\xi$ and $\delta>0$ for $\eta>-\xi$, which results in an integration contour as depicted in figure \ref{fig::phase_3}.

\begin{figure}[h]
	\begin{center}
	\begin{tikzpicture}
		\draw[thick,dashed] (-7.5,0,0) -- (-6.5,0);
		\draw[thick,->] (-6.5,0) -- (6,0);
		\draw[thick] (-5,0.1)--(-5,-0.1) node[below]{$-1$};
		\draw[thick] (-1.5,-0.1)--(-1.5,0.1) node[above]{$-y$};
		\draw[thick] (0,-0.1)--(0,0.1) node[above]{$0$};
		\draw[thick] (1.5,-0.1)--(1.5,0.1 )node[above]{$y$};
		\draw[thick] (5,-0.1)--(5,0.1) node[above]{$1$};
		\draw[thick] (3.5,-0.1)--(3.5,0.1) node[above]{$\xi$};
		
		\draw[thick,dashed, blue] (-7.5,0.25) -- (-6.5,0.25);
		\draw[thick, blue] (-6.5,0.25) -- (-5.25,0.25);
		\draw[thick, blue] (-4.75,0.25) -- (-3.5,0.25);
		\draw[thick, blue] (-3.5,0.25) -- (-3.5,-0.25);
		\draw[thick, blue] (-3.5,-0.25) -- (-1.75,-0.25);
		\draw[thick, blue] (-1.25,-0.25) -- (1.25,-0.25);
		\draw[thick, blue] (1.75,-0.25) -- (3.5,-0.25);
		\draw[thick, blue] (3.5,-0.15) -- (3.5,-0.35);

		\draw[thick,blue] (-5.25,0.25) arc (180:0:0.25);	
		\draw[thick,blue] (-1.75,-0.25) arc (-180:0:0.25);
		\draw[thick,blue] (1.25,-0.25) arc (-180:0:0.25);

		\draw[thick,decorate,decoration={brace,mirror}](-7.5,-0.8) -- (-6.25,-0.8) node[below]{$e^{i\pi s_{13}}$} -- (-5,-0.7);
		\draw[thick,decorate,decoration={brace,mirror}](-5,-0.8) -- (-3.25,-0.8) node[below]{$1$} -- (-1.5,-0.7);
		\draw[thick,decorate,decoration={brace,mirror}](-1.5,-0.8) -- (0,-0.8) node[below]{$e^{i\pi s_{23}}$} -- (1.5,-0.7);
		\draw[thick,decorate,decoration={brace,mirror}](1.5,-0.8) -- (2.5,-0.8) node[below]{$e^{i\pi s_{23}}e^{i\pi s_{2\overline3}}$} -- (3.5,-0.7);
	\end{tikzpicture}
	\end{center}
	\caption{Phase $\Pi(y,\xi,\eta)$ for $y<\xi<1$.}\label{fig::phase}
\end{figure}
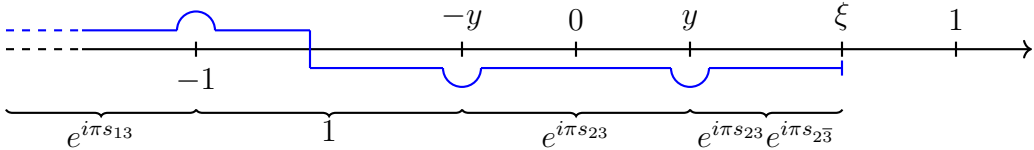

We can now do an example to illustrate the general procedure. To be concrete, we consider the region $y<\xi<1$. If $y<\xi<1$ the behaviour of the imaginary parts in the $\eta$ terms in \eqref{eq::phase2} at the branch points is
\begin{IEEEeqnarray*}{ll}
	\eta\approx-1:\quad&\delta=\xi+\eta\approx\xi-1<0\ ,\\
	\eta\approx-y:&\delta\approx\xi-y>0\ ,\\
	\eta\approx y:&\delta\approx\xi+y>0\ ,\\
	\eta\approx1:&\delta\approx\xi+1>0\ .
\end{IEEEeqnarray*}
Moreover, for this example we have that for the $\xi$-dependent terms the real part of $1-\xi+i\varepsilon\delta$ is positive and the real parts of all other $\xi$-dependent terms are negative. In order to have all real parts of the $\xi$ dependent terms positive, we can change the signs in the latter case if we do the same simultaneously for the corresponding $\eta$-terms, cf.\ \eqref{eq::sign_z}, i.e.\
\begin{IEEEeqnarray}{rCl}
	\mathcal{A}\Big|_{y<\xi<1}&\sim&\int_0^1\mathrm{d}y\int^1_{y}\mathrm{d}\xi\int^\xi_{-\infty}\mathrm{d}\eta\,F(y,\xi,\eta)2^{s_{1\overline1}}|2y|^{s_{2\overline2}}|1-y|^{2s_{12}}|1+y|^{2s_{1\overline2}}|\xi-\eta|^{s_{3\overline3}}\nonumber\\
	&&\times(-\xi+1+i\varepsilon\delta)^{s_{13}}(\eta+1-i\varepsilon\delta)^{s_{13}}(\xi+1-i\varepsilon\delta)^{s_{1\overline3}}(-\eta+1+i\varepsilon\delta)^{s_{1\overline3}}\nonumber\\
	&&\times(\xi-y-i\varepsilon\delta)^{s_{23}}(-\eta-y+i\varepsilon\delta)^{s_{23}}(\xi+y-i\varepsilon\delta)^{s_{2\overline3}}(-\eta+y+i\varepsilon\delta)^{s_{2\overline3}}\ .\label{eq::phase3}
\end{IEEEeqnarray}
Now we can use \eqref{eq::sign_z} again to make the real parts of the $\eta$ dependent terms in the Koba-Nielsen factor positive (and take the limit $\varepsilon\to0$). Given the signs in table \ref{tab::phase} and the signs for $\delta$ depicted in figure \ref{fig::phase_3} we see that the integrand in \eqref{eq::phase3} is given by
\begin{IEEEeqnarray}{rCl}
	\mathcal{A}\Big|_{y<\xi<1}&\sim&\int_0^1\mathrm{d}y\int^1_{y}\mathrm{d}\xi\int^\xi_{-\infty}\mathrm{d}\eta\,F(y,\xi,\eta)\Pi(y,\xi,\eta)2^{s_{1\overline1}}|2y|^{s_{2\overline2}}|1-y|^{2s_{12}}|1+y|^{2s_{1\overline2}}|\xi-\eta|^{s_{3\overline3}}\nonumber\\
	&&\times|1-\xi|^{s_{13}}|1+\eta|^{s_{13}}|1+\xi|^{s_{1\overline3}}|1-\eta|^{s_{1\overline3}}|y-\xi|^{s_{23}}|y+\eta|^{s_{23}}|y+\xi|^{s_{2\overline3}}|y-\eta|^{s_{2\overline3}}\nonumber\\\label{eq::phase4}
\end{IEEEeqnarray}
with a phase $\Pi(y,\xi,\eta)$ which is depicted in figure \ref{fig::phase}.
\noindent From this analysis we can conclude that only if the real parts of the corresponding $\xi$ and $\eta$ dependent terms have opposite signs do we get a contribution to the phase of this integration region. Hence, we can conclude that the phase obtained by this procedure is consistent with
\begin{IEEEeqnarray}{rCl}
	\Pi(y,\xi,\eta)&=&e^{i\pi s_{13}\Theta(-(1-\xi)(1+\eta))}e^{i\pi s_{1\overline3}\Theta(-(1+\xi)(1-\eta))} e^{i\pi s_{23}\Theta(-(y-\xi)(y+\eta))}\nonumber\\
	&&\times e^{i\pi  s_{2\overline3}\Theta(-(y+\xi)(y-\eta))} e^{i\pi  s_{3\overline3}\Theta(-(\xi-\eta))}\ ,
\end{IEEEeqnarray}
where we have added $e^{i\pi  s_{3\overline3}\Theta(-(\xi-\eta))}$ for completeness. This factor accounts for the contribution coming from $(\xi-\eta)^{s_{3\overline3}}$ for $\xi<\eta$ such that we can write 
\begin{IEEEeqnarray}l
	(\xi-\eta)^{s_{3\overline3}}=|\xi-\eta|^{s_{3\overline3}}e^{i\pi  s_{3\overline3}\Theta(-(\xi-\eta))}
\end{IEEEeqnarray}
using \eqref{eq::sign_z} for all values of $\xi$ and $\eta$.
\begin{table}[H]
	\begin{center}
		\begin{tabular}{r c|c|c|c}
			          & $\eta<-1$ & $-1<\eta<-y$ & $-y<\eta<y$ & $y<\eta<\xi$ \\
			$1+\eta$: &   $<0$    &     $>0$     &    $>0$     & $>0$             \\ \hline
			$1-\eta$:  &   $>0$    &     $>0$     &    $>0$     & $>0$             \\ \hline
			$-y-\eta$: &   $>0$    &     $>0$     &    $<0$     & $<0$             \\ \hline
			$y-\eta$:  &   $>0$    &     $>0$     &    $>0$     & $<0$
		\end{tabular}
	\end{center}
	\caption{Real part of the $\eta$ dependent terms in \eqref{eq::phase3} for $y<\xi<1$.}\label{tab::phase}
\end{table}
	\section{Invariance of a correlator under $PSL(2,\mathbb R)$ transformation}\label{sec::psl2r}
Let us start with some general remarks about a correlation function\footnote{We will only consider the chiral (holomorphic) fields. For the anti-chiral (antiholomorphic) fields similar results hold.} of a CFT on the Riemann sphere $S^2=\mathbb C\cup\infty$ following \cite{BLT}. The structure of a correlation function is restricted: A correlation function is the vacuum expectation value of a radially ordered product of fields. Because the ground state is invariant under $SL(2,\mathbb C)$ transformations\footnote{The vacuum is not invariant under the action of all generators, we only need to consider the globally well defined generators, which are $L_{-1},L_0$ and $L_1$.} a general CFT correlator with $n$ fields $\phi_i(z_i)$ has to satisfy
\begin{IEEEeqnarray}r
	\langle\phi'_1(z_1)\cdots\phi'_n(z_n)\rangle_{S^2}=\langle\phi_1(z_1)\cdots\phi_n(z_n)\rangle_{S^2}\ ,\label{eq::inv}
\end{IEEEeqnarray}
where we used that $\phi'(z)=U\phi(z)U^{-1}$ is the $SL(2,\mathbb C)$ transformed field $\phi(z)$ and the transformation matrix $U$ leaves the in and out ground state invariant. The three globally defined generators of $SL(2,\mathbb C)$ act on the fields $\phi(z)$ as 
\begin{IEEEeqnarray}l
	\begin{IEEEeqnarraybox}[][c]{CCCClCl}
		\IEEEstrut
		L_{-1}&:&\text{translations}&\quad&U=e^{b L_{-1}}&\qquad&\phi'(z)=\phi(z+b)\ ,\\
		L_{0}&:&\begin{matrix}
			\text{dilatations and}\\\text{rotations}
		\end{matrix}&\quad&U=e^{\ln a L_{0}}&\qquad&\phi'(z)=a^h\phi(az)\ ,\\
		L_{1}&:&\begin{matrix}
			\text{special conformal}\\\text{transformations}
		\end{matrix}&\quad&U=e^{cL_1}&\qquad&\phi'(z)=\left(\tfrac{1}{1-cz}\right)^{2h}\phi\left(\tfrac{z}{1-cz}\right)\ ,
	\IEEEstrut
	\end{IEEEeqnarraybox}\label{eq::trafo_generator}
\end{IEEEeqnarray}
where $h$ is the conformal dimension of $\phi(z)$. This transformation behaviour under global $SL(2,\mathbb C)$ transformations of the correlator of $n$ primary fields $\phi_i(z_i)$ with conformal weight $h_i$ leads to the following constraints
\begin{IEEEeqnarray}{l}
	\sum_{i=1}^n\frac{\partial}{\partial z_i}\langle\phi_1(z_1)\cdots\phi_n(z_n)\rangle_{S^2}=0\ ,\label{eq::61}\\
	\sum_{i=1}^n\left(z_i\frac{\partial}{\partial z_i}+h_i\right)\langle\phi_1(z_1)\cdots\phi_n(z_n)\rangle_{S^2}=0\ ,\\
	\sum_{i=1}^n\left(z_i^2\frac{\partial}{\partial z_i}+2z_ih_i\right)\langle\phi_1(z_1)\cdots\phi_n(z_n)\rangle_{S^2}=0\ .\label{eq::62}
\end{IEEEeqnarray}
Conversely, if a correlator satisfies the conditions \eqref{eq::61}--\eqref{eq::62}, the correlator is invariant under $SL(2,\mathbb{C})$ transformations \cite{BLT}.\par
On the Riemann sphere we can separate holomorphic and antiholomorphic fields and discuss their correlation functions separately, but for other topologies this can be more subtle: Due to the boundary of the disk $D_2$ the holomorphic and antiholomorphic fields (and also the holomorphic and antiholomorphic parts of the same field) interact with each other such that we cannot separate them any more. In the following we want to derive similar conditions to \eqref{eq::61}--\eqref{eq::62} for conformal primaries $\phi_i(z_i,\overline z_i)$ on the unit disk. The purpose is to find the conditions the correlator in \eqref{eq::correlator} has to fulfil to be invariant under conformal transformations. Relations between the kinematic factors, which are imposed by these conditions, will allow us to map one vertex operator position fixing to another.\par
The conformal Killing group of the disk\footnote{Here, we actually mean the upper half plane $\mathbb{H}_+$, since the conformal Killing group of the disk is $SU(1,1)$. However, we have mapped the scattering process on the disk to the upper half plane and use these two expressions synonymously.} is $PSL(2,\mathbb R)$, which means that transformation parameters have to be real numbers. Hence, the conformal transformation for the holomorphic and antiholomorphic components on the disk are the same. The infinitesimal transformations of the worldsheet coordinates are given by
\begin{IEEEeqnarray}l
	\begin{IEEEeqnarraybox}[][c]{rCl}
		\IEEEstrut
		z'&=&z+\epsilon(z)\ ,\\
		\overline z'&=&\overline z+\overline\epsilon(\overline z)=\overline z+\epsilon(\overline z)\ ,
		\IEEEstrut
	\end{IEEEeqnarraybox}\label{eq::trafo_z}
\end{IEEEeqnarray}
where we used that $\overline \epsilon = \epsilon$, because the conformal Killing group of the unit disk is $PSL(2,\mathbb{R})$. We can then compute the infinitesimal $PSL(2,\mathbb R)$ transformation of a primary field:
\begin{IEEEeqnarray}{rCl}
	\delta_\epsilon\phi(z,\overline z)&=&\phi'(z,\overline z)-\phi(z,\overline z)=\phi'(z'-\epsilon(z),\overline z'-\epsilon(\overline z))-\phi(z,\overline z)\ .
\end{IEEEeqnarray}
Above we have used \eqref{eq::trafo_z} and are now going to perform a Taylor expansion of $\phi'(z'-\epsilon(z),\overline z'-\epsilon(\overline z))$ in the infinitesimal parameters $\epsilon(z)$ and $\epsilon(\overline z)$. Up to first order in $\epsilon$ we obtain
\begin{IEEEeqnarray}{rCl}
	\delta_\epsilon\phi(z,\overline z)&=&\phi'(z',\overline z')-\epsilon(z)\partial\phi(z,\overline z)-\epsilon(\overline z)\overline\partial\phi(z,\overline z)-\phi(z,\overline z)+\mathcal{O}(\epsilon^2)\ .
\end{IEEEeqnarray}
We can now use the general transformation property of a primary field under conformal transformations, which is given by $\phi'(z',\overline z')=\left(\frac{\partial z'}{\partial z}\right)^{-h}\left(\frac{\partial \overline z'}{\partial \overline z}\right)^{-h}\phi(z,\overline z)$, where $(h,\overline h)$ are the conformal weights of the holomorphic and antiholomrphic components of $\phi$, so that 
\begin{IEEEeqnarray}{rCl}
	\delta_\epsilon\phi(z,\overline z)&=&\left[(1+\partial\epsilon(z))^{-h}(1+\overline\partial\epsilon(\overline z))^{-\overline h}-\left(1+\epsilon(z)\partial+\epsilon(\overline z)\overline\partial\right)\right]\phi(z,\overline z)+\mathcal{O}(\epsilon^2)\ ,
\end{IEEEeqnarray}
which then becomes
\begin{IEEEeqnarray}{rCl}
	\delta_{\epsilon}\phi(z,\overline z)&=&-\left(h\partial\epsilon(z)+\overline h\overline\partial\epsilon(\overline z)+\epsilon(z)\partial+\epsilon(\overline z)\overline\partial\right)\phi(z,\overline z)+\mathcal{O}(\epsilon^2)\ .\label{eq::travo_phi}
\end{IEEEeqnarray}
The infinitesimal conformal transformations can be written as
\begin{IEEEeqnarray}{l}
	\epsilon(z)=\epsilon_{-1}+\epsilon_0z+\epsilon_1z^2.\label{eq::con_trafo}
\end{IEEEeqnarray}
We can take this explicit form of $\epsilon$ and plug it into \eqref{eq::travo_phi}. After computing all derivatives and rearranging the terms the transformation of a conformal primary takes the form \cite{Polchinski1}
\begin{IEEEeqnarray}{rCl}
	\delta_{\epsilon}\phi\left(z,\overline z\right)&=&-\left[\epsilon_{-1}\left(\partial+\overline\partial\right)+\epsilon_0\left(h+z\partial+\overline h+\overline z\overline \partial\right)+\epsilon_1\left(2hz+z^2\partial+2\overline h\overline z+\overline z^2\overline \partial\right)\right]\phi\left(z,\overline z\right). \nonumber \\  \label{eq::ex_travo_phi}
\end{IEEEeqnarray}

As already stated above the correlator of $n$ primary fields is invariant under conformal transformations, which was displayed in \eqref{eq::inv}. If we write down the analogue of \eqref{eq::inv}
\begin{IEEEeqnarray}r
	\langle\phi'_1(z_1,\overline z_1)\cdots\phi'_n(z_n,\overline z_n)\rangle_{D_2}=\langle\phi_1(z_1,\overline z_1)\cdots\phi_n(z_n,\overline z_n)\rangle_{D_2}\label{eq::inv_disk}
\end{IEEEeqnarray}
and then subtract both sides we find that the variation with respect to a conformal transformation has to vanish. Hence, we obtain for infinitesimal $\epsilon$
\begin{IEEEeqnarray}{rCl}
	0&=&\delta_\epsilon\langle\phi_1(z_1,\overline z_1)\cdots\phi_n(z_n,\overline z_n)\rangle_{D_2}=\sum_{i=1}^n\langle\phi_1(z_1,\overline z_1)\cdots\delta_\epsilon\phi_i(z_i,\overline z_i)\cdots\phi_n(z_n,\overline z_n)\rangle_{D_2}\ .\nonumber\\\label{eq::inv_2}
\end{IEEEeqnarray}
We can then use the explicit form of the infinitesimal variation of $\phi$, which is given in \eqref{eq::ex_travo_phi} and compute the right side of equation \eqref{eq::inv_2}.
\begin{IEEEeqnarray}{rCl}
	0&=&\sum_{i=1}^n\left[\epsilon_{-1}\left(\partial_i+\overline\partial_i\right)+\epsilon_{0}\left(h_i+z_i\partial_i+\overline h_i+\overline z_i\overline \partial_i\right)\right.\nonumber\\
	&&\left.+\epsilon_{1}\left(2h_iz_i+z_i^2\partial_i+2\overline h_i\overline z_i+\overline z_i^2\overline \partial_i\right)\right]\langle\phi_1\left(z_1,\overline z_1\right)\cdots\phi_n\left(z_n,\overline z_n\right)\rangle_{D_2}\ .
\end{IEEEeqnarray}
In general, the coefficients $\epsilon_{-1},\epsilon_0$ and $\epsilon_1$ of the transformation \eqref{eq::con_trafo} are not vanishing. Hence, we find three conditions for the correlator on the disk to be invariant under conformal transformations. Explicitly, after equating coefficients for $\epsilon_{-1},\epsilon_0$ and $\epsilon_1$ this results in the following equations
\begin{IEEEeqnarray}{rCl}
	0&=&\sum_{i=1}^n\left(\partial_i+\overline\partial_i\right)\langle\phi_1(z_1,\overline z_1)\cdots\phi_n(z_n,\overline z_n)\rangle_{D_2}\ ,\label{eq::04}\\
	0&=&\sum_{i=1}^n\left(h_i+z_i\partial_i+\overline h_i+\overline z_i\overline \partial_i\right)\langle\phi_1(z_1,\overline z_1)\cdots\phi_n(z_n,\overline z_n)\rangle_{D_2}\ ,\\
	0&=&\sum_{i=1}^n\left(2h_iz_i+z_i^2\partial_i+2\overline h_i\overline z_i+\overline z_i^2\overline \partial_i\right)\langle\phi_1(z_1,\overline z_1)\cdots\phi_n(z_n,\overline z_n)\rangle_{D_2}\ .\label{eq::05}
\end{IEEEeqnarray}
We can interpret this result by comparing it with the conditions \eqref{eq::61}--\eqref{eq::62} previously found for a CFT on the Riemann sphere. To do so, let us first rewrite the primary field $\phi(z,\overline z)\to\phi(z)\overline\phi(\overline z)$, where $\phi(z)$ is the holomorphic and $\overline\phi(\overline z)$ is the antiholomorphic component of $\phi(z,\overline z)$, which depend only on $z$ and $\overline z$, respectively. Hence, we can write the correlator of $n$ such fields as
\begin{IEEEeqnarray}l
	\langle\phi_1(z_1,\overline z_1)\cdots\phi_n(z_n,\overline z_n)\rangle_{D_2}\to\langle\phi_1(z_1)\overline\phi_1(\overline z_1)\cdots\phi_n(z_n)\overline\phi_n(\overline z_n)\rangle_{D_2}\ .
\end{IEEEeqnarray}
By doing so we have to treat both components as independent fields that interact with all other $2n-1$ fields. Thus, we have a correlator of $2n$ independent conformal primaries with conformal weights $h_i$ and $\overline h_i$, respectively. From this point of view the conditions \eqref{eq::04}--\eqref{eq::05} seem natural, because they describe the conditions a correlator of $2n$ primary field has to satisfy to be conformally invariant similar to the ones for a CFT on the Riemann sphere.\\
\\\textbf{An example:}\\
To illustrate that the constraints \eqref{eq::04}--\eqref{eq::05} are suitable for our purpose, we want to check that a correlation function of $n$ plane wave factors (or $n$ tachyons) satisfies those. Moreover, by doing so we illustrate that for our discussion in appendix \ref{sec::correlator} the Koba-Nielsen factor is not involved in finding the constraints \eqref{eq::c} such that those constraints only arise from the zero mode correlator of three closed strings on the disk.\par
The Koba-Nielsen factor of $n$ closed strings scattering off a D$p$-brane is given by
\begin{IEEEeqnarray}{l}
	\KN=\biggl\langle\prod_{i=1}^{n}e^{ip_i\cdot X(z_i,\overline z_i)} \biggl\rangle_{D_2}=\prod_{i=1}^{n}|z_i-\overline z_i|^{\frac{\alpha'}{2}p_iDp_i}\prod_{\substack{i,j=1\\i<j}}^n|z_i-z_j|^{\alpha'p_i\cdot p_j}|z_i-\overline z_j|^{\alpha'p_i\cdot D\cdot p_j} .
\end{IEEEeqnarray}
Each plane wave factor $e^{ip_i\cdot X(z_i,\overline z_i)}$, which can be split in a holomorphic part $e^{ip_i\cdot X(z_i)}$ and an antiholomorphic part $e^{ip_i\cdot D\cdot X(\overline z_i)}$, has conformal weight $(\frac{\alpha'p_i^2}{4},\frac{\alpha'p_i^2}{4})$. We start with the first equation \eqref{eq::04}, where we just need to perform the derivatives with respect to the holomorphic and antiholomorphic coordinates $z_i$ and $\overline z_i$ of the plane wave factors, which are given by
\begin{IEEEeqnarray}{rCl}
	\partial_i\KN&=&\frac12\biggl[\frac{\alpha'p_i{\cdot }D{\cdot} p_i}{z_i-\overline z_i}+\sum_{\substack{j=1\\i\neq j}}^{n}\left(\frac{\alpha'p_i{\cdot}p_j}{z_i-z_j}+\frac{\alpha'p_i{\cdot}D{\cdot}p_j}{z_i-\overline z_j}\right)\biggl]\KN\ ,\\
	\overline\partial_i\KN&=&\frac12\biggl[-\frac{\alpha'p_i{\cdot}D{\cdot}p_i}{z_i-\overline z_i}+\sum_{\substack{j=1\\i\neq j}}^{n}\left(\frac{\alpha'p_i{\cdot}p_j}{\overline z_i-\overline z_j}+\frac{\alpha'p_i{\cdot}D{\cdot}p_j}{\overline z_i-z_j}\right)\biggl]\KN\ ,
\end{IEEEeqnarray}
respectively. Adding up all of these derivatives and rearranging them we find 
\begin{IEEEeqnarray}{rCl}
	\sum_{i=1}^{n}\left(\partial_i+\overline\partial_i\right)\KN
	&=& \sum_{\substack{i,j=1\\i\neq j}}^n\left[\frac{\alpha'p_i{\cdot}p_j}{2}\left(\frac{1}{z_i-z_j}+\frac{1}{\overline{z}_i-\overline{z}_j}\right)+\frac{\alpha'p_i{\cdot}D{\cdot}p_j}{2}\left(\frac{1}{z_i-\overline z_j}+\frac{1}{\overline{z}_i-z_j}\right)\right]\KN\nonumber \\
	&=& 0 \ .
\end{IEEEeqnarray}
The above expression vanishes, because we find every term twice but with a different overall sign, i.e. we have for instance $\frac{1}{z_i-z_j}$ and $\frac{1}{z_j-z_i}$ for fixed $i$ and $j$, which add up to zero.\par
For the second constraint we need to include the conformal weight of the plane wave factors and also have to multiply the derivatives by the corresponding coordinate. This leads to a more complicated expression
\begin{IEEEeqnarray}{l}
	\sum_{i=1}^{n}\left(\frac{\alpha'p_i^2}{4}+z_i\partial_i+\frac{\alpha'p_i^2}{4}+\overline z_i\overline\partial_i\right)\KN=\sum_{i=1}^{n}\left(\frac{\alpha'p_i^2}{2}+\frac{\alpha'p_i{\cdot}D{\cdot}p_i}{2}\right)\KN \nonumber\\
	\hspace{2cm} +\sum_{\substack{i,j=1\\i\neq j}}\biggl[\frac{\alpha'p_i{\cdot}p_j}{2}\left(\frac{z_i}{z_i-z_j}+\frac{\overline z_i}{\overline{z}_i-\overline{z}_j}\right) +\frac{\alpha'p_i{\cdot}D{\cdot}p_j}{2}\left(\frac{z_i}{z_i-\overline z_j}+\frac{\overline z_i}{\overline{z}_i-z_j}\right)\biggl]\KN\nonumber\\
	\hspace{2cm} =\sum_{i,j=1}\frac{\alpha'}{2}(p_i{\cdot}p_j+p_i{\cdot}D{\cdot}p_j)\KN=0\ ,\label{eq::200}
\end{IEEEeqnarray}
where we used that we find terms like $\frac{z_i}{z_i-z_j}$ and $\frac{z_j}{z_j-z_i}$ for fixed $i$ and $j$ such that their sum adds up to $\frac{z_i}{z_i-z_j}+\frac{z_j}{z_j-z_i}=1$. Then, in \eqref{eq::200} we end up with an expression that vanishes by momentum conservation so that also the second constraint is satisfied by the Koba-Nielsen factor of $n$ closed strings.\par
For the last equation we have
\begin{IEEEeqnarray}{l}
	\sum_{i=1}^{n}\left(2z_i\frac{\alpha'p_i^2}{4}+z_i^2\partial_i+2\overline z_i\frac{\alpha'p_i^2}{4}+\overline z_i^2\overline\partial_i\right)\KN=\sum_{i=1}^{n}\frac12\biggl[\sum_{j=1}^{n}p_i{\cdot}(p_j+D{\cdot}p_j)(z_i+\overline z_i)\biggl]\KN=0\ . \nonumber\\
\end{IEEEeqnarray}
Hence, the Koba-Nielsen factor $\KN$ of $n$ closed strings scattered off a D$p$-brane satisfies all three constraints \eqref{eq::04}--\eqref{eq::05} and is therefore invariant under conformal transformations.
	\section{Calculation on the double cover}
\label{sec:symmetrization}

In this appendix we would like to give some arguments why it is plausible that the calculation on the double cover misses certain poles of the result on the disk. To this end, we consider the closed string 2-point function on the disk, which can for instance be found in the formula between (4) and (5) in \cite{Hashimoto:1996bf}. The $s$-channel poles arise from the second term and we will focus on this term in the following.

For ${\rm Re}(s)>0$ and ${\rm Re}(t)>-1$, we have
\begin{eqnarray}
\int_0^1 dy \left[\frac{4 y}{(1 + y)^2}\right]^s \left[ \frac{(1 - y)^2}{(1 + y)^2} \right]^t \frac{1 - y}{y (1 + y)} &=& 4^s \frac{\Gamma (s) \Gamma(2 + 2 t)}{\Gamma(2 + s + 2 t)}\  _2F_1\left[\genfrac{}{}{0pt}{}{1 + 2 s + 2 t,\ s}{ 2 + s + 2 t}; -1\right] \nonumber \\
&& \label{expr1a} \\
&=& 4^s \Gamma (s) \frac{\Gamma (2 + 2 t) \Gamma(3/2 + s + t)}{\Gamma(2 + 2s + 2 t) \Gamma(3/2 + t)}  \label{expr1b}\\
&=& 4^s \Gamma (s) 4^{-s} \frac{\Gamma (1+t)}{\Gamma (1+s+t)}  \label{expr1c}\\
&=& t \frac{ \Gamma (s) \Gamma (t)}{\Gamma (1+s+t)} \stackrel{s \rightarrow 0}{\longrightarrow} \frac1s + {\cal O}(s^0) \label{expr1d}\ ,
\end{eqnarray}
where we used mathematica to get \eqref{expr1a} (and mathematica gives the condition that we imposed above), \eqref{expr1b} follows from 
\be
_2F_1\left[\genfrac{}{}{0pt}{}{a,\ b}{ 1+a-b}; -1\right] = \frac{\Gamma (1+a-b) \Gamma (1+\tfrac12 a)}{\Gamma (1+a) \Gamma (1+\tfrac12 a-b)}
\ee
and \eqref{expr1c} is obtained with mathematica again. In \eqref{expr1d} we performed the low energy limit, i.e.\   we expanded for small $s$ (independently of $t$). Doing so, we find an $s$-channel pole. 

Let us now compare this with the result if we integrate $y$ over $[-1,1]$, as is done on the double cover. For ${\rm Re}(s)>0$ and ${\rm Re}(t)>-1$ and {\it additionally} ${\rm Re}(s+t)<0$ (conditions imposed again by mathematica), we have
\begin{IEEEeqnarray}l
\hspace{-1.2cm}\int_{-1}^1 dy \left[\frac{4 y}{(1 + y)^2}\right]^s \left[ \frac{(1 - y)^2}{(1 + y)^2} \right]^t \frac{1 - y}{y (1 + y)} =\\
= 4^s \Gamma (s) \left[(-1)^{(1 + s)} \frac{\Gamma(-2 (s + t))}{\Gamma(-s - 2 t)}\ _2F_1 \left[\genfrac{}{}{0pt}{}{-1 - 2 t,\ s}{ -s - 2 t}; -1\right] \right. \nonumber \\
 \left. \hspace{1.2cm} + \frac{\Gamma (2 + 2 t)}{\Gamma(2 + s + 2 t)}\ _2F_1 \left[\genfrac{}{}{0pt}{}{1 + 2 s + 2 t,\ s}{ 2 + s + 2 t}; -1\right] \right]  \nonumber\\
=  4^s \Gamma (s) \left[(-1)^{(1 + s)} \frac{\Gamma(-2 (s + t)) \Gamma (\tfrac12 - t)}{\Gamma (- 2 t) \Gamma (\tfrac12 - s - t)} \right.  \nonumber \\
\left. \hspace{1.2cm} + \frac{\Gamma (2 + 2 t) \Gamma(3/2 + s + t)}{\Gamma(2 + 2s + 2 t) \Gamma(3/2 + t)} \right]  \nonumber \\
 =  4^s \Gamma (s) \left[(-1)^{(1 + s)} 4^{-s} \frac{\Gamma(- s - t)}{\Gamma (- t)} + 4^{-s} \frac{\Gamma (1+t)}{\Gamma (1+s+t)}  \right] \nonumber \\
= t \frac{ \Gamma (s) \Gamma (t)}{\Gamma (1+s+t)} - (-1)^{s} \frac{ \Gamma (s) \Gamma(- s - t)}{\Gamma (- t)} \nonumber \\
 \stackrel{s \rightarrow 0}{\longrightarrow} \frac1s - \frac1s + {\cal O}(s^0) = {\cal O}(s^0) \ .
\end{IEEEeqnarray}
Obviously, there is no s-channel pole anymore in this case. The reason is the following: The origin of the $1/s$-pole in \eqref{expr1d} is the term $y^{s-1}$ in the integrand of \eqref{expr1a}. Due to the odd power of this term (for $s \rightarrow 0$) the corresponding singularity at $y=0$ disappears when integrating over both positive and negative values of $y$. 

Another way of seeing that the $s$-channel poles disappear on the double cover is the following. The result on the double cover can be obtained from the disk level result by symmetrization, cf.\ the discussion at the end of section \ref{sec::integration}. If one starts with formula (1.2) in \cite{6pt} and symmetrizes $5 \leftrightarrow 6$ by hand, the first term on the right hand side would vanish because of the antisymmetry of $T_{56}$. Hence, also the pole in $s_5$ would vanish.

	\newpage
	\bibliographystyle{mystyle}
	\bibliography{./ref}
	\markboth{Bibliography}{Bibliography}
\end{document}